\documentclass[twocolumn,english,aps,english,floatfix,groupedaddress,superscriptaddress]{revtex4-1}
\usepackage[T1]{fontenc}
\usepackage[latin9]{inputenc}
\usepackage{color}
\usepackage{booktabs}
\usepackage{amsmath}
\usepackage{amssymb}
\usepackage{graphicx}
\usepackage{esint}
\usepackage{braket}
\usepackage{graphics}
\usepackage{times}
\usepackage{bm}
\usepackage{hyperref}
\usepackage{bpchem}

\definecolor{RoyalBlue}{cmyk}{1, 0.80, 0, 0}
\makeatletter

\providecommand{\tabularnewline}{\\}

 \@ifundefined{textcolor}{}
 {%
   \definecolor{BLACK}{gray}{0}
   \definecolor{WHITE}{gray}{1}
   \definecolor{RED}{rgb}{1,0,0}
   \definecolor{GREEN}{rgb}{0,1,0}
   \definecolor{BLUE}{rgb}{0,0,1}
   \definecolor{CYAN}{cmyk}{1,0,0,0}
   \definecolor{MAGENTA}{cmyk}{0,1,0,0}
   \definecolor{YELLOW}{cmyk}{0,0,1,0}
 }

\hypersetup{
colorlinks=true,
breaklinks=true, 
hyperfootnotes=true,
urlcolor=RoyalBlue, 
citecolor=RoyalBlue, 
linkcolor=RoyalBlue, 
pdftitle = {Reason is but choosing: On the alternatives for bath correlators and spectral densities from mixed quantum-classical simulations},
pdfauthor = {Stephanie Valleau, Alex Eisfeld, Alan Aspuru-Guzik}
}

\makeatother

\usepackage{babel}
\begin{document}

\title{On the alternatives for bath correlators and spectral densities from
mixed quantum-classical simulations}

\author{{St\'{e}phanie Valleau$^1$,
Alexander Eisfeld$^2$,
and Al\'{a}n Aspuru-Guzik$\, $}}

\email{aspuru@chemistry.harvard.edu}

\selectlanguage{english}%

\address{{Department of Chemistry and Chemical Biology, Harvard University, Cambridge, Massachusetts 02138, USA 
\\ $^2$Max Planck Institute for the Physics of Complex Systems, N\"{o}thnitzer Stra{\ss}e 38, 01187 Dresden, Germany } }
\begin{abstract}
We investigate on the procedure of extracting a ``spectral density''
from mixed QM/MM calculations and employing it in open quantum systems
models. In particular, we study the connection between the energy
gap correlation function extracted from ground state QM/MM and the
bath spectral density used as input in open quantum system approaches.
We introduce a simple model which can give intuition on when the ground
state QM/MM propagation will give the correct energy gap. We also
discuss the role of higher order correlators of the energy-gap fluctuations
which can provide useful information on the bath. Further, various
semiclassical corrections to the spectral density, are applied and
investigated. Finally, we apply our considerations to the photosynthetic
Fenna-Matthews-Olson complex. For this system, our results suggest
the use of the Harmonic prefactor for the spectral density rather
than the Standard one, which was employed in the simulations of the
system carried out to date.
\end{abstract}
\maketitle

\section{Introduction\label{sec:Introduction}}

In the study of the dynamics of large systems such as photosynthetic
complexes, reduced models, which provide information on a small set
of system degrees of freedom at the price of tracing out the rest
of the bath degrees of freedom, have become very popular. Amongst
these methods, which are open quantum systems approaches, one can
find various Quantum Master Equations \cite{Seogjoo2007,Mohseni2008,Plenio2008,Rebentrost2009b,Cao2009,Ishizaki2009_3,Ishizaki2011,Sarovar2010,abramavicius2010,Cao2010,Moix2011,Kreisbeck2011,Mazziotti2011,Ritschel2011,Rebentrost2011,Singh_Brumer2011,singh_brumer_2011,Pachon_Brumer_2012,Vlaming2011,Saikin2011,Mohseni2011,Zhu2011}
and Stochastic Schrödinger Equations \cite{RoEiWo09_058301_,DeVega_2011},
which often rely on describing the system-bath interaction through
a two-time bath correlation function or a bath spectral density. Therefore,
it is of considerable interest to obtain these quantities. While a
full quantum mechanical treatment of such large systems is out of
reach, one viable approach is to use a mixed quantum-classical approach
for the nuclear-electronic degrees of freedom. However, there is no
unique way of obtaining a bath correlation function or a bath spectral
density when resorting to quasiclassical theories. In this work, we
provide criteria that can be helpful to choose an appropriate strategy
for this task. 

As a case study, we consider the Fenna-Matthews-Olson (FMO) light-harvesting
pigment protein complex found in green sulfur bacteria. For this system,
recent efforts have been undertaken to extract the bath spectral densities
from mixed quantum-classical calculations \cite{Olbrich2011,Shim2012,Olbrich2011JCPL}.
The FMO complex has a trimeric structure, where each monomer contains,
within the protein scaffolding, eight bacteriochlorophyll (BChl) molecules,
which can transport electronic excitation energy. Up to recently,
it was thought that only seven of the BChl's actually were present
and most of the previous studies have focused on that case. Shim's
results \cite{Shim2012} which we employ in this work, are indeed
based on the case of one monomer with seven BChl molecules. Experimentally,
it has been possible to extract a spectral density for the BChl with
the lowest transition energy \cite{Wendling2000,Adolphs2006}. However,
one can expect that each BChl has a different spectral density, due
to its specific protein environment. 

One theoretical approach to obtain the spectral densities from a microscopic
description is a mixed quantum mechanics/classical mechanics (QM/MM)
model \cite{Mercer1999}. In this approach, the nuclear degrees of
freedom are treated classically and the relevant system quantities
are calculated quantum mechanically. Then, from the microscopic description,
spectral densities and correlation functions can be extracted and
employed in the reduced models. 

A specific QM/MM approach, which has become popular in recent years
in the context of photosynthetic complexes \cite{Olbrich2010,Olbrich2011,Shim2012,Damjanovic2002}
and has been employed for FMO \cite{Olbrich2011,Shim2012}, consists
in propagating the nuclei in the ground electronic state of the FMO
complex, thus the change in the classical forces due to excitation
of the BChls is ignored. The bath correlation function and spectral
densities are then extracted from the energy gap trajectories, i.e.\ 
the electronic transition energies which depend on the time dependent
nuclear configuration. This transition energy is calculated using
quantum chemistry, for example TDFT \cite{Shim2012} or semi-empirical
approaches~\cite{Olbrich2011}. One thus obtains a time dependent
energy gap two-time correlation function. Usually, a spectral density
(SD) is derived from the time correlation function, to characterize
the frequency dependent coupling of the electronic transitions to
the environmental degrees of freedom. In the previous investigations
on the FMO complex \cite{Olbrich2011JCPL,Olbrich2011,Shim2012}, the
spectral densities differ by orders of magnitude respect to each other
and also with respect to the SD extracted from experiment \cite{Wendling2000,Adolphs2006}.

In this work we revisit the data of Shim \cite{Shim2012}. We shed
light on the connection between the mixed QM/MM gap correlation function
and the open quantum system bath correlation function using a simple
model. The mixed QM/MM gap correlation function is real. However,
in general, the full quantum correlation function will have an imaginary
part. We employ different semiclassical \emph{a posteriori} corrections
to recover this part and compare the resulting spectral densities.
Much work has been carried out on these\emph{ a posteriori} semiclassical
corrections \cite{Egorov1997b,Egorov1999,Egorov1999a,Kim2002,Ramirez2004,Skinner2001},
but the question of which approximation is best remains open. Towards
answering this question, we show that a simple model of shifted harmonic
Born-Oppenheimer surfaces leads to two of the semiclassical \emph{a
posteriori} corrections, each obtained with a different phase space
probability distribution. We thus establish the link to a microscopic
picture. This model of shifted harmonic potential surfaces is of particular
interest, since the spectral density used in the open quantum system
approaches emerges from such a description. 

Finally, we will investigate whether the results of the QM/MM fulfill
the requirements needed to employ the extracted spectral density in
open quantum system methods. These methods rely on the validity of
assumptions such as linear coupling of the system to the bath and
a bath of harmonic oscillators. We attempt to investigate whether
these assumptions are valid in our case by evaluating higher order
correlators of the energy gap time traces. In addition, we compare
the spectral densities obtained at different temperatures. In most
open quantum system approaches, when the harmonic bath approximation
is employed, the spectral density is temperature independent. Thus,
one can use this invariance as a criteria for choosing which of the
applied \emph{a posteriori }corrections is most reasonable to be employed
in these methods. In particular, our findings suggest that the best
\emph{a posteriori} semiclassical approximation for FMO is the Harmonic
\cite{Frommhold1993,Berens1981A} correction rather than the Standard
\cite{Berne1970,Oxtoby1981,Litovitz1976} one, which has so far been
employed in the context of the simulation of exciton dynamics in photosynthetic
complexes \cite{Damjanovic2002,Olbrich2011,Shim2012,Olbrich2011JCPL}.
Together, these aspects provide a clearer microscopic picture of the
complex approximations involved in combining ground state QM/MM and
open quantum system approaches. 

The paper is structured as follows: we begin by introducing the general
quantum two-time correlation function in Section \ref{sec:quantum_autocorrelation_function}.
We introduce its time symmetries and its Fourier transform and subsequently
we define the spectral density. A brief summary of the general \emph{a
posteriori} semiclassical approximations to the quantum Fourier transform
of the correlator from the classical Fourier transform is given in
Section \ref{sec:semiclassical_corrections}. In Section \ref{sec:egap_ct_simple_model},
we introduce the concept of an energy gap correlation function for
two-level systems as models for molecules coupled to a bath and show
how this leads to a quantum bath correlation function and spectral
density which are consistent with the open quantum system approach.
In Section \ref{sec:Analytical-consideration-for-harmonic-surfaces},
we show that one can introduce a microscopic model which leads to
some of the same prefactors described in the general case in Section
\ref{sec:semiclassical_corrections}. Finally, we investigate the
conditions of linear system-bath coupling and harmonic bath in Section
\ref{sec:higher_order_correlators}. In particular, we evaluate high-order
multi-time correlation functions for the bath. These considerations
are applied to our specific QM/MM calculations for FMO in Section
\ref{sec:Results}. We conclude in Section \ref{sec:Conclusions}
by summarizing our findings.

\section{The quantum correlation function and the spectral density\label{sec:quantum_autocorrelation_function}}

In this section, we introduce the definition of the quantum two-time
bath correlation function. The generic Hamiltonian of a system coupled
to a bath, in the absence of external fields, can be expressed as
\begin{equation}
\hat{H}=\hat{H}_{{\rm S}}\left(\mathbf{q\mathrm{,}p}\right)+\hat{H}_{{\rm B}}\left(\mathbf{Q\mathrm{,}P}\right)+\hat{H}_{{\rm SB}}\left(\mathbf{q\mathrm{,}p},\mathbf{Q},\mathbf{P}\right),\label{eq: general_H}
\end{equation}
where $\hat{H}_{{\rm S}}$ is the system Hamiltonian, $\hat{H}_{{\rm B}}$
is the bath Hamiltonian, $\hat{H}_{{\rm SB}}$ is the system-bath
Hamiltonian. In addition, $\mathbf{\left(q\mathrm{,}p\right)}=\left(q_{j},p_{j}\right)$
and $\left(\mathbf{Q\mathrm{,}P}\right)=\left(Q_{k},P_{k}\right)$,
indicate the generalized multidimensional conjugated coordinates for
the system and the bath respectively. The indexes $j=1,...,f$ and
$k=1,...,F$ run over the system ($f$) and bath ($F$) degrees of
freedom respectively. The system-bath Hamiltonian can be written as
a function of the system, $\hat{A}$, and bath, $\hat{B},$ operators:
\begin{equation}
\hat{H}_{{\rm SB}}\left(\mathbf{q\mathrm{,}p},\mathbf{Q},\mathbf{P}\right)=\sum_{m}\hat{A}_{m}\left(\mathbf{q\mathrm{,}p}\right)\otimes\hat{B}_{m}\left(\mathbf{Q},\mathbf{P}\right).\label{eq: HSB}
\end{equation}

The influence of the bath on the system can be described by time-correlation
functions. We will mostly focus on the two-time bath correlation function
\begin{equation}
C_{nm}(t-t')={\rm tr_{B}}\{\hat{B}_{n}(t,\mathbf{Q},\mathbf{P})\hat{B}_{m}(t',\mathbf{Q},\mathbf{P})\hat{\rho}_{{\rm B}}\}.\label{eq: Correlator_quantum}
\end{equation}
Here, $\hat{B}_{m}(t,\mathbf{Q},\mathbf{P})=e^{i\hat{H}_{{\rm B}}t/\hbar}\hat{B}_{m}(\mathbf{Q},\mathbf{P})e^{-i\hat{H}_{{\rm B}}t/\hbar}$,
and 
\begin{equation}
\hat{\rho}_{{\rm B}}=\frac{e^{-\beta\hat{H}_{{\rm B}}}}{{\rm tr_{B}}\{e^{-\beta\hat{H}_{{\rm B}}}\}},\label{eq: bath_density_matrix}
\end{equation}
where $\beta=1/\left(k_{{\rm B}}T\right)$ and $T$ is the temperature.
In the following, we will be interested only in the $n=m$ correlators,
which we will indicate as $C(\tau)$ with $\tau=t-t'$, dropping the
subscript notation for simplicity. In section \ref{sec:higher_order_correlators},
we will briefly discuss higher order correlators.

The correlator defined above is in general complex and one can show,
see e.g.~\cite{Yan2005,Berne1970}, that it has the following symmetries
with respect to time,
\begin{equation}
C(-t)=C^{*}(t)=C(t-i\beta\hbar).\label{eq:symmetries}
\end{equation}

\subsection{Fourier transform of the time correlation function and symmetries
of the correlator \label{sec:fourier_transform_tdmd}}

We define $G(\omega)$, the Fourier transform of the time correlation
function 
\begin{equation}
G(\omega)\equiv\mathcal{F}[C(t)](\omega)=\int_{-\infty}^{\infty}e^{i\omega t}C(t)dt.\label{eq:Gomega}
\end{equation}
The function $G(\omega)$ is in general, temperature-dependent, real
and positive. In this work, we will refer to it as the Temperature-Dependent
Coupling Density (TDCD). 

It will be convenient to split $G(\omega)$ into a symmetric and antisymmetric
component which originate respectively from the real and imaginary
parts of $C(t)$,
\begin{eqnarray}
G(\omega) & = & G_{\mathrm{sym}}(\omega)+G_{\mathrm{asym}}(\omega),\label{eq:symandasym}\\
G_{{\rm sym}/{\rm asym}}(\omega) & = & \frac{1}{2}\left(G\left(\omega\right)\pm G\left(-\omega\right)\right).
\end{eqnarray}
In this definition, we have followed the convention of Ref.~\cite{Egorov1999}.
Note that in the literature there exist other definitions, e.g.~the
corresponding equations in Ref.~\cite{May2004}, differ by a factor
of $2$ from the ones used here %
\footnote{This means that when comparing to these definitions we need to multiply
$G_{{\rm asym}}(\omega)$ by two.%
}. The detailed-balance condition, which follows directly from the
second time symmetry in Eq.~\ref{eq:symmetries}, implies that the
overall TDCD is related to its asymmetric~\footnote{Alternatively, one can employ the symmetric part to obtain, $G(\omega)=2/(1+exp(-\beta\hbar\omega)G_{\mathrm{sym}}(\omega)$$=\left(1-{\rm coth\left(\beta\hbar\omega/2\right)}\right)G_{\mathrm{sy}m}(\omega).$%
} part by
\begin{eqnarray}
G(\omega) & = & \frac{2}{1-e^{-\beta\hbar\omega}}G_{\mathrm{asym}}(\omega)\label{eq: Gw_asym_detailed _balance}\\
 & = & \left(1+{\rm coth\left(\beta\hbar\omega/2\right)}\right)G_{\mathrm{asym}}(\omega).
\end{eqnarray}
It will be convenient to abbreviate $G_{\mathrm{asym}}(\omega)$ by
defining
\[
J(\omega)\equiv G_{\mathrm{asym}}(\omega).
\]

Using Eq.~\ref{eq: Gw_asym_detailed _balance} and the definition
of $G(\omega)$, Eq.~\ref{eq:Gomega}, one can express the correlation
function as a function of $J(\omega)$,
\begin{equation}
C(t)=\frac{1}{2\pi}\int_{-\infty}^{\infty}d\omega\, e^{-i\omega t}\Big(\coth(\beta\hbar\omega/2)+1\Big)\, J(\omega).\label{eq:relation_between_CandJ}
\end{equation}

\subsection{The spectral density\label{sec:spectral_density}}

Another quantity which is often of interest is the so-called ``spectral
density''. The spectral density describes the frequency dependent
coupling of the system to the bath. There are different definitions
of spectral density in the literature (for example $J(\omega)$ is
sometimes refered to as the spectral density). We follow the convention
of defining the spectral density as a positive frequency function
\begin{equation}
j(\omega)=\Theta(\omega)\, J(\omega)/\pi.\label{eq: sd_asym}
\end{equation}
Here $\Theta(\omega)$ is the Heavyside function, which is one for
positive arguments and zero for negative ones. The scaling by $\pi$
has been introduced for later convenience. Note that 
\begin{equation}
J(\omega)=\pi\cdot\big(j\left(\omega\right)-j\left(-\omega\right)\big).\label{eq: Gasym_sd}
\end{equation}

\subsection{General semiclassical a posteriori approximations \label{sec:semiclassical_corrections}}

For systems of more than a few degrees of freedom, and in general,
it is difficult to calculate the exact correlation function, and therefore
its Fourier transform, by using a fully quantum mechanical treatment.
However, using classical mechanics one can obtain its classical counterpart
with much less effort. Therefore, it is common to attempt to construct
the quantum spectral density from the classical one. 

We define the fully-classical correlation function as the classical
$\hbar\rightarrow0$ limit of Eq. \ref{eq: Correlator_quantum},
\begin{equation}
\begin{aligned}C^{\mathrm{cl}}(t) & =\int d\mathbf{Q}d\mathbf{P}\, B\!\left(t,\mathbf{Q},\mathbf{P}\right)\, B\!\left(0,\mathbf{Q},\mathbf{P}\right)\mathcal{\, W}\!\left(\mathbf{Q},\mathbf{P}\right).\end{aligned}
\label{eq:classical_ct}
\end{equation}
Here $\mathcal{W}\left(\mathbf{Q},\mathbf{P}\right)$ is the classical
bath phase-space density, defined as

\begin{equation}
\mathcal{W}\left(\mathbf{Q},\mathbf{P}\right)=\frac{e^{-\beta H_{{\rm {\rm B}}}\left(\mathbf{Q},\mathbf{P}\right)}}{\int d\mathbf{Q}d\mathbf{P}e^{-\beta H_{{\rm B}}\left(\mathbf{Q},\mathbf{P}\right)}}.\label{eq: rho_b_classical}
\end{equation}
and the quantum bath operators $\hat{B}$ in Eq.~\ref{eq: Correlator_quantum}
have been substituted by classical functions of the phase space variables
$B\left(t,\mathbf{Q},\mathbf{P}\right)$. 

The classical TDCD is defined as 
\begin{equation}
G^{\mathrm{cl}}(\omega)=\mathcal{F}[C^{{\rm cl}}(t)](\omega)=\int_{-\infty}^{\infty}e^{i\omega t}C^{{\rm cl}}(t)dt.\label{eq: classical_Gt}
\end{equation}
Note that $C^{\mathrm{cl}}(t)$ is a real and symmetric function in
contrast to its quantum counterpart. This is also the case in the
mixed QM/MM simulations employed for FMO \cite{Shim2012,Olbrich2011}.
The QM/MM correlation function obtained is real and no information
about the important imaginary part of the quantum correlator is available
\emph{a priori}. 

It is now desirable to be able re-construct, at least partially, the
exact quantum spectral density from the classical one, through a simple
description. Ideally, such a correction should be applied \emph{a
posteriori }and should not require extensive additional computation.
Much work has been carried out in this direction, see e.g.~\cite{Egorov1997b,Egorov1999,Egorov1999a,Kim2002,Ramirez2004,Skinner2001}.
As described in Ref.~\cite{Egorov1999}, one can define various semiclassical
approximations to the full quantum mechanical $G(\omega)$ starting
from its classical counterpart $G^{\mathrm{cl}}(\omega)$. We report
each of these approximations in Table \ref{tab:G_J_prefactor}, second
column.

These corrections all originate from expansions in $\hbar$ and use
of the symmetry properties of the two-time correlation function and
its Fourier transform. Note that if one expands the quantum correlator
$C(t)$ in powers of $\hbar$, the first term is real and symmetric
and corresponds to $C^{\mathrm{cl}}(t)$. 
\begin{table*}
\caption{\label{tab:G_J_prefactor} Column two: Various expressions for obtaining
a semiclassical temperature-dependent coupling density TDCD $G(\omega)$
from the classical $G^{{\rm cl}}(\omega)$ as discussed in, e.g. \cite{Egorov1999}.
Column three: Expressions for obtaining the semiclassical asymmetric
TDCD $J(\omega)$ from the classical $G^{{\rm cl}}(\omega)$. These
follow from the expressions in column two and from detailed balance
(Eq. \ref{eq: Gw_asym_detailed _balance}).}

\centering{}%
\begin{tabular}{lll}
\toprule 
\emph{Method} & \emph{Expression for $G(\omega)$} & \emph{Expression for $J(\omega)=G_{{\rm asym}}(\omega)$}\tabularnewline
\midrule
Standard \cite{Berne1970,Litovitz1976,Oxtoby1981} & $G^{{\rm std}}(\omega)=\mathrm{\frac{2}{1+e^{-\beta\hbar\omega}}\mathit{G}^{{\rm cl}}(\omega)}$ & $J^{{\rm std}}(\omega)=\mathrm{tanh\left(\frac{\beta\hbar\omega}{2}\right)\mathit{G}^{cl}(\omega)}$\tabularnewline
Harmonic \cite{Frommhold1993,Berne1970,Berens1981A} & $G^{{\rm harm}}(\omega)=\frac{\beta\hbar\omega}{1-e^{-\beta\hbar\omega}}G^{{\rm cl}}(\omega)$ & $J^{{\rm harm}}(\omega)=\frac{\beta\hbar\omega}{2}G^{{\rm cl}}(\omega)$\tabularnewline
Schofield \cite{Schofield1960} & $G^{{\rm scho}}(\omega)=e^{\beta\hbar\omega/2}G^{{\rm cl}}(\omega)$ & $J^{{\rm scho}}(\omega)=\mathrm{sinh}(\frac{\beta\hbar\omega}{2})G^{{\rm cl}}(\omega)$\tabularnewline
Egelstaff \cite{Egelstaff1962} & $G^{{\rm egel}}(\omega)=e^{\beta\hbar\omega/2}\int_{-\infty}^{\infty}e^{i\omega t}C^{{\rm cl}}\left(\sqrt{t^{2}+(\beta\hbar/2)^{2}}\right)dt\;\;$ & $J^{{\rm egel}}(\omega)={\rm sinh}(\frac{\beta\hbar\omega}{2})\mathcal{F}\left[G^{{\rm cl}}(\sqrt{t^{2}+(\beta\hbar/2)^{2}})\right](\omega)$\tabularnewline
Schofield-Harmonic \cite{Egorov1999}\ \ \  & $G^{{\rm s-h}}(\omega)=e^{\beta\hbar\omega/4}\sqrt{\frac{\beta\hbar\omega}{1-e^{-\beta\hbar\omega}}}G^{{\rm cl}}(\omega)$ & $J^{{\rm s-h}}(\omega)=\sqrt{\frac{\beta\hbar\omega}{2}\mathrm{sinh}(\frac{\beta\hbar\omega}{2})}G^{{\rm cl}}(\omega)$\tabularnewline
\bottomrule
\end{tabular}
\end{table*}
The assumption that $C(t)=C^{{\rm cl}}(t)$, which leads to the standard
approximation, is in general not correct. In fact, since both of the
correlation functions are obtained after thermal averaging, we see
that they must differ at least by their respective partition functions. 

At low frequencies, $\omega\beta\hbar\equiv\omega_{\mathsf{b}}<1$
(i.e.~$\hbar\omega<k_{B}T$) all approximations give nearly identical
results and give the same value for $\omega_{\mathsf{b}}=0$.

The various approximations for $J(\omega)$, and thus for the spectral
density, can straightforwardly be derived from those of $G(\omega)$
by using Eq.~\ref{eq: Gw_asym_detailed _balance}. The resulting
expressions are reported in column three of Table \ref{tab:G_J_prefactor}
and the prefactors follow the same trend as those for $G(\omega)$
as a function of frequency.

Now, given all the functional forms described above, the question
is how to choose the most appropriate one. For the FMO complex, it
is unclear at first sight which one would be the best. In Section
\ref{sec:Analytical-consideration-for-harmonic-surfaces}, we will
investigate a model to elucidate the origin of these prefactors. This
will help to discriminate between these corrections. In Section \ref{sec:Results},
we will apply all of the corrections listed in Table \ref{tab:G_J_prefactor}
to our energy gap traces and discuss the differences between each
approach.

\section{Energy gap correlation function for a simple model \label{sec:egap_ct_simple_model}}

In the mixed QM/MM calculations for photosynthetic systems \cite{Damjanovic2002,Olbrich2011,Shim2012},
the nuclear trajectories are propagated in the electronic ground state
using MD with short time steps. For a set of longer times steps within
these trajectories, the electronic transition energies of the BChl
molecules are computed using an electronic structure calculation method.
Because it is computationally costly to calculate the electronic states
for the full set of seven/eight coupled BChls simultaneously \cite{Olbrich2011},
the system was divided into seven/eight subsystems for which the electronic
states were calculated separately. Thus, in these calculations no
excited state interactions are included explicitly %
\footnote{Approximations such as the transition density cube \cite{Krueger_Scholes_1998}
can be employed to obtain couplings between the local excited states.
More sophisticated models that include polarization effects \cite{Scholes_curutchet_2007,Neugebauer_Curutchet2010}
can also be employed for this purpose.%
}. The Hamiltonian of the coupled BChls is then written as $H=\sum_{n=1}^{N}H_{n}+\sum_{n<m}V_{nm}$
where $H_{n}$ denotes the Hamiltonian of BChl $n$ and $V_{nm}$is
the Coloumb (transition dipole-dipole) interaction between them. To
establish a connection to the open quantum system approach, each BChl
is treated as an electronic two level system. These two-level systems
and the electronic interaction between them are taken to be the system
part. The coupling to internal nuclear degrees of freedom and the
surrounding protein will then lead to fluctuations of these quantities
in time (for more details see e.g.~\cite{Damjanovic2002}). From
the time dependence of the transition energy between electronic ground
and excited state for each BChl, a classical ground-excited state
energy-gap correlation function can be obtained. In turn, spectral
densities can be extracted from the energy-gap correlation functions.

The gap correlation function, as obtained from the MD simulations,
is a quantity which up to the previous section, has not been connected
to the open quantum system approach described in Sec.~\ref{sec:quantum_autocorrelation_function}.
In this Section, we will explore a simple model with Born-Oppenheimer
(BO) surfaces which can clarify the connection.

\subsection{Quantum correlation function and energy gap correlation function
for a molecule \label{sub:Quantum-autocorrelation-function-molecule}}

Lets us begin by considering a single molecule (BChl) treated in the
Born-Oppenheimer approximation. The molecule is modeled as a two-level
system with an electronic adiabatic ground $\left|g\right\rangle $
and excited $\left|e\right\rangle $ state. We can think of the BO-surfaces
as having the dependence of the environment (protein and other BChls)
already included, ignoring however the resonant dipole-dipole interaction.
The approximation of two levels is reasonable in the limit where the
next excited state is very far in energy space from the first. Usually,
non-adiabatic couplings can be also neglected, as chosen in our calculations. 

Given this model, we investigate how the general correlation function,
Eq.~\ref{eq: Correlator_quantum}, is related to the energy gap correlation
function. 

We write the full Hamiltonian formally as 
\begin{equation}
\hat{H}=\hat{H}_{{\rm g}}(\mathbf{Q}\mathbf{\mathrm{,}P})\left|g\right\rangle \!\!\left\langle g\right|+\hat{H}_{{\rm e}}(\mathbf{Q\mathrm{,}P})\left|e\right\rangle \!\!\left\langle e\right|,\label{eq:2LH}
\end{equation}
where $\hat{H}_{{\rm g}}(\mathbf{Q})$ and $\hat{H}_{{\rm e}}(\mathbf{Q})$
are the nuclear Hamiltonians for the ground and excited state in the
BO approximation. In mass scaled coordinates $\left(Q_{j}=\sqrt{m_{j}}q_{j}\,;\, P_{j}=p_{j}/\sqrt{m_{j}}\right)$,
the Hamiltonians can be expressed as $\begin{aligned}\hat{H}_{{\rm g}}(\mathbf{Q\mathbf{\mathrm{,}P}}) & =\sum_{j=1}^{F}P_{j}/2+V_{{\rm g}}(\mathbf{Q})\end{aligned}
$ and $\hat{H}_{{\rm e}}(\mathbf{Q}\mathbf{\mathrm{,}P})=\hat{H}_{{\rm g}}(\mathbf{Q\mathbf{\mathrm{,}P}})+\hat{\Delta}_{{\rm eg}}(\mathbf{Q})$,
where $V_{{\rm g}}(\mathbf{Q})$ denotes the grounds state potential
energy surface. For later purpose, we have expressed the excited state
nuclear Hamiltonian with respect to the ground state potential by
introducing the energy gap operator,
\begin{eqnarray}
\hat{\Delta}_{{\rm eg}}(\mathbf{Q}) & = & \hat{H}_{{\rm e}}(\mathbf{Q\mathbf{\mathrm{,}P}})-\hat{H}_{{\rm g}}(\mathbf{Q,{\bf P}})\label{eq:gap_twolevel_BO}\\
 & = & \hbar\omega_{{\rm eg}}+\lambda_{0}+V_{{\rm e}}(\mathbf{Q})-V_{{\rm g}}(\mathbf{Q}).\nonumber 
\end{eqnarray}
This operator quantifies the energy difference between the excited
state and the ground state surface. A coordinate independent constant
energy difference $\hbar\omega_{eg}+\lambda_{0}$ has been explicitly
written down, so that the remaining part $V_{{\rm e}}(\mathbf{Q})-V_{{\rm g}}(\mathbf{Q})$
does not contain any coordinate independent contributions. This division
and the meaning of $\hbar\omega_{eg}$ and $\lambda_{0}$ will become
clear in Section \ref{sub:Quantum-autocorrelation-function-harmonic}.

The total Hamiltonian can be rewritten as 
\begin{eqnarray}
\hat{H} & = & \hat{H}_{{\rm g}}\cdot\hat{\mathbb{I}}+\left(\hbar\omega_{{\rm eg}}+\lambda_{0}\right)\left|e\right\rangle \!\!\left\langle e\right|+\hat{\Delta}\left|e\right\rangle \!\!\left\langle e\right|,\label{eq: newH_2LS-1}
\end{eqnarray}
where we have defined the \emph{reduced} gap operator $\hat{\Delta}\equiv\hat{\Delta}_{{\rm eg}}-\hbar\omega_{{\rm eg}}-\lambda_{0}$.

To establish a connection to the open quantum system model, as presented
in Sec. \ref{sec:quantum_autocorrelation_function}, we choose 
\begin{eqnarray}
\hat{H}_{{\rm B}} & = & \hat{H}_{{\rm g}}(\mathbf{Q\mathbf{\mathrm{,}P}})\\
\hat{H}_{{\rm SB}} & = & \hat{\Delta}(\mathbf{Q})\left|e\right\rangle \!\!\left\langle e\right|\\
\hat{H}_{{\rm S}} & = & \left(\hbar\omega_{{\rm eg}}+\lambda_{0}\right)\left|e\right\rangle \!\!\left\langle e\right|,
\end{eqnarray}
where we have set the energy of the electronic ground state $\ket{g}$
to zero. From the form of $\hat{H}_{{\rm SB}}$ we identify the system
operator $\hat{A}_{{\rm e}}=\left|e\right\rangle \!\!\left\langle e\right|$
and the bath operator $\hat{B}=\hat{\Delta}_{{\rm eg}}(\mathbf{Q})$.
We can now define the usual bath correlation function as 
\begin{equation}
C(t)={\rm tr_{B}}\left\{ \hat{\Delta}(t)\hat{\Delta}(0)\hat{\rho}_{{\rm B}}\right\} ,\label{eq:gap_autocorrelation_function-1}
\end{equation}
where we have dropped the dependence on bath coordinates in the notation
for simplicity. $\hat{\Delta}$ can be thought of as a ``gap'' operator,
that is, as a measure of the energy difference between the ground
and excited state at a given nuclear configuration. From now on we
will indicate reduced gap correlation functions as 
\begin{equation}
\alpha(t)\equiv{\rm tr_{B}}\left\{ \hat{\Delta}(t)\hat{\Delta}(0)\hat{\rho}_{{\rm B}}\right\} ,\label{eq:reduced_gap_autocorr_alpha}
\end{equation}
to distinguish them from the general bath correlation function $C(t)$.
Eq. \ref{eq:reduced_gap_autocorr_alpha} corresponds to the full quantum
gap correlation function that one would obtain, e.g., from a quantum
simulation on the FMO complex, considering only two electronic levels
per molecule and after including the protein environment.

\subsection{Quantum correlation function and energy gap correlation function
for harmonic surfaces \label{sub:Quantum-autocorrelation-function-harmonic}}

While the approach outlined in the previous section is applicable
to arbitrary potential surfaces, in most of the open quantum system
approaches used to describe the FMO complex, the bath is taken as
an (infinite) set of harmonic oscillators for the environment of each
BChl. Each oscillator coordinate is then assumed to be linearly coupled
to the electronic excitation of the BChls, i.e.\ $H_{{\rm SB}}=|e\rangle\!\langle e|\otimes\sum_{j}\tilde{\kappa_{j}}Q_{j}$
where $\tilde{\kappa_{j}}$ is a coupling constant. 

\begin{figure}
\centering{}\includegraphics[width=0.75\columnwidth]{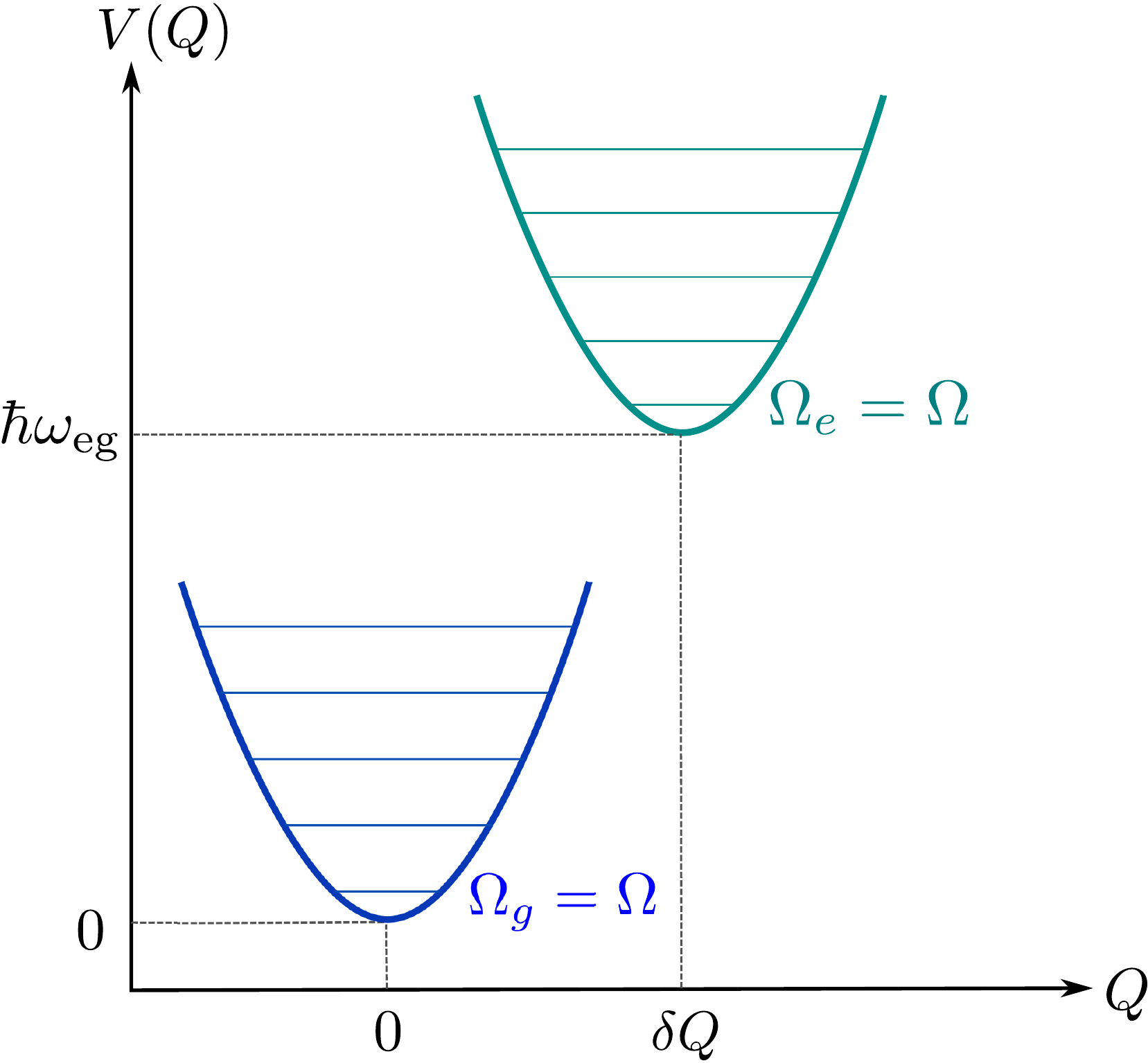}\caption{Shifted identical harmonic Born-Oppenheimer surfaces, $\Omega$ is
the frequency of each harmonic potential and $\delta Q$ is the coordinate
shift between the minima of the ground and excited state potentials.
This model is the one employed in Sec.~\ref{sub:Quantum-autocorrelation-function-harmonic}
to derive classical and semiclassical expressions for the Fourier
transform of the bath correlation function $G(\omega)$ and for the
spectral density. \label{fig:Shifted-two Harmonic}}
\end{figure}

To establish the connection between the reduced gap operator and this
system-bath interaction, we now consider identical shifted harmonic
potential surfaces, as sketched in Fig. \ref{fig:Shifted-two Harmonic}.
The nuclear Hamiltonians defined in the general case in the previous
Section \ref{sub:Quantum-autocorrelation-function-molecule} become,
$\hat{H}_{{\rm g}}(\mathbf{Q\mathbf{\mathrm{,}P}})=\frac{1}{2}\sum_{j=1}^{F}(P_{j}^{2}+\Omega_{j}^{2}Q_{j}^{2})$
and $\hat{H}_{{\rm e}}(\mathbf{Q}\mathbf{\mathrm{,}P})=\hbar\omega_{{\rm eg}}+\frac{1}{2}\sum_{j}(P_{j}^{2}+\Omega_{j}^{2}(Q_{j}-\delta Q_{j})^{2})$
where $\Omega_{j}$ is the frequency of the $j$-th oscillator. This
model for a finite small number of vibrational modes of the chromophores,
has been successfully employed to describe the optical properties
of molecular aggregates \cite{Sp09_4267_,ScFi84_269_,GuZuCh09_154302_,RoEiDv11_054907_}.
These Hamiltonians can be rewritten as function of $a_{j}^{\dagger}$
and $a_{j}$, the ground state bosonic creation and annihilation operators
which are related to the conjugated coordinates by $Q_{j}=\sqrt{\hbar/\left(2\Omega_{j}\right)}(a_{j}^{\dagger}+a_{j})$
and $P_{j}=i\sqrt{\hbar\Omega_{j}/2}(a_{j}^{\dagger}-a_{j})$. One
obtains
\begin{equation}
\begin{aligned}\hat{H}_{{\rm g}} & =\sum_{j}\hbar\Omega_{j}a_{j}^{\dagger}a_{j}\\
\hat{H}_{{\rm e}} & =\hat{H}_{{\rm g}}+\hbar\omega_{{\rm eg}}+\lambda_{0}-\sum_{j}\kappa_{j}(a_{j}^{\dagger}+a_{j}).
\end{aligned}
\end{equation}
 Here, the constant shift $\lambda_{0}$, previously introduced in
Eq.~\ref{eq:gap_twolevel_BO}, corresponds to the frequently empoyed
reorganization energy $\lambda_{0}\equiv\lambda_{R}=\sum_{j}\frac{1}{2}\Omega_{j}^{2}\delta Q^{2}=\sum_{j}\hbar\Omega_{j}X_{j}$.
We have also introduced the so-called Huang-Rhys factor \cite{MeOs95__}:
$X_{j}=\frac{\Omega_{j}}{2\hbar}\delta Q_{j}^{2}$ and a (frequency
dependent) coupling constant $\kappa_{j}=\hbar\Omega_{j}\sqrt{X_{j}}$.
Note that the total Hamiltonian is now in the standard form of an
open quantum system model, as in Eq.~\ref{eq: general_H}, with the
relevant quantities given in Table~\ref{tab:Expression_HO_BO}. In
particular the reduced energy gap operator is given by
\begin{equation}
\hat{\Delta}=-\sum_{j}\kappa_{j}(a_{j}^{\dagger}+a_{j}).\label{eq:Red_gap_creation}
\end{equation}

\begin{table*}
\caption{Expressions of the system bath quantities for the case of two Born-Oppenheimer
harmonic surfaces as sketched in Fig. \ref{fig:Shifted-two Harmonic}.\label{tab:Expression_HO_BO}}

\centering{}%
\begin{tabular}{ll}
\toprule 
\emph{Quantity} & \emph{Expression}\tabularnewline
\midrule
System Hamiltonian & $\hat{H}_{{\rm S}}=\left(\hbar\omega_{{\rm eg}}+\lambda_{R}\right)\left|e\right\rangle \!\!\left\langle e\right|$ \tabularnewline
System-bath Hamiltonian & $\hat{H}_{{\rm SB}}=\left|e\right\rangle \!\!\left\langle e\right|\hat{\Delta}(\mathbf{Q})$\tabularnewline
Bath Hamiltonian & $\hat{H}_{{\rm B}}=\hat{H}_{{\rm g}}=\sum_{j}\hbar\Omega_{j}a_{j}^{\dagger}a_{j}$\tabularnewline
Energy gap operator & $\hat{\Delta}_{{\rm eg}}(\mathbf{Q})=\hbar\omega_{{\rm eg}}+\lambda_{{\rm R}}-\sum_{j}\sqrt{2\hbar\Omega_{j}^{3}X_{j}}Q_{j}$ \tabularnewline
Reduced energy gap operator & $\hat{\Delta}(\mathbf{Q})=\hat{\Delta}_{{\rm eg}}(\mathbf{Q})-(\hbar\omega_{{\rm eg}}+\lambda_{{\rm R}})$\tabularnewline
Reorganization energy & $\lambda_{0}=\lambda_{{\rm R}}=\sum_{j}\frac{1}{2}\Omega_{j}^{2}\delta Q_{j}^{2}$\tabularnewline
Coupling constant & $\kappa_{j}=\hbar\Omega_{j}\sqrt{X_{j}}$\tabularnewline
Huang-Rhys factor & $X_{j}=\Omega_{j}\delta Q_{j}^{2}/\left(2\hbar\right)$\tabularnewline
Unitless constant & $\zeta_{j}=\hbar\Omega_{j}/\left(k_{B}T\right)$\tabularnewline
\bottomrule
\end{tabular}
\end{table*}

From this expression of the energy gap operator one obtains the quantum
two-time bath correlation function (see e.g.~\cite{May2004}) 
\begin{equation}
\alpha(t)=\int_{0}^{\infty}\!\!\! j(\omega)\left[\mathrm{coth}\left(\frac{\hbar\omega\beta}{2}\right)\mathrm{cos}\left(\omega t\right)-i\mathrm{sin}\left(\omega t\right)\right]d\omega\label{eq:harmonic_2point_corr}
\end{equation}
with the temperature independent spectral density
\begin{align}
j(\omega)= & \sum_{j=1}^{F}\kappa_{j}^{2}\,\delta(\omega-\Omega_{j}).\label{eq:discrete_j_harmonic}
\end{align}
Note that from the definition Eq.~\ref{eq: Gasym_sd} we have $J(\omega)=\pi\left(j(\omega)-j(-\omega)\right)=\pi\sum_{j=1}^{F}\kappa_{j}^{2}\,\big[\delta(\omega-\Omega_{j})-\delta(\omega+\Omega_{j})\big]$,
which is also temperature independent. 

To establish a connection to the classical correlator, which is real
and symmetric, we note that $j(\omega)$ can be obtained from the
real part of $\alpha(t)$ via
\begin{equation}
j(\omega)=\frac{2}{\pi}\mathrm{tanh}\!\!\left(\frac{\hbar\omega\beta}{2}\right)\int_{0}^{\infty}{\rm Re}\{\alpha(t)\}\mathrm{cos}\left(\omega t\right)dt.\label{eq:jwfromrealCt_cosineTRansf-1}
\end{equation}
When using this expression to obtain the spectral density from QM/MM
simulations one often assumes that $C^{{\rm cl}}(t)\approx{\rm Re}\{\alpha(t)\}$,
following the Standard approximation. Then, after a Fourier transform
and use of symmetry relations for $G(\omega)$ one finds the following
expression,
\begin{eqnarray}
j(\omega) & = & \frac{1}{\pi}\tanh\!\!\left(\frac{\hbar\omega\beta}{2}\right)G^{{\rm cl}}(\omega).\label{eq:jwfromrealCt_cosineTRansf}
\end{eqnarray}
This is the expression (up to the constant prefactor $1/\hbar$) used
in Refs.~\cite{Shim2012,Damjanovic2002}, to obtain spectral densities.

\section{Classical and semiclassical limits of the correlators and spectral
densities for harmonic surfaces \label{sec:Analytical-consideration-for-harmonic-surfaces}}

As outlined in the previous section, the harmonic model allows for
a simple analytic solution in the quantum mechanical case. Now we
will show that the system also has a solution in the classical case.
In particular, in this section, we will introduce a model to construct
exact relations between the classical gap-correlation and the quantum
one. To this end, we will consider classical dynamics in the ground
state BO potentials within an initial value representation of the
initial state which is consistent with the mixed QM/MM approach. For
each initial value, we calculate a trajectory and the corresponding
reduced classical energy gap between the two surfaces, i.e. $\Delta(\mathbf{Q}(t),\mathbf{P}(t))$.
We then average over many trajectories.

\subsection{Classical equations of motion\label{sub:Classical-Equations-of-motion}}

The classical equation of motion of the $j$-th harmonic bath coordinate
is $\ddot{Q_{j}}+\Omega_{j}^{2}Q_{j}=0$. Solving this differential
equation with the initial condition $\left(Q_{j0},P_{j0}\right)=\left(Q_{j}(t=0),P_{j}(t=0)\right)$
yields the time dependent coordinate trajectories 
\begin{eqnarray}
Q_{j}(t) & = & Q_{j}(t;Q_{j0},P_{j0})\nonumber \\
 & = & Q_{j0}\cos(\Omega_{j}t)+\frac{P_{j0}}{\Omega_{j}}\sin(\Omega_{j}t).\label{eq:Q(t)_class}
\end{eqnarray}
For each trajectory, the energy gap is then given by 
\begin{align}
\Delta(t) & =\Delta(t;Q_{j0},P_{j0})\nonumber \\
 & =-\sum_{j}(\Omega_{j}^{2}\delta Q_{j})\, Q_{j}(t;Q_{j0},P_{j0})\label{eq:Delta(t)harmonic}
\end{align}
where the parametric dependence of $Q_{j}$ and $\Delta$ on the initial
conditions $\left(Q_{j0},P_{j0}\right)$ has been explicitly indicated.

\subsection{Energy gap correlator\label{sub:Energy-gap-correlator}}

The evaluation of the reduced gap correlation function, Eq.~\ref{eq:reduced_gap_autocorr_alpha},
in the classical limit, results in the following expression
\begin{eqnarray}
\alpha(t) & = & \sum_{jk}\int d\mathbf{P}_{0}d\mathbf{Q}_{0}\mathcal{W}(\mathbf{Q}_{0},\mathbf{P}_{0})\times\nonumber \\
 &  & \Delta(t;Q_{j0},P_{j0})\Delta(0;Q_{k0},P_{j0})\label{eq:classical_alpha}
\end{eqnarray}
where $\mathcal{W}(\mathbf{Q}_{0},\mathbf{P}_{0})$ is the initial
distribution and $d\mathbf{P}_{0}d\mathbf{Q}_{0}$ denotes the set
of all coordinates, i.e.\ $d\mathbf{Q_{0}}=dQ_{10}\cdots dQ_{M0}$.
For harmonic potential surfaces, Eq~\ref{eq:classical_ct}, is time-evolved
following Eq.~(\ref{eq:Delta(t)harmonic}). In this Section, we will
investigate two different choices for the initial distribution, namely
a Boltzmann distribution, as in Ref.~\ref{eq: classical_Gt}, and
a Wigner distribution which resembles the quantum thermal state. We
will refer to the two cases as the classical limit and the semi-classical
limit, respectively.

\subsection{Classical and semiclassical correlation functions\label{sub:Classical-and-Semiclassical-autocorr}}

\subsubsection{Classical limit}

To obtain the classical limit of the correlator, we choose the Boltzmann
distribution for the initial coordinates which corresponds to a purely
classical thermal state. The distribution is defined as follows
\begin{equation}
\mathcal{W}^{\mathrm{boltz}}(\mathbf{Q}_{0},\mathbf{P}_{0})=\prod_{j}\,\mathcal{W}_{j}^{\mathrm{boltz}}(Q_{j0},P_{j0}),
\end{equation}
with $\mathcal{W}_{j}^{\mathrm{boltz}}(Q_{j0},P_{j0})=\frac{\beta\Omega_{j}}{2\pi}e^{-\frac{\beta}{2}(P_{j0}^{2}+\Omega_{j}^{2}Q_{j0}^{2})},$
and it is normalized to one, i.e.,$\int dP_{j0}dQ_{j0}\mathcal{W}_{j}^{\mathrm{boltz}}(Q_{j0},P_{j0})=1$.
Note that $(\Omega_{j}^{2}\delta Q_{j})^{2}=2\hbar X_{j}\Omega_{j}^{3}$.
Using Eq.~\ref{eq:classical_alpha} and the Boltzmann distribution
for initial positions and momenta, we obtain, 
\begin{align}
\alpha_{\mathrm{boltz}}(t)= & \sum_{j}(\hbar\Omega_{j})^{2}X_{j}\cos\!\left(\Omega_{j}t\right)\ \Big(\frac{2}{\zeta_{j}}\Big).\label{eq:alpha_boltzmann}
\end{align}
 Here we have introduced the abbreviation $\zeta_{j}\equiv\hbar\Omega_{j}/\left(k_{B}T\right)$.

\subsubsection{Semiclassical limit}

In order to obtain the semiclassical limit, we take the quantum Wigner
distribution for the initial coordinates and use it in Eq. \ref{eq:classical_alpha}.
The Wigner distribution is given by 
\begin{equation}
\mathcal{W}^{\mathrm{wig}}(\mathbf{Q}_{0},\mathbf{P}_{0})=\prod_{j}\,\mathcal{W}_{j}^{\mathrm{wig}}(Q_{j0},P_{j0}),
\end{equation}
 where we have used the compact notation
\[
\begin{aligned}\mathcal{W}_{j}^{\mathrm{wig}}(Q_{j0},P_{j0}) & \equiv2\tanh\!\!\left(\frac{\zeta_{j}}{2}\right)e^{-\tanh(\zeta_{j}/2)\Big(\frac{\Omega_{j}}{\hbar}Q_{j0}^{2}+\frac{1}{\hbar\Omega_{j}}P_{j0}^{2}\Big)}.\end{aligned}
\]
The normalization of the Wigner distribution is chosen such that $\int\frac{dP_{j0}dQ_{j0}}{2\pi\hbar}\mathcal{W}_{j}^{\mathrm{wig}}(Q_{j0},P_{j0})=1$.
The resulting expression of the energy gap correlation function is
\begin{align}
\alpha_{\mathrm{wig}}(t)= & \sum_{j}(\hbar\Omega_{j})^{2}X_{j}\cos(\Omega_{j}t)\ {\rm coth\!\!\left(\frac{\zeta_{j}}{2}\right)}.\label{eq:alpha_wigner}
\end{align}

\subsection{Classical and semiclassical spectral densities\label{sub:Semiclassical-and-classical-sd}}

After a Fourier transform of the classical correlators in Eq.~\ref{eq:alpha_boltzmann}
and \ref{eq:alpha_wigner} we obtain, for the Boltzmann distribution
\begin{equation}
G_{{\rm boltz}}(\omega)=\pi\sum_{j}\left(\frac{2k_{{\rm B}}T}{\hbar\omega}\right)\kappa_{j}^{2}\left(\delta\left(\omega-\Omega_{j}\right)+\delta\left(\omega+\Omega_{j}\right)\right),
\end{equation}
and for the Wigner distribution
\begin{equation}
G_{{\rm wig}}(\omega)=\pi\sum_{j}{\rm coth\!\!}\left(\frac{\hbar\omega}{2k_{{\rm B}}T}\right)\kappa_{j}^{2}\left(\delta\left(\omega-\Omega_{j}\right)+\delta\left(\omega+\Omega_{j}\right)\right).\label{eq:}
\end{equation}
Here $\kappa_{j}=\hbar\Omega_{j}\sqrt{X_{j}}$ as in Tab. \ref{tab:Expression_HO_BO}.
Now, using Eq.~\ref{eq:discrete_j_harmonic} for the spectral density
$j(\omega)$ in the quantum case, and using $J(\omega)=\pi(j(\omega)-j(-\omega))$
we can write
\begin{eqnarray}
G_{{\rm boltz}}(\omega) & = & \frac{2k_{{\rm B}}T}{\hbar\omega}J(\omega)\label{eq: gboltzjw}\\
G_{{\rm wig}}(\omega) & = & \coth\!\!\left(\frac{\hbar\omega}{2k_{{\rm B}}T}\right)J(\omega).\label{eq: gwigjw}
\end{eqnarray}
 By inverting these equations the exact quantum $J(\omega)$ can be
expressed in terms of the classical $G_{{\rm boltz}}(\omega)=\int_{-\infty}^{\infty}e^{i\omega t}\alpha_{\mathrm{boltz}}(t)dt$
or the semiclassical $G_{{\rm wig}}(\omega)=\int_{-\infty}^{\infty}e^{i\omega t}\alpha_{\mathrm{wig}}(t)dt$

\begin{eqnarray}
J_{{\rm boltz}}(\omega) & = & \frac{\hbar\omega}{2k_{B}T}G_{{\rm boltz}}(\omega)\label{eq:Jw_wig_boltz}\\ 
J_{{\rm wig}}(\omega) & = & \tanh\!\!\left(\frac{\hbar\omega}{2k_{B}T}\right)G_{{\rm wig}}(\omega).
\end{eqnarray}

We see that in our harmonic model the semiclassical Wigner distribution
yields the same prefactor as for the Standard approximation described
in Section \ref{sec:semiclassical_corrections} while the Boltzmann
distribution gives the same prefactor as the Harmonic approximation,
also described in Section \ref{sec:semiclassical_corrections}.

\section{Models for system-bath coupling - higher order correlators\label{sec:higher_order_correlators}}

As discussed in the introduction, there has been a lot of interest
in modeling the exciton dynamics of the FMO complex using open quantum
system approaches. These usually require as input a bath two-time
correlation function or (equivalently) a spectral density and they
rely on the assumption of linear coupling to the bath and on a bath
described by harmonic oscillators %
\footnote{Note that there are two common approximations for which the information
on the system bath coupling is entirely described by the two-time
bath correlation function, namely linear response theory and second
order perturbation theory in system bath coupling. %
}.

In the previous Section~\ref{sub:Quantum-autocorrelation-function-harmonic},
we have discussed that this model corresponds to shifted adiabatic
BO surfaces of identical curvature. We have shown that in this case,
the energy gap two-time correlation function for a classical ground-state
propagation is directly proportional to the quantum one and we have
extracted the appropriate (frequency dependent) proportionality constant.
For other shapes of the potential surfaces involved, one will in general
obtain different proportionality constants, although the delta-peaks
of the spectral densities can be located at the same energies (the
positions are determined by the shape of the ground state potential). 

It is not clear, \emph{a priori,} if the approximation of shifted
harmonic surfaces (or equivalently linear coupling to a harmonic bath)
is a good one for the system under consideration. To gain some insight
on this question, from an analysis of QM/MM trajectories, one possibility
is to consider higher order correlators. If the approximation of linearly
coupled harmonic oscillators is inadequate, one expects that higher
order correlators will have a significant relative weight. 

We proceed to discuss some properties of correlations of the bath
gap operator, Eq.~\ref{eq:gap_twolevel_BO}. The energy gap operators
can be described by a function of the bath coordinates and expanded
in terms of these as
\begin{equation}
\hat{\Delta}=\sum_{i}\xi_{i}^{(0)}+\sum_{i}\xi_{i}^{(1)}Q_{i}+\sum_{ij}\xi_{ij}^{(2)}Q_{i}Q_{j}+...\:.\label{eq:expansion of gap operators}
\end{equation}
 When only terms up to first order in $Q$ are significant, as in
the case of the Harmonic surfaces in the linear system bath coupling
limit, Tab.~\ref{tab:Expression_HO_BO}, we can write the two-time
correlation function as
\begin{equation}
\alpha(t,0)=\left\langle \hat{\Delta}(t)\hat{\Delta}(0)\right\rangle =\sum_{ij}\xi_{i}^{(1)}\xi_{j}^{(1)}\Big<Q(t)Q_{j}(0)\Big>.\label{eq:two_point_linear}
\end{equation}
Here, we have excluded the zeroth-order term which corresponds, e.g.,
to a reorganization energy, and is usually renormalized into the system
Hamiltonian. The angular brackets $\left\langle ...\right\rangle ={\rm tr}_{{\rm B}}\left\{ ...,\hat{\rho}_{{\rm B}}\right\} $
indicate thermal averaging over the bath degrees of freedom. Similarly,
the three-time correlation function becomes 
\begin{eqnarray}
\alpha(t',t,0) & = & \left\langle \hat{\Delta}(t')\hat{\Delta}(t)\hat{\Delta}(0)\right\rangle \nonumber \\
 & = & \sum_{ijk}\xi_{i}^{(1)}\xi_{j}^{(1)}\xi_{k}^{(1)}\left\langle Q_{i}(t')Q_{j}(t)Q_{k}(0)\right\rangle .\label{eq:alpha_t_t_0}
\end{eqnarray}
In the case of a harmonic bath, the three-time correlation function
will vanish, and in general any odd permutation of the harmonic bath
coordinates will vanish. 

However, if one considers the case where one retains the second order
term in Eq.~\ref{eq:expansion of gap operators}, the two-time correlator
will become:
\begin{eqnarray}
\alpha(t,0) & = & \sum_{ijkl}\Xi_{ij}\Xi_{kl}\left\langle {\bf Q}_{ij}(t){\bf Q}_{kl}(0)\right\rangle ,\label{eq: alphat0}
\end{eqnarray}
where we have defined $\Xi_{ij}$ and $\mathbf{Q}_{ij}(t)$ as
\[
\Xi_{ij}=\begin{cases}
0 & ;\; i=j=0\\
\xi_{i}^{(1)} & ;\; j=0\;\wedge\; i\neq0\\
\xi_{j}^{(1)} & ;\; i=0\;\wedge\; j\neq0\\
\xi_{ij}^{(2)} & ;\; i,j\neq0
\end{cases}
\]

\[
{\bf Q}_{ij}(t)=\begin{cases}
0 & ;\; i=j=0\\
Q_{i}(t) & ;\; j=0\;\wedge\; i\neq0\\
Q_{j}(t) & ;\; i=0\;\wedge\; j\neq0\\
Q_{i}(t)\cdot Q_{j}(t) & ;\; i,j\neq0
\end{cases}.
\]
Analogously the three-time correlator becomes
\begin{equation}
\begin{aligned}\alpha(t',t,0) & =\sum_{ijklmn}\Xi_{ij}\Xi_{kl}\Xi_{mn}\left\langle {\bf Q}_{ij}(t'){\bf Q}_{kl}(t){\bf Q}_{mn}(0)\right\rangle .\end{aligned}
\label{eq: 3pt_2nd_order_gap}
\end{equation}

If the bath is harmonic, it is straightforward to show that all terms
with an odd number of coordinate operators in the averages will vanish.
Yet, we see that in general, unless the coupling to the bath coordinates
is linear and the bath consists of Harmonic oscillators, the three-point
correlator will not vanish. It may therefore be necessary to go beyond
the simple description using only the two-time correlator.

\section{Application to the FMO complex\label{sec:Results}}

In this section, we apply the approximations discussed in Section~\ref{sec:semiclassical_corrections},
to the energy gap trajectories obtained from the mixed QM/MM simulations
for the FMO complex of \emph{Prosthecochloris aestuarii} as carried
out recently by us in Ref.~\cite{Shim2012}. The nuclear trajectories
were obtained by classical MD using the AMBER 99 force field. An isothermal-isobaric
(NPT) ensemble was employed in the MD simulations. For the calculation
of the energy gap, snapshots of the nuclear coordinates were taken
at every 4 fs. For each ground state configuration, the gap was obtained
by computing the energy corresponding to the $Q_{y}$ transition of
the BChl's using time-dependent time-dependent density functional
theory with BLYP functional within the Tamm-Dancoff approximation. 

The calculations were carried out at 77 and 300K and both temperature
were treated on the same footing. We do not expect there to be additional
sampling problems for the low temperatures because, up to current
knowledge, FMO does not undergo any major conformational changes in
this temperature range. More details on the computation can be found
in Ref.~\cite{Shim2012}.

The calculation of the SD from the time dependent gap energy is based
on the model described in Section~\ref{sec:egap_ct_simple_model}.
The actual MD simulation might deviate from this model e.g.~because
the thermostat could influence the dynamical evolution and thus the
correlation function. We plan to investigate this aspect in future
work. For now we will assume that the thermostat doesn't influence
the dynamics and that the models introduced in Section~\ref{sec:egap_ct_simple_model}
provide a reasonable description of a two level molecule treated in
the QM/MM approach.

\subsection{TDCD and spectral density from mixed QM/MM with a posteriori semiclassical
corrections\label{sub:TDCD-and-Spectral-results}}

Using the energy gap trajectories obtained in Ref.$\,$\cite{Shim2012},
we evaluated the different semiclassical approximations as reported
in Tab.~\ref{tab:G_J_prefactor}. We denote the time-points at which
the energy gap is calculated by $t_{i}$ and the corresponding energy
gap by $X_{i}$ where $i=0\dots N-1$ runs over the $N$ the time-points.
As in Ref.~\cite{Shim2012} we evaluated the correlator by using
a discrete representation, which implements the $k$-th element of
the two-time correlator as
\begin{equation}
C_{k}=\frac{1}{(N-k)}\sum_{i=1}^{N-k}\left(X_{i}-\bar{X}\right)\left(X_{i+k}-\bar{X}\right)\label{eq:  numeric two point}
\end{equation}
where $\bar{X}$ is the mean. Here, one assumes that the $N-k$ values
$X_{i}$ give a faithful initial distribution which reproduces the
Boltzmann distribution. To minimize spurious effects in the Fourier
transform, we multiplied the time trace by a Gaussian of variance
$\sigma_{{\rm gaussian}}^{2}=0.09\cdot t_{{\rm max}}^{2}=2.304\cdot10^{5}{\rm \: fs^{2}}$
with $t_{{\rm max}}=1600\,{\rm fs}$, the length of the correlation
function (as reported in \cite{Shim2012}). The Gaussian is normalized
to have unitary area in frequency domain %
\footnote{The variance in frequency domain is $\sigma_{\omega}^{2}=4.3403\cdot10^{-6}\frac{1}{{\rm fs}^{2}}=122.33\,{\rm cm^{-1}}$.
This give a FWHM of 26 ${\rm cm}^{-1}$ %
} following our definition of the Fourier transform in Eq.~\ref{eq:Gomega},
so that in frequency domain this corresponds to a convolution with
a Gaussian with a FWHM of $26\:{\rm cm^{-1}}$. Next, we computed
the different semiclassical quantities of Table$\:$\ref{tab:G_J_prefactor}
using our initial time trace. 

\begin{figure}
\noindent \begin{centering}
\includegraphics[angle=-90,width=1\columnwidth]{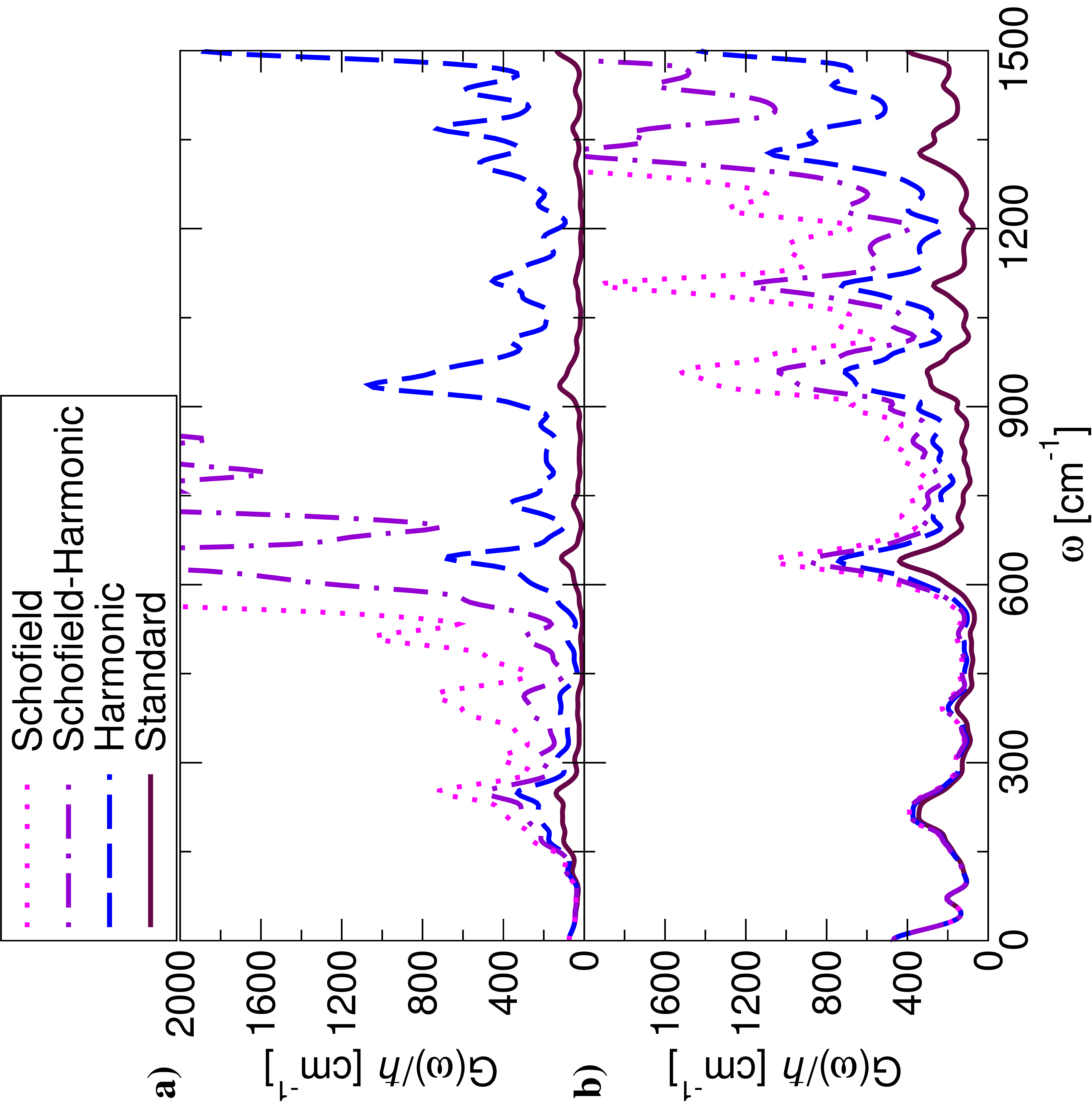}
\par\end{centering}

\caption{\label{fig:TDMD_site_1}Positive frequency part of the temperature-dependent
coupling densities $G(\omega)$ obtained\textcolor{black}{, as described
in the Section \ref{sub:TDCD-and-Spectral-results}} with each of
the Standard, Harmonic, Schofield and Schofield-Harmonic corrections
as (see Tab.~\ref{tab:G_J_prefactor}, column two). In panel a) results
are at 77K and in panel b) 300K. }
\end{figure}

In Figure~\ref{fig:TDMD_site_1}, we show the temperature-dependent
coupling densities TDCDs (as defined in Eq.~\ref{eq:Gomega}), for
site 1 of the FMO complex (site 1 at 77K and 300K) evaluated using
the different approximations listed in Table$\:$\ref{tab:G_J_prefactor}
column two.\textcolor{magenta}{{} }We notice how, as expected, there
are little differences between the approximations at low frequencies.
Only at higher frequencies the TDCD differs significantly for each
approach. The Egelstaff approximation incorrectly predicts a negative
spectral density for low frequencies in this case and was therefore
not shown in the plots.\textcolor{magenta}{{} }

From the general definition of each semiclassical correction, it isn't
clear which one is most accurate. To better reason on which one to
choose, we will look at the temperature dependence of the spectral
density. Further, we will compare to experimental results and finally
we will evaluate the three point correlator (Sec.\ref{sub:Higher-order-results}).

\subsection{Analysis of prefactors in terms of temperature dependence of the
spectral density\label{sub:Analysis-of-prefactors-T-dep}}

\begin{figure*}
\begin{centering}
\includegraphics[width=1\columnwidth]{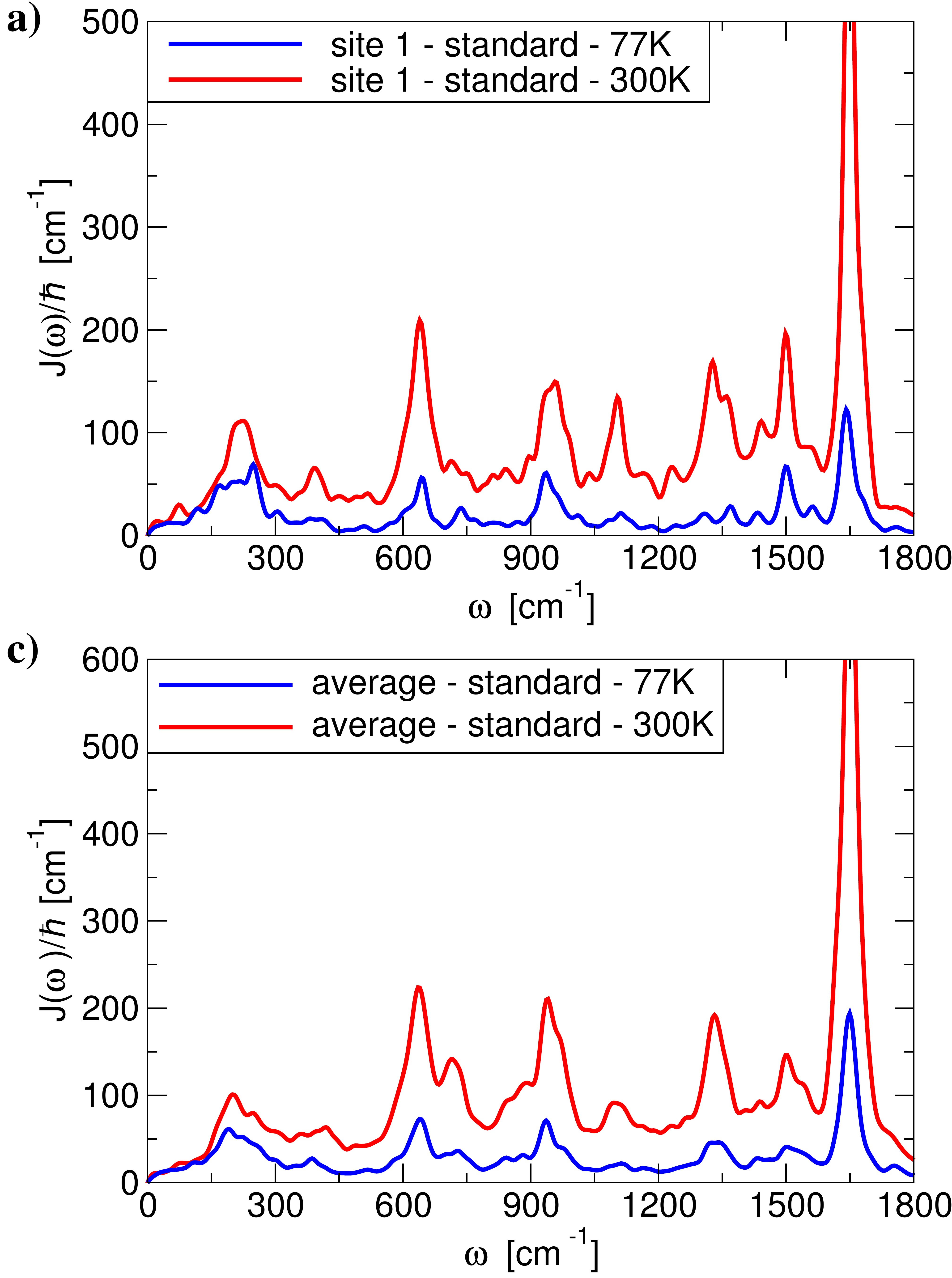}\includegraphics[width=1\columnwidth]{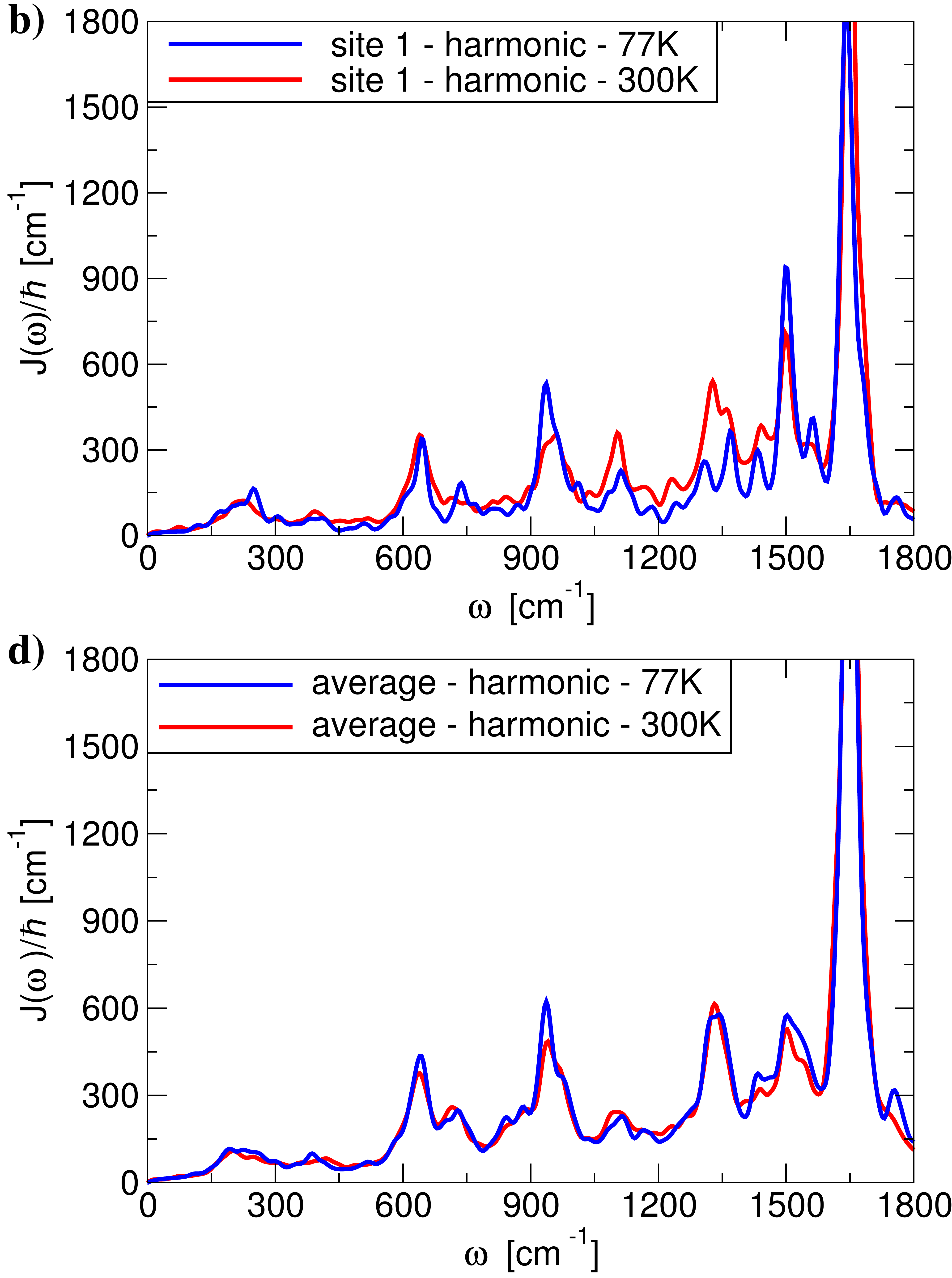}
\par\end{centering}

\caption{Panel a) comparison of the asymmetric component of temperature-dependent
coupling density $J(\omega)\equiv G_{{\rm asym}}(\omega)\,;$ for
site 1 of the Fenna-Matthews-Olson complex, obtained with the Standard
approximation (Tab.$\,$\ref{tab:G_J_prefactor}, first line, third
column) at 77K and at 300K. Panel b) comparison of $J(\omega)$ obtained
for site 1 with the Harmonic approximation (Tab.$\,$\ref{tab:G_J_prefactor},
second line, third column) at 77K and at 300K. We see clearly that
the Harmonic prefactor gives a roughly temperature independent $J(\omega)$,
while large differences are seen using the Standard prefactor.\label{fig: Temperature_dependence}}
\end{figure*}

From our discussion in Section \ref{sec:Introduction}, we recall
that many open quantum system approaches rely on the assumption of
linear coupling to a bath of harmonic oscillators. This leads to a
temperature-independent spectral density $j(\omega)$, as discussed
in Section \ref{sub:Quantum-autocorrelation-function-harmonic}. Inspection
of Fig.~\ref{fig:TDMD_site_1} shows that for all but the Harmonic
approximation the TDCD (from which one obtains $J(\omega)$ which
is directly proportional to the spectral density $j(\omega)$) obtained
from the QM/MM is not similar at different temperatures. This is more
apparent at higher frequencies. To gain further insight into this
temperature dependence, in Fig.~\ref{fig: Temperature_dependence},
we compare the asymmetric TDCD ($J(\omega)=\pi j(\omega)\,;\,\omega>0$)
obtained using the Standard (panel a) and the Harmonic (panel b) approximations
for site 1 of the FMO complex. Results for all sites at both temperatures
are reported in the Supplementary Information, Appendix~\ref{sec:supplementary}.
and the corresponding data files for the Harmonic $J(\omega)$ can be 
downloaded from Ref.~\cite{Supplementary_download}.
One clearly sees that for the Standard correction there is a huge
difference between the 77K and the 300K results. However, in the case
of the Harmonic correction the spectral densities obtained at the
two temperatures nicely lie on top of each other, as one would require
for a temperature-independent spectral density. This result suggests
that the Harmonic correction is the appropriate one to employ to obtain
spectral densities to be used in open quantum system models which
assume linear coupling to a bosonic bath. 

Note, that the good agreement at both temperatures for the Harmonic
correction might be purely accidental or due to the fact that the
MD is not fully converged. We would need to run much longer QM/MM
trajectories to improve the statistics and check the convergence of
the distributions. This lack of statistics could also explain the
fact that for the SD averaged over all chomophores (panels c) and
d)), the agreement between both temperatures is slightly better than
for the individual sites. 

Finally we would like to remark that a temperature dependence of the
reorganization energy has been observed in the context of electron
transfer (ET) donor-acceptor energy gap spectral densities \cite{LebardMatyushov2008,WarshelParson2001}. 

\subsection{Comparison to experimental spectral density}

\begin{figure*}
\begin{centering}
\includegraphics[clip,width=1\columnwidth]{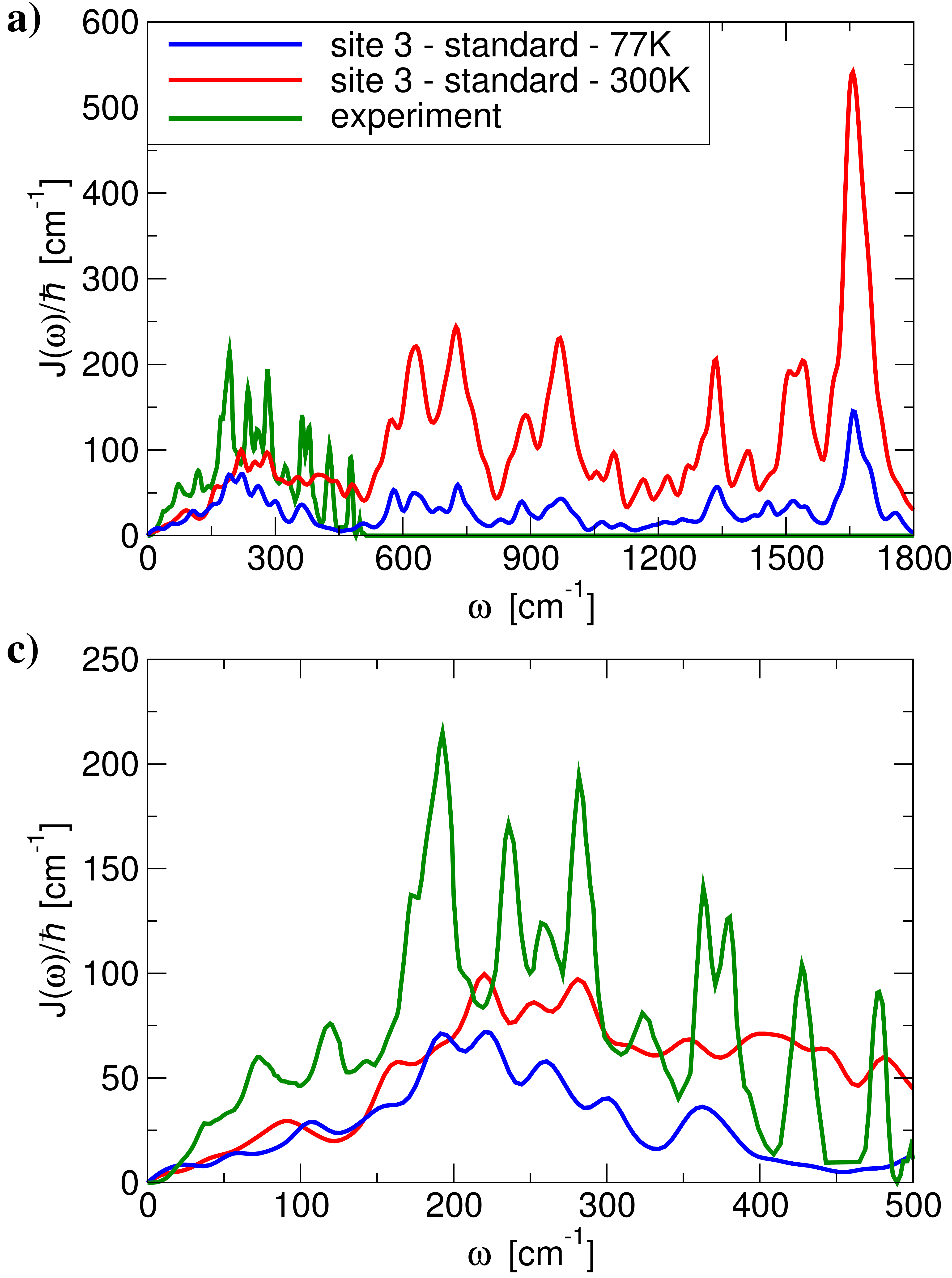}\includegraphics[clip,width=1\columnwidth]{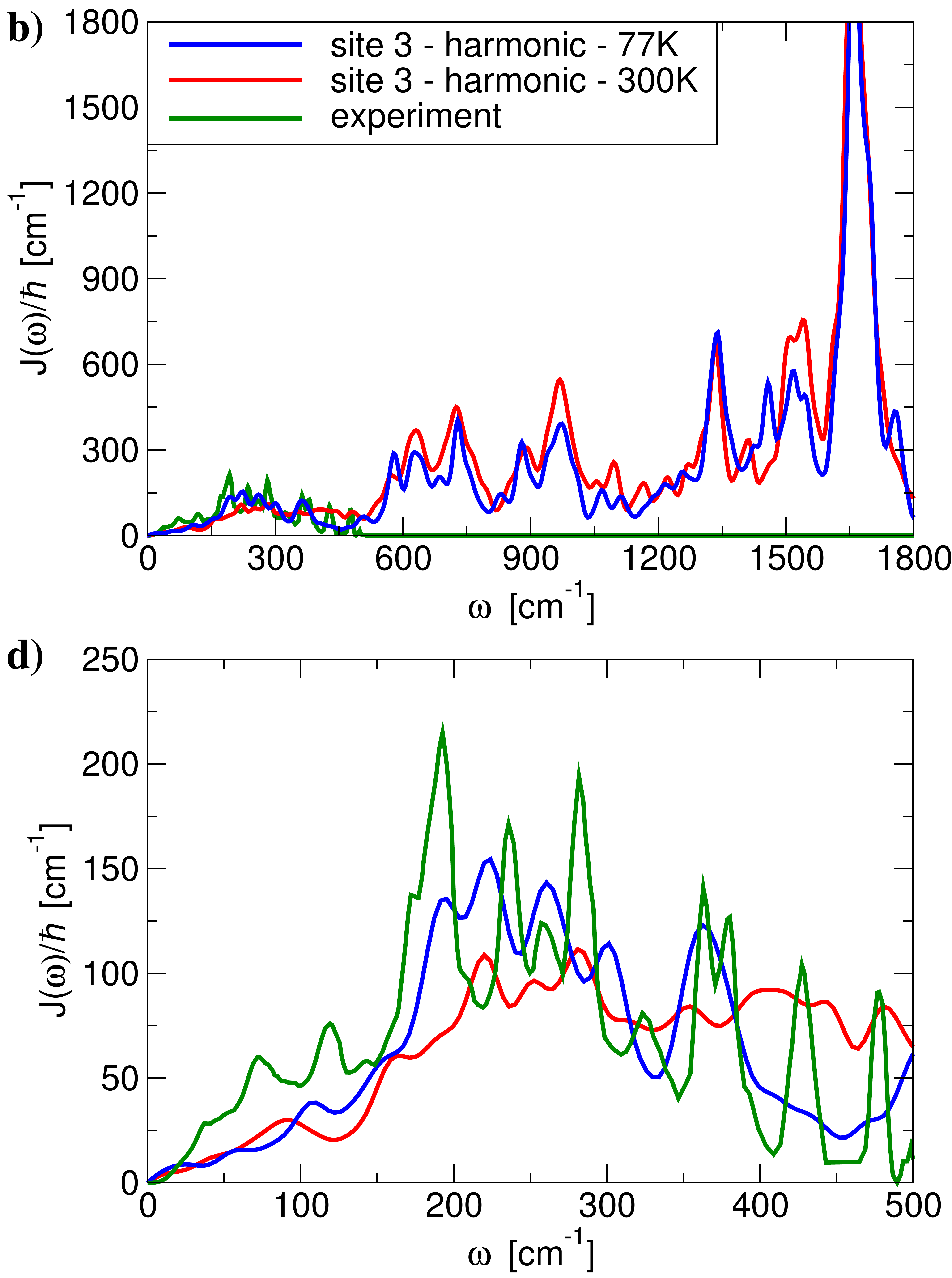}
\par\end{centering}

\caption{ Panel a) shows $J(\omega)$ for site 3 of the FMO complex calculated
with the Standard approximation at 77 and 300K and the green curve
corresponds to the experimental spectral density rescaled by $\pi$
to obtain $J(\omega)$ as defined in Eq. \ref{eq: Gasym_sd} \cite{Adolphs2006,Wendling2000}(More
details on the experimental spectral density are given in the text).
Panel b) shows $J(\omega)$ for site 3 calculated with the Harmonic
approximation at 77 and 300K and again the green curve corresponds
to the experimental spectral density \cite{Adolphs2006,Wendling2000}.
The agreement with the experimental (green) spectral density is slightly
better for the Harmonic approximation than for the Standard approximation.
Panels c) and d) correspond to the same quantities as those of panels
a) and b) in the low frequency region, here we note that both approximations
are roughly equivalent for $\frac{\hbar\omega}{k_{{\rm B}}T}<1$ (e.g
at $T=77\,{\rm K}$ for $\omega<55\,{\rm cm^{-1}}$and at $T=300\:{\rm K}$
for $\omega<200\,{\rm cm^{-1}}$). Further, the spectral density,
as defined in Eq.~\ref{eq: Gasym_sd} can be obtained by dividing
$J(\omega)$ by $\pi$. \label{fig:Compare-to-exp} }
\end{figure*}

In Fig.~\ref{fig:Compare-to-exp} panel a) and b) we compare the
asymmetric TDCD for site 3 (Standard and Harmonic correction), with
the asymmetric TDCD obtained from fluorescence line narrowing (FLN)
experiments \cite{Wendling2000}. We focus on the low frequency part
(up to \textasciitilde{}500 ${\rm cm^{-1}}$), which is relevant for
energy transfer in the FMO complex. The FLN results are obtained from
the lowest excitonic peak of the FMO absorption spectrum which is
believed to be generated almost entirely by BChl 3. Therefore, we
compare the experiment to the theoretical spectral density obtained
from the QM/MM for BChl 3. 

The experimental spectral density shown in Fig.~\ref{fig:Compare-to-exp}
is based on the dotted curve $\tilde{j}^{{\rm exp}}(\omega)$ of Fig.~2
of Ref.~\cite{Adolphs2006}, which is in good agreement with the
one-phonon vibrational profile (OPVP) of Ref.~\cite{Wendling2000},
because of the small total Huang-Rhys factor. Note that the extraction
of the OPVP uses the same model of shifted harmonic potential surfaces
as we did in Section \ref{sub:Quantum-autocorrelation-function-harmonic}.
Thus it corresponds to a SD which is suitable as input in the open
system approaches. In this harmonic model the profile $\tilde{j}{}^{{\rm {\rm {\rm exp}}}}(\omega)$
is related to our definition of the spectral density by $j^{{\rm {\rm {\rm exp}}}}(\omega)=(\hbar\omega)^{2}\,\tilde{j}{}^{{\rm {\rm {\rm exp}}}}(\omega)$.
The positive frequency part of the asymmetric TDCD, $J(\omega)$ is
obtained from the spectral density, as defined in Eq.~\ref{eq: Gasym_sd},
by $J^{{\rm {\rm {\rm exp}}}}(\omega)=\pi\, j^{{\rm {\rm {\rm exp}}}}(\omega)$.

From panel e) and f) of Fig.~\ref{fig:Compare-to-exp}, we see that
the magnitude and overall lineshape of both the Standard and the Harmonic
correction are in good agreement with the FLN data, in contrast with
previous results \cite{Shim2012} %
\footnote{The standard asymmetric component $J(\omega)=G_{{\rm asym}}(\omega)$
shown in Fig.~\ref{fig:Compare-to-exp} panel a), corresponds to
the spectral density in Ref.~\cite{Shim2012} if one multiplies the
Shim result by $\pi\cdot{\rm tanh}(\omega\beta\hbar/2)/{\rm tanh}(\nu\beta\hbar/2)$
to obtain $J(\omega)$ (here $\nu=\omega/(2\pi)$). We would like
to point out that a different convention was employed in \cite{Shim2012}. %
}. 

A closer inspection of the curves in panels c) and d) of Fig.~\ref{fig:Compare-to-exp}
shows that the width of the peaks obtained from the QM/MM simulation
is much broader than that obtained from the FLN data. As described
in Section \ref{sub:TDCD-and-Spectral-results}, this broadening is
due to the finite length of the numerical correlator, and to the convolution
with a gaussian function in frequency domain, which results in a broadening
of FWHM 26$\,{\rm cm}^{-1}$. Also, the position of the peaks do not
perfectly coincide. There might be various reasons for this discrepancy:
The trajectories might be too short, the quantum chemical calculations
of the transition gap are not accurate enough, or the thermostat leads
to some spurious effects. One has also to keep in mind that there
are uncertainties in the experimental data as well. The experimental
data (in particular at higher frequencies) probably do not represent
the actual spectral density of BChl 3 (excitonic effects might play
a relevant role, and it was difficult to extract the lineshape from
the representation of Refs.~\cite{Adolphs2006} and \cite{Wendling2000}).

Nevertheless, this good agreement in magnitude and overall lineshape
makes us confident, that the QM/MM procedure can indeed be useful
to extract spectral densities. 

Finally, it seems that the Harmonic correction describes the FLN data
slightly better in terms of amplitude, respect to the Standard correction. 

\subsection{Higher-order correlation function\label{sub:Higher-order-results}}

\begin{figure*}
\noindent \begin{raggedright}
\textbf{a)\includegraphics[angle=-90,width=1\columnwidth]{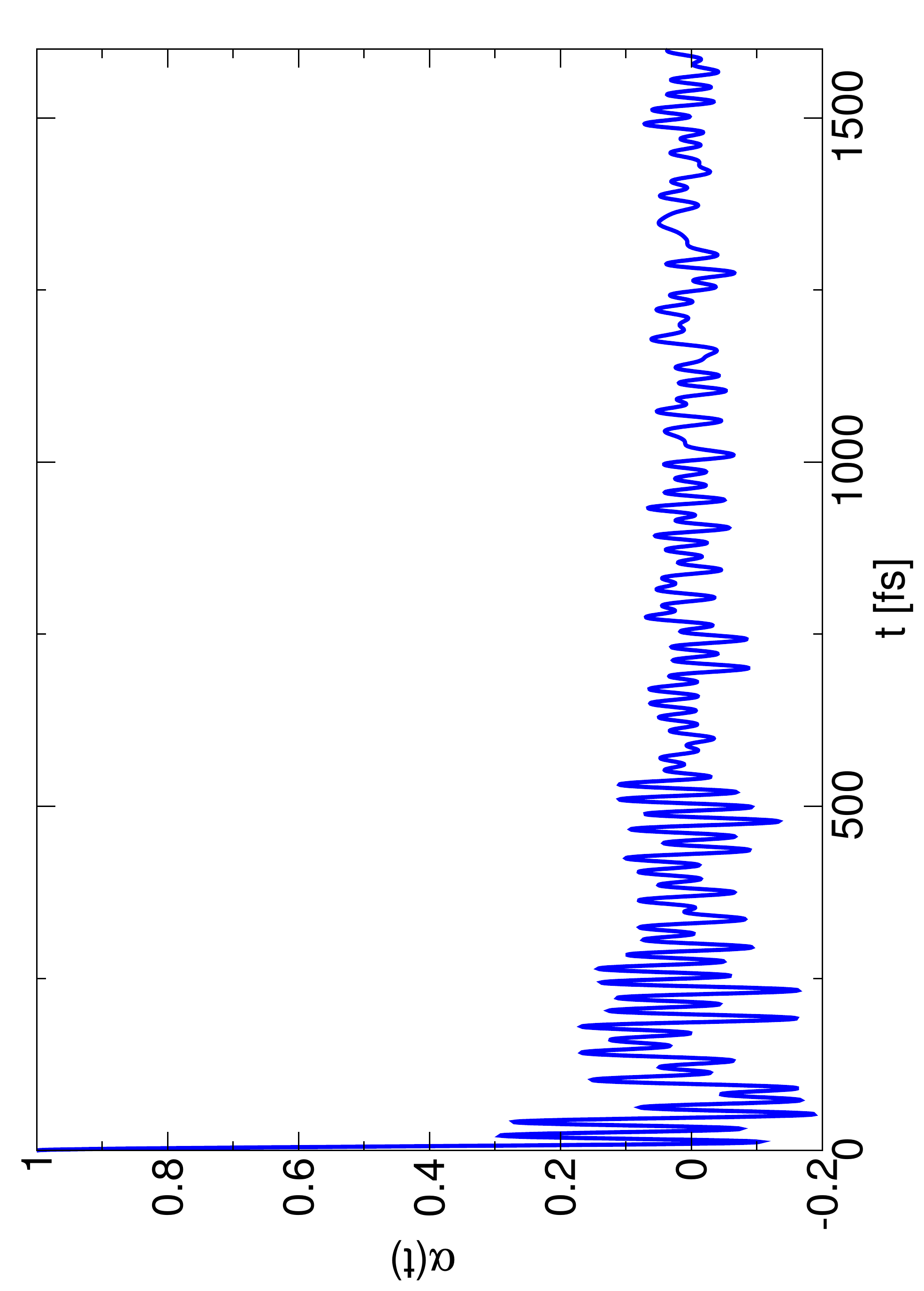}b)\includegraphics[angle=-90,width=1\columnwidth]{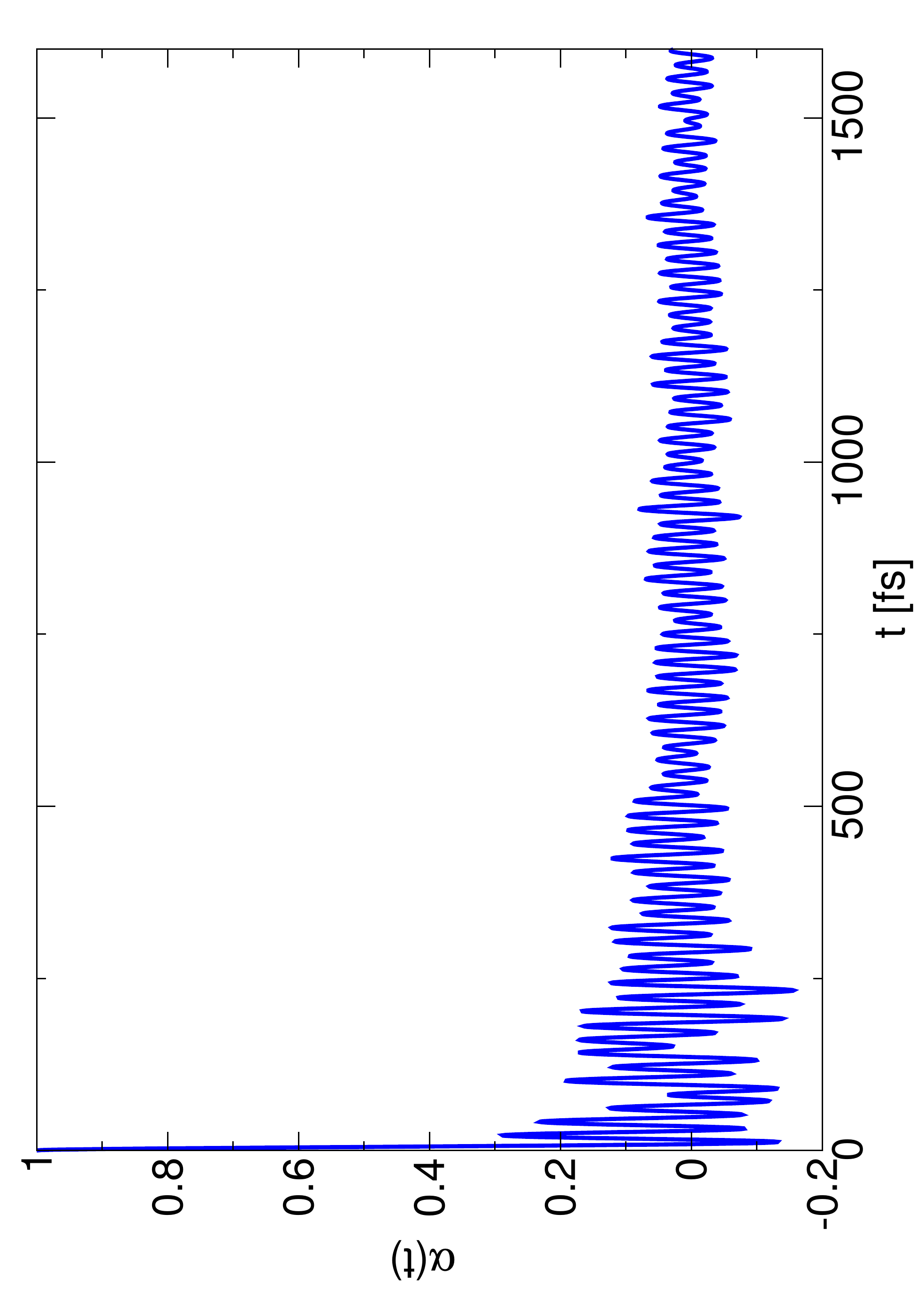}}
\par\end{raggedright}

\noindent \begin{raggedright}
\textbf{c)\includegraphics[width=8.5cm]{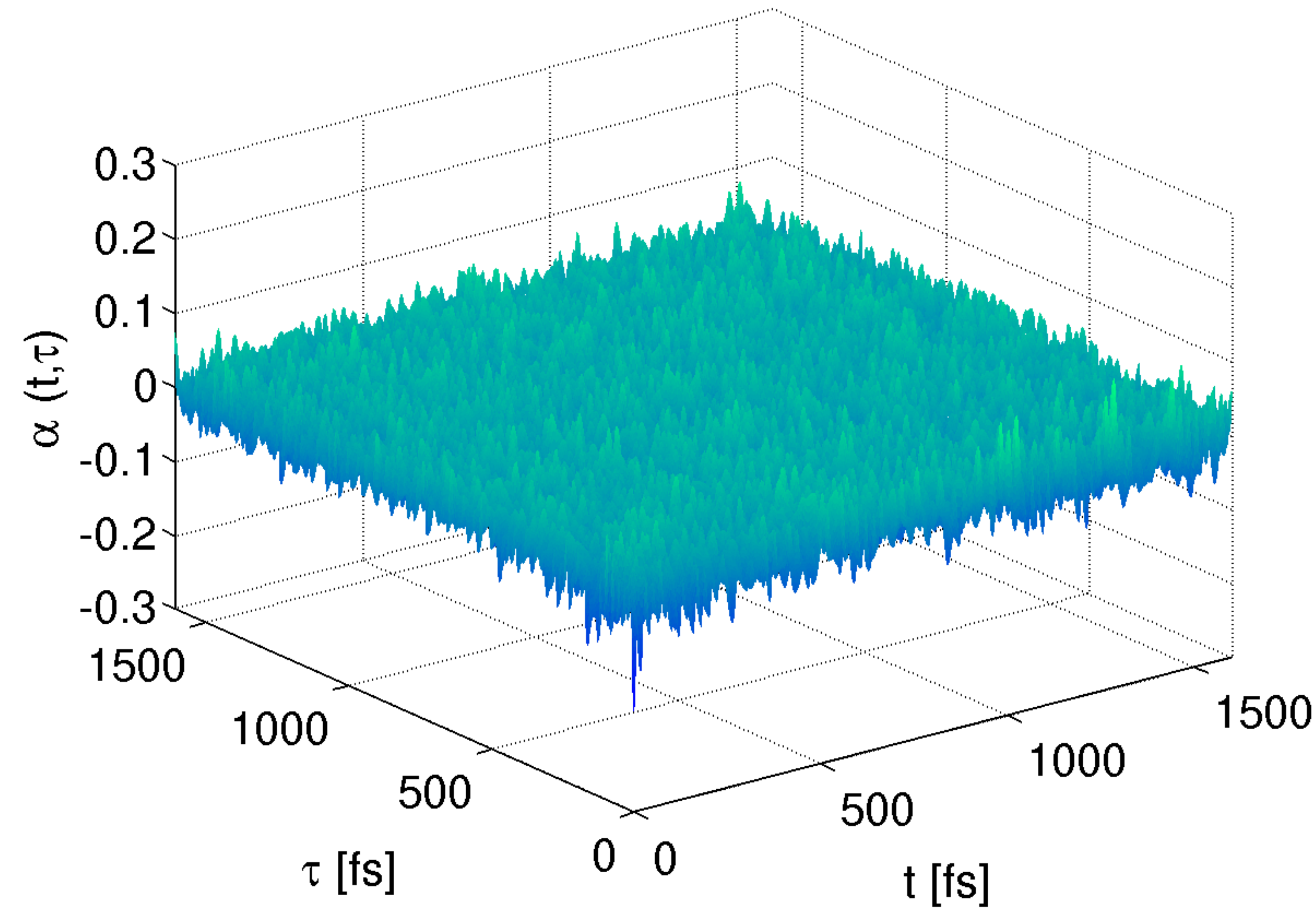}d)\includegraphics[width=8.5cm]{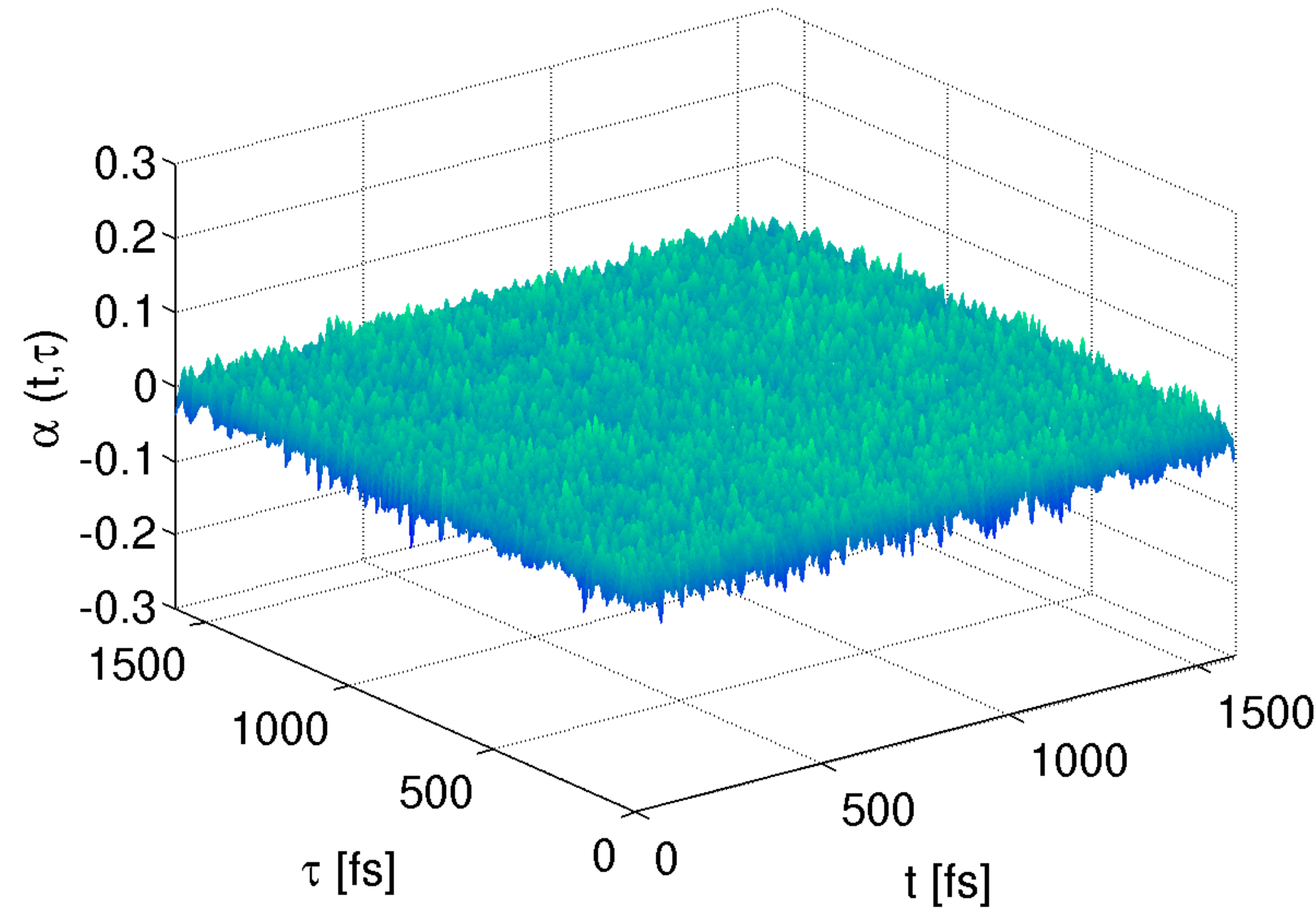}}
\par\end{raggedright}

\caption{Panel a): Two-time correlation function of the energy gap fluctuations
of site 1 of the FMO complex, normalized by $s^{2}$ (the variance)
at 77K after evaluating it as in Eq. \ref{eq:  numeric two point}.
Panel b): Two-time correlation function for site 1 at 300K. Panel
c) Three-time correlation function of the energy gap fluctuations
of site 1 of the FMO complex, as defined in Eq.~\ref{eq: numeric_three_point}
normalized by $s^{3}$ at 77K. Panel d) Three-time correlation function
for site 1 at 300K. \label{fig:twoandthreepointcorrelator-1} }
\end{figure*}

From the theory of discrete processes, similarly to Eq. \ref{eq:  numeric two point},
we see that the $(k,j)$-th element of the three-time correlator is
\begin{equation}
\begin{aligned}C(k,j) & =\frac{1}{(N-k-j)}\,\,\sum_{i=1}^{N-k-j}\left(\Delta X_{i}\right)\left(\Delta X_{i+k}\right)\left(\Delta X_{i+k+j}\right)\end{aligned}
\label{eq: numeric_three_point}
\end{equation}
with $\Delta X_{i}\!=\! X_{i}\!-\!\bar{X}$ where $\bar{X}$ is the
mean and $N$ is the number of time points (as defined in Sec.~\ref{sub:TDCD-and-Spectral-results}).
We compare the two-time and the three-time correlators by dividing
them by increasing powers of the standard deviation $s\equiv\sqrt{m^{(2)}}$,
thus we use Eq.~\ref{eq:  numeric two point} for the two-time correlation
function and divide it by $s^{2}$ and we divide Eq.~\ref{eq: numeric_three_point}
by $s^{3}$. The results for site 1 of the FMO complex at 77 and 300K
are reported in Fig.~\ref{fig:twoandthreepointcorrelator-1}. For
the two-time correlator, Fig.~\ref{fig:twoandthreepointcorrelator-1}
panels a) and b), we see correlations up to at least 1000 time steps,
while for the three-time correlator, panels c), d), e) and f), we
see a rather noisy profile with values about one/two orders of magnitude
smaller than the largest value of the two-time correlations. This
is observed for all sites and temperatures (Results for all sites
can be found in the Supplementary information, Appendix~\ref{sec:supplementary}).

This means that since we find a small three-time correlator, the linear
coupling to a harmonic bath assumption is probably good. In fact,
as described in Sec.~\ref{sec:higher_order_correlators} this case
corresponds to linear coupling to the bath and Gaussian correlated
bath operators. Of course, the statistics of the three-time correlator
is not great due to the finite length of the time trajectories, but
we think that the general tendency is correct. One should also keep
in mind that there may be fortuitous cases in which the three-time
correlator is roughly zero and the bath is not harmonic. Further,
this comparison is based on the order of magnitude of the correlations,
the three-time correlator is only much smaller. It may be that for
some modes of the system, certain frequencies, present in the three-time
correlator's two dimensional Fourier transform give a more important
contribution to the dynamics than other frequencies present in the
spectral density. Nonetheless, the above result encourages the idea
that the assumption of linear coupling and harmonic bath is valid.
This, in turn, implies that one should use the Harmonic semiclassical
correction in Sec.~\ref{sec:semiclassical_corrections}, which is
also consistent with the prefactor found in \ref{sec:egap_ct_simple_model}.

On a final note, to confirm with certainty that the bath is Harmonic,
one should evaluate higher order correlators, beyond the three-time
correlator. However, to obtain a statistically relevant estimate,
much longer time dependent energy gap trajectories, which are expensive
in terms of the QM/MM propagation, would be required. Work in this
direction is being carried out in our groups.

\section{Conclusions\label{sec:Conclusions}}

In this work, we have investigated the connection between the gap
correlation function extracted from ground state QM/MM and the bath
spectral density used as input in many open quantum system approaches.

One important point is that the classical bath correlation function
is real while the quantum mechanical one is generally complex. There
exist several semiclassical \emph{a posteriori} corrections which
aim to fix this and we have employed them on our time traces to recover
a part of the imaginary component.

The discussed prefactors originate from general expansions in orders
of $\hbar$ and do not include information on the specific type of
system-bath coupling, etc. We have investigated two simple models
and found that the prefactors obtained correspond to two of the general
semiclassical expressions. Thus, we have linked the semiclassical
limits with a microscopic potential energy surface picture.

We have shown that the gap-correlation function extracted from ground
state QM/MM only corresponds to the fully quantum excited state calculations
in the case of shifted parabolas. This model for a few vibrational
modes of the chromophores has been successfully used to describe the
optical properties of molecular aggregates. Including only a finite
number of internal vibrations is probably a good approximation for
molecules in the gas phase or suprafluid Helium nanodroplets \cite{RoEiDv11_054907_}.
However, for molecules in solution or when a protein environment is
present it is no longer a good approximation to include only only
a few (undamped) modes. In particular, one has to take into account
the interaction of the vibrations with the environment in addition
to the direct interaction of the electronic excitation with the environment.
For this general situation, it is no longer clear whether the model
of shifted harmonic potential surfaces is indeed a good description
of the system. 

Therefore, we have investigated whether the approximation of harmonic
bath and linear coupling is accurate for our QM/MM calculations for
the FMO photosynthetic complex by computing the next higher order
correlator beyond the two-time correlator. The three-point correlator
seems to give a small contribution which, while not being conclusive,
suggests to us that the Harmonic/linear coupling model is a good approximation.
The evaluation of the four-time correlation function would be useful
to bolster this claim. 

The analysis of the temperature dependence of prefactors for the spectral
density also suggests that the Harmonic approximation is preferred
to use for the FMO complex, and perhaps other photosynthetic complexes,
rather than the Standard one when employing it in Open Quantum system
approaches. 

Having made these choices, the theoretical results are in reasonably
good agreement with the experimental spectral density. These result
in a much better agreement than in our previous work, which underestimated
the magnitude of the spectral density \cite{Shim2012} and than other
QM/MM calculations \cite{Olbrich2011} which overestimate the coupling
to the bath by one order of magnitude.

Finally, we have explained the link between bath correlation function
and gap correlation function and\emph{ }found models under which the
gap correlation function can actually be viewed as a general open
quantum system bath correlation function. 
\begin{acknowledgments}
We thank Dr. Semion Saikin, for his useful advice and Gerhard Ritschel
for going over the manuscript. A. E. acknowledges financial support
from the DFG under Contract No. Ei 872/1-1. S. V. and A. A. G. acknowledge
support from the Center for Excitonics, an Energy Frontier Research
Center funded by the U.S. Department of Energy, Office of Science
and Office of Basic Energy Sciences under Award Number DE-SC0001088
as well as support from the Defense Advanced Research Projects Agency
under award number N66001-10-1-4060. S. V. and A. A.-G. acknowledge
support from the Harvard Quantum Optics center.
\end{acknowledgments}

\begin{appendix}

\section{Supplementary information}\label{sec:supplementary}

\subsection{Two and three-time correlation functions for all site of the fmo
complex}

\begin{figure*}[H]
\begin{centering}
\textbf{a)\includegraphics[width=1\columnwidth]{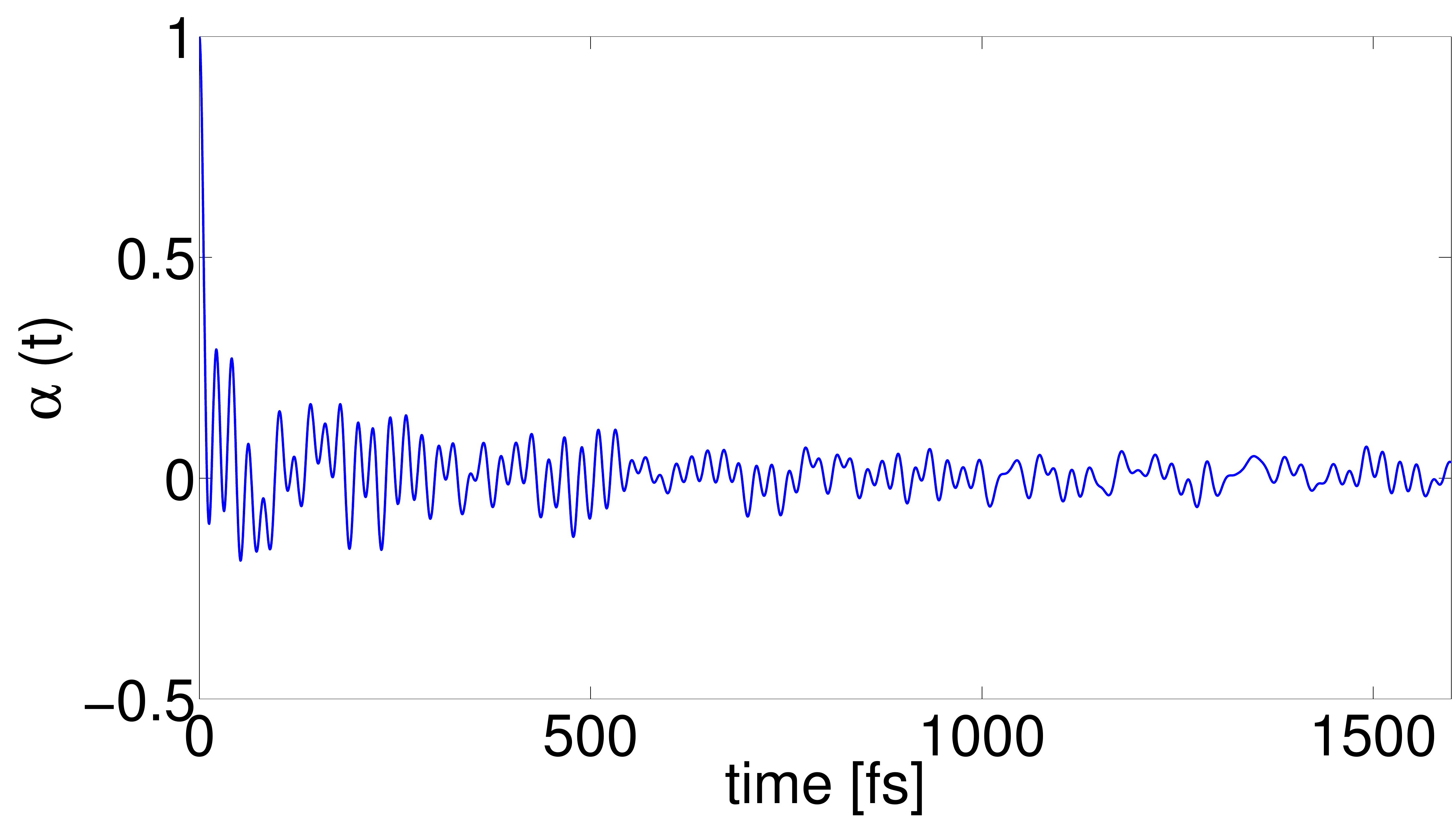}b)\includegraphics[width=1\columnwidth]{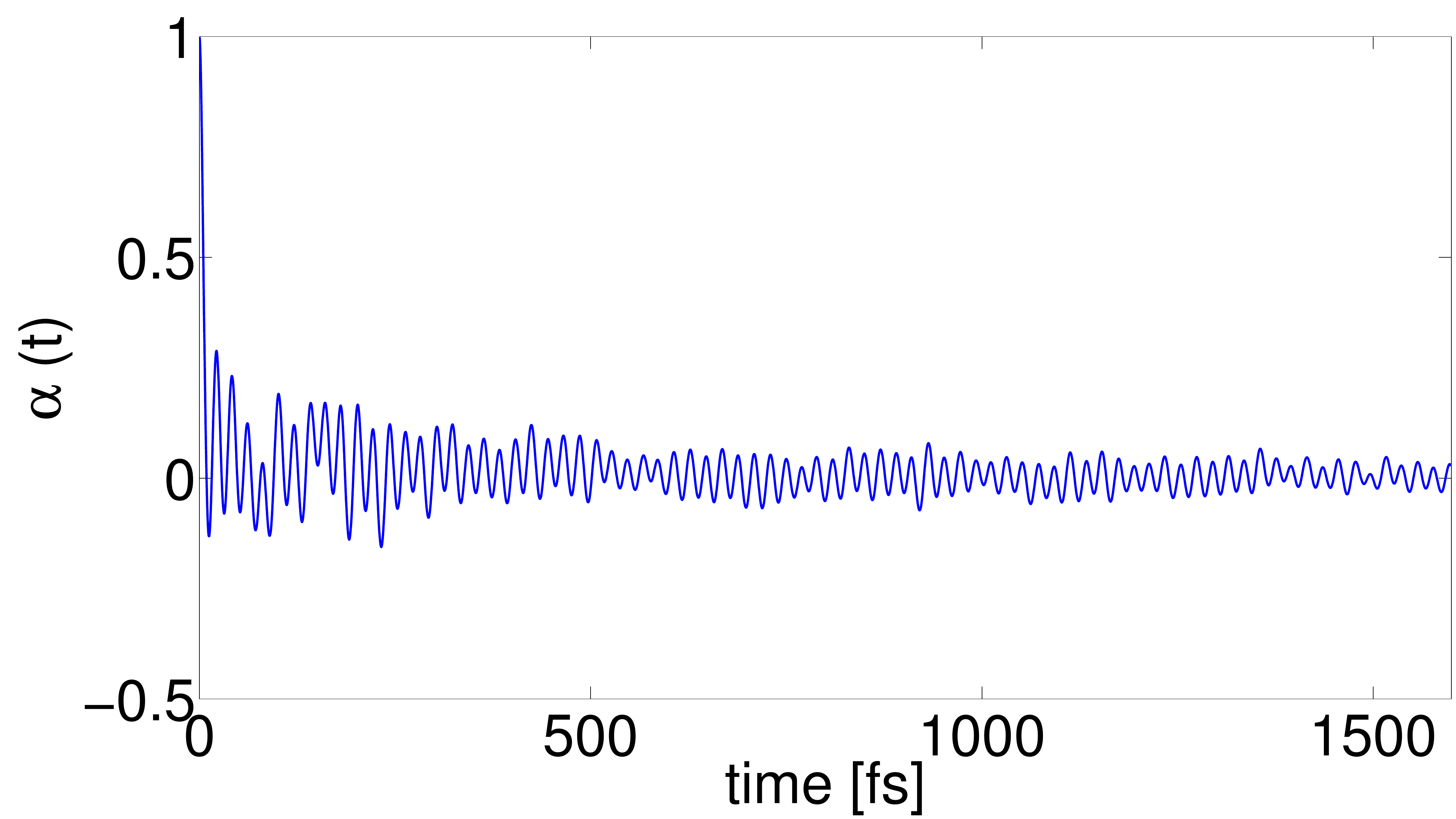}}
\par\end{centering}

\begin{centering}
\textbf{c)\includegraphics[width=1\columnwidth]{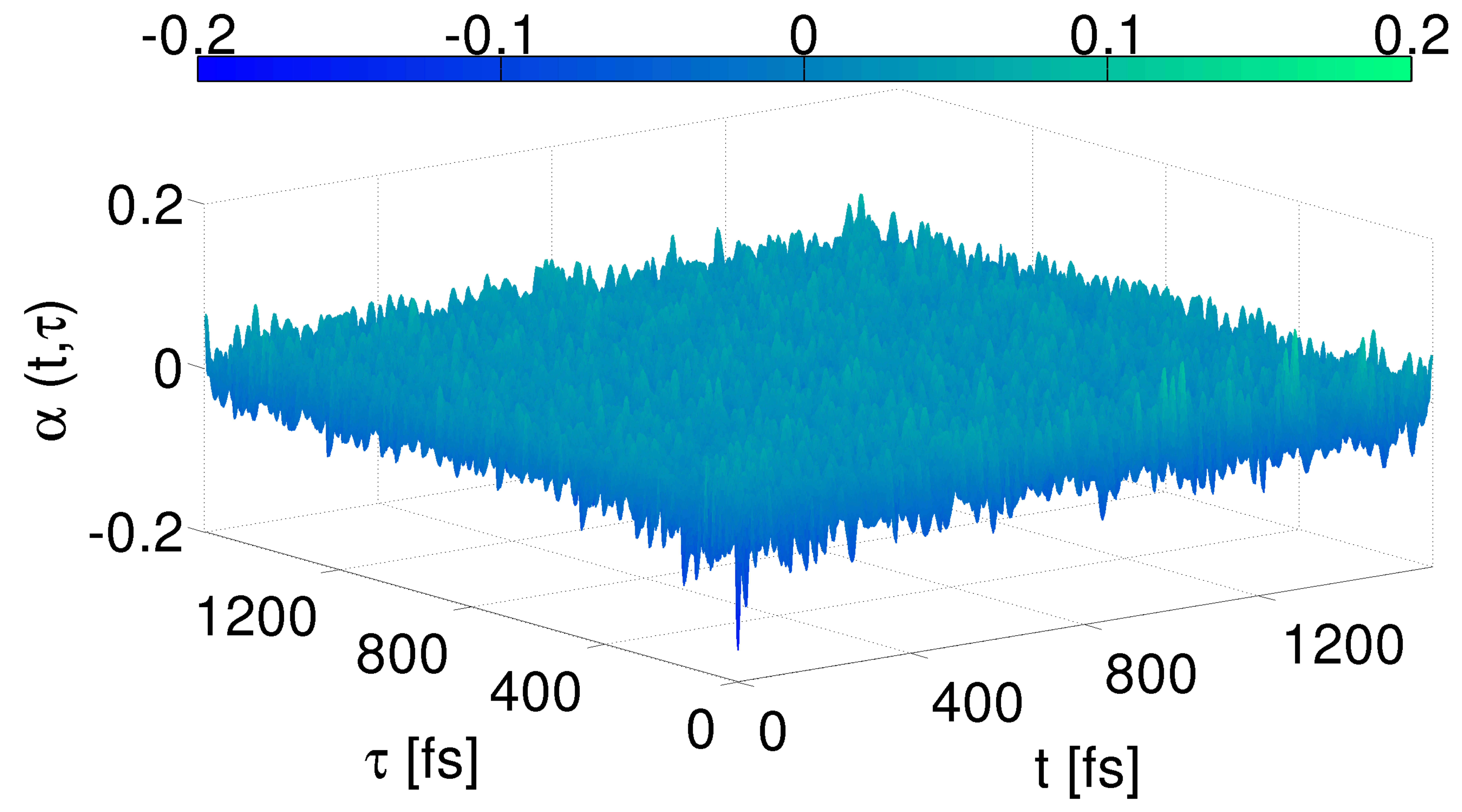}d)\includegraphics[width=1\columnwidth]{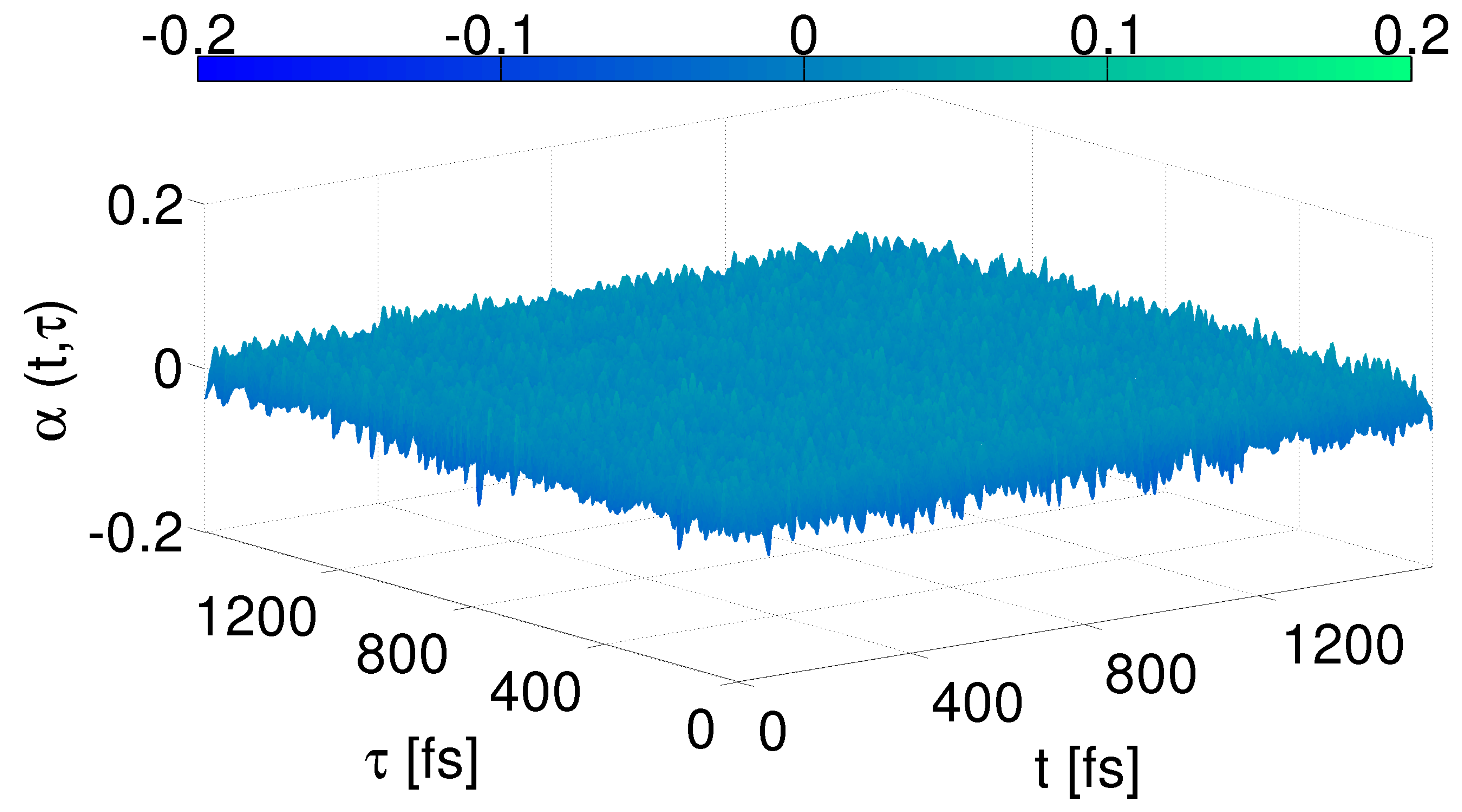}}
\par\end{centering}

\caption{Panel a): Two-time correlation function of the energy gap fluctuations
of site 1 of the FMO complex, normalized by the variance $s^{2}$
at 77K as discussed in the text. Panel b): Two-time correlation function
for site 1 at 300K. Panel c) Three-time correlation function of the
energy gap fluctuations of site 1 of the FMO complex, as discussed
in the text and multiplied by $s^{3}$ at 77K. Panel d) Three-time
correlation function for site 1 at 300K. \label{fig:corr_s_1} }
\end{figure*}

\begin{figure*}[H]
\begin{centering}
\includegraphics[width=1\columnwidth]{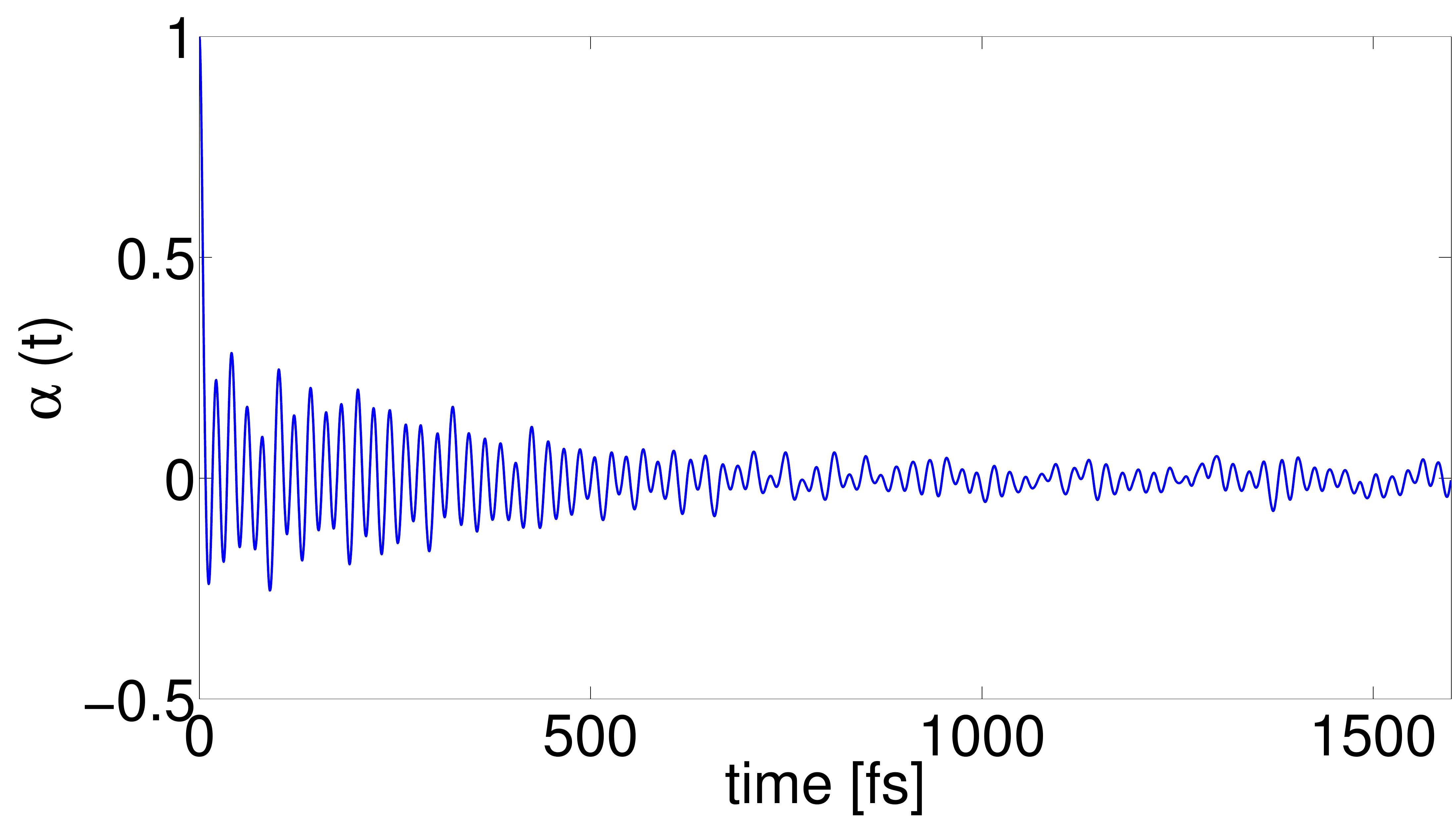}\includegraphics[width=1\columnwidth]{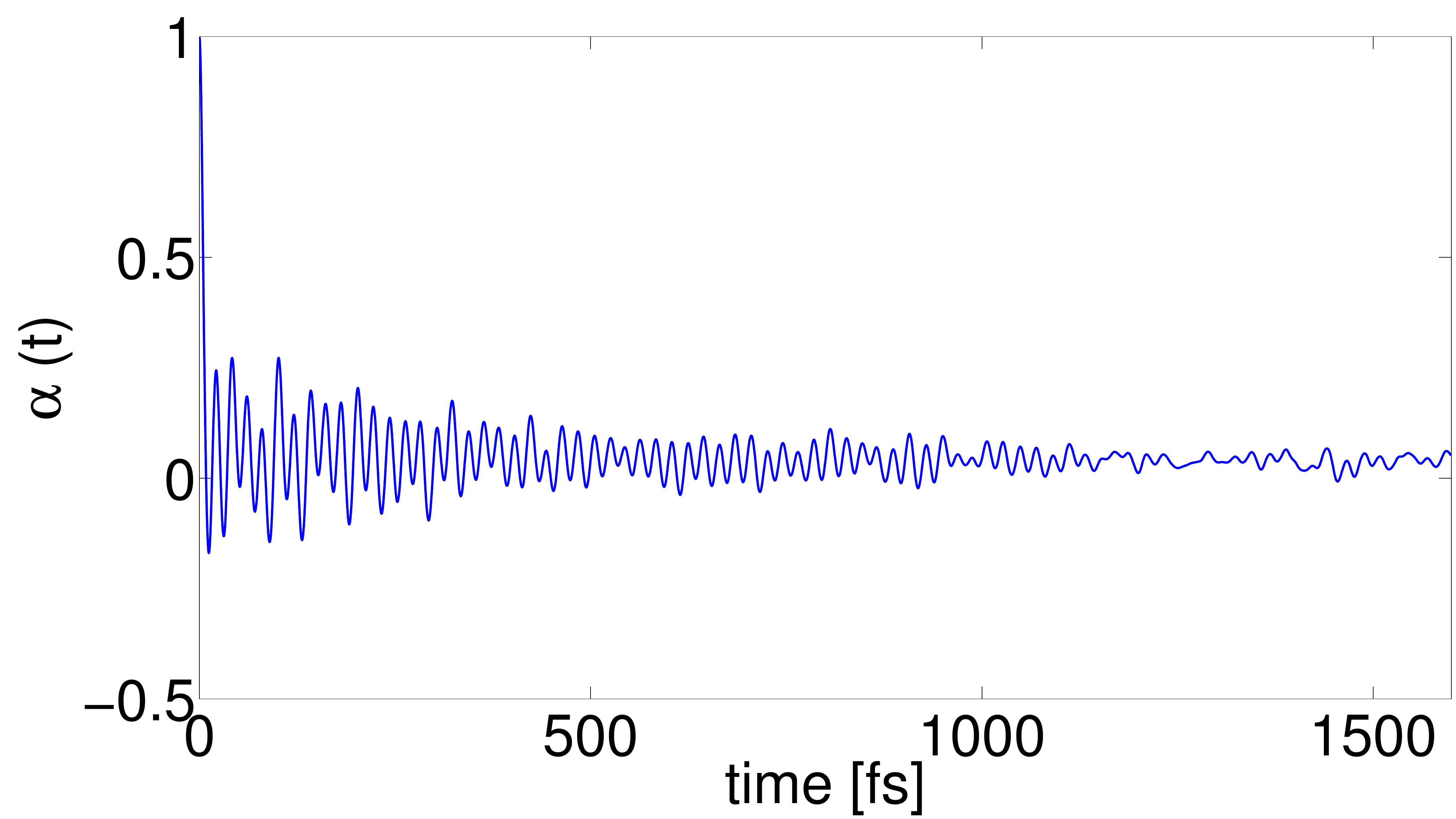}
\par\end{centering}

\begin{centering}
\includegraphics[width=1\columnwidth]{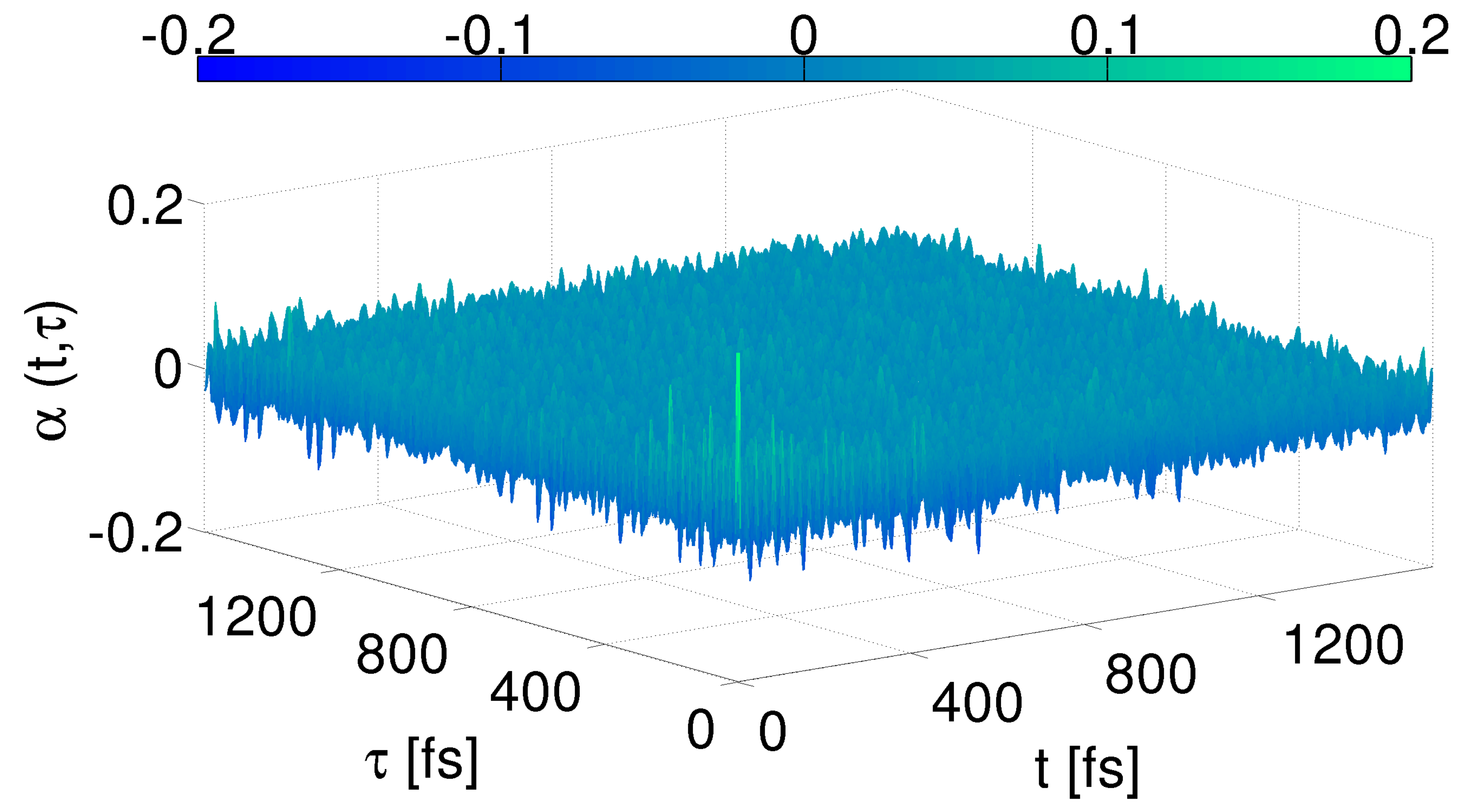}\includegraphics[width=1\columnwidth]{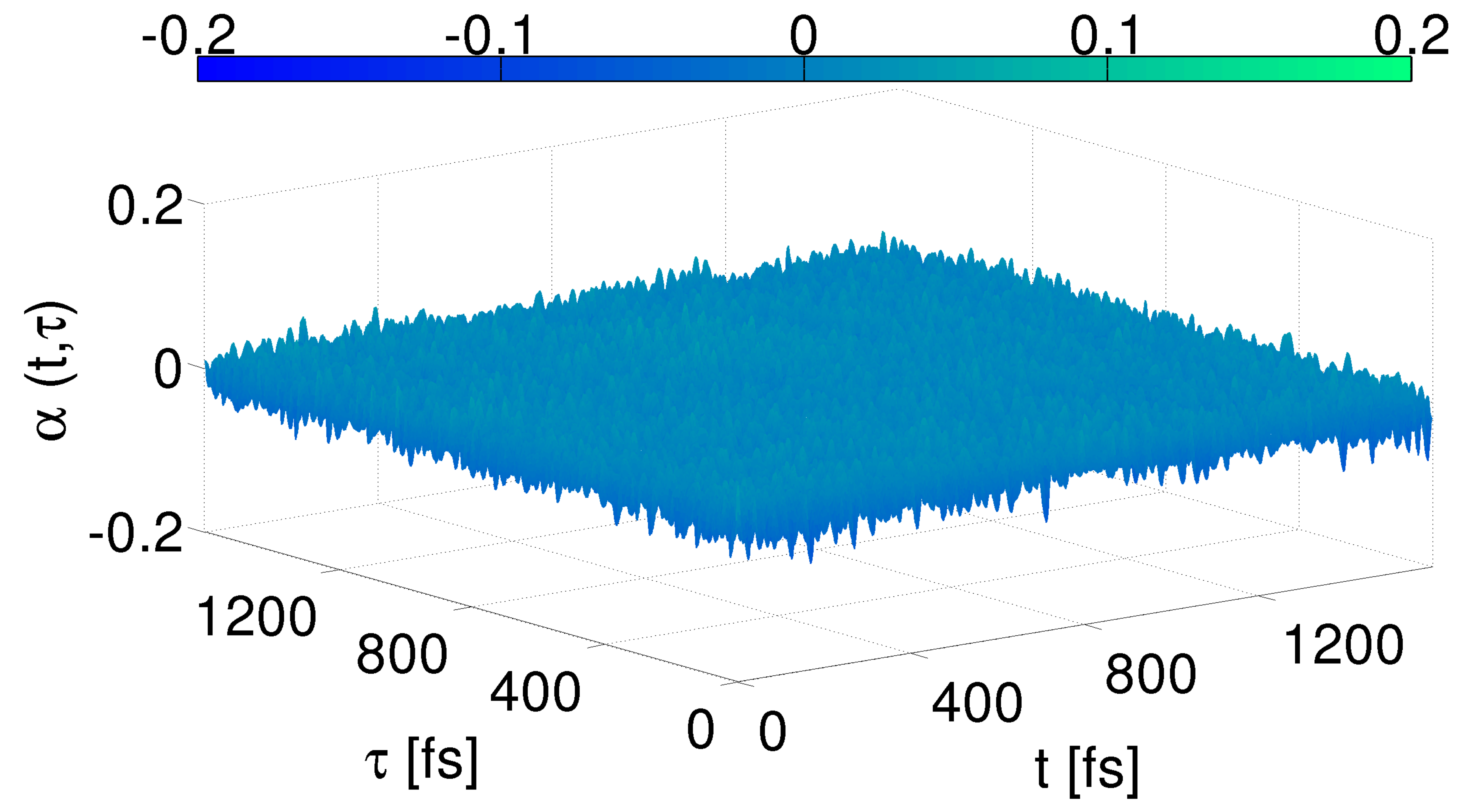}
\par\end{centering}

\caption{Same as for Fig. \ref{fig:corr_s_1} but for site 2 of the FMO complex.
\label{fig: corr_s_2} }
\end{figure*}

\begin{figure*}[H]
\begin{centering}
\includegraphics[width=1\columnwidth]{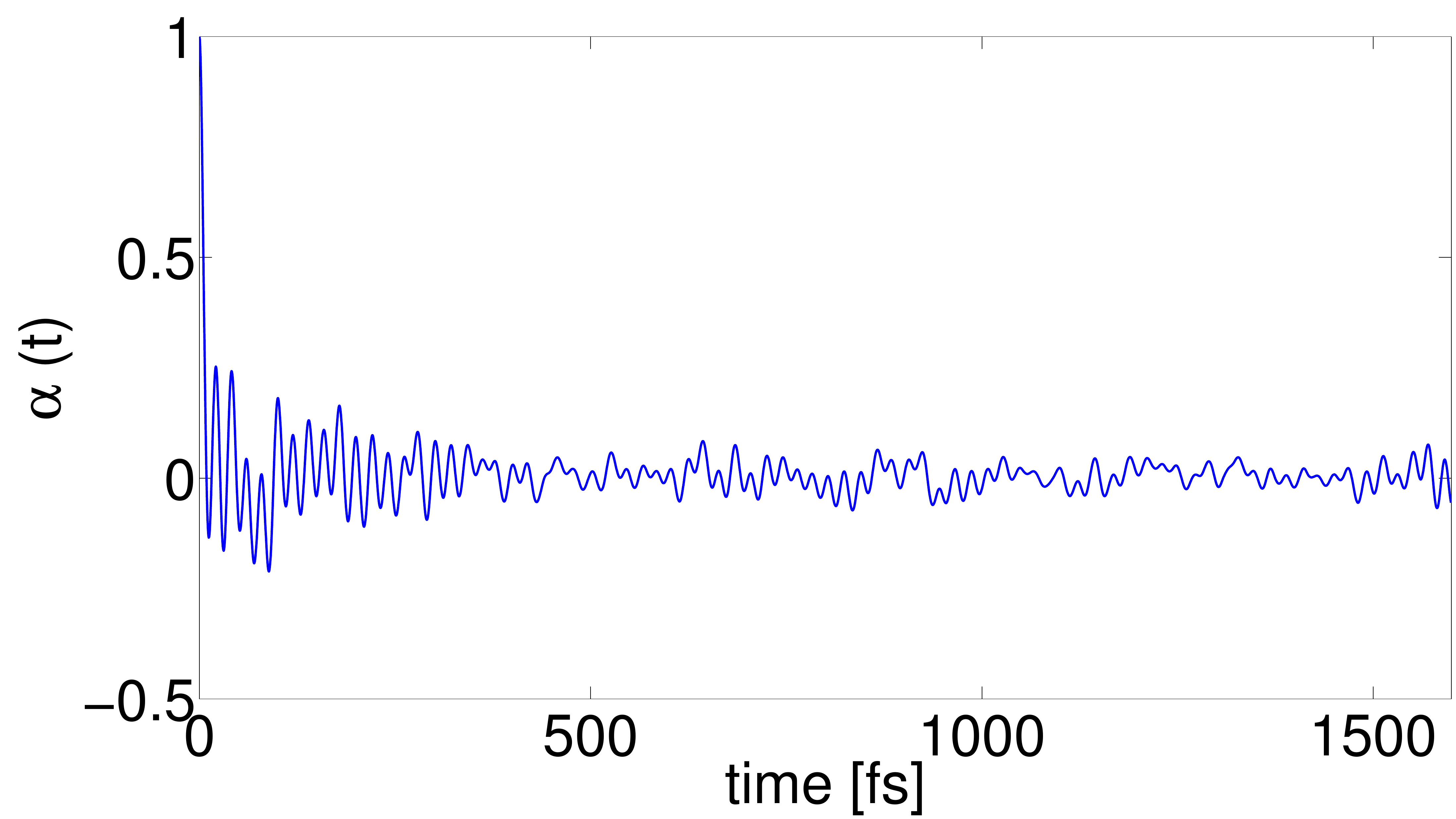}\includegraphics[width=1\columnwidth]{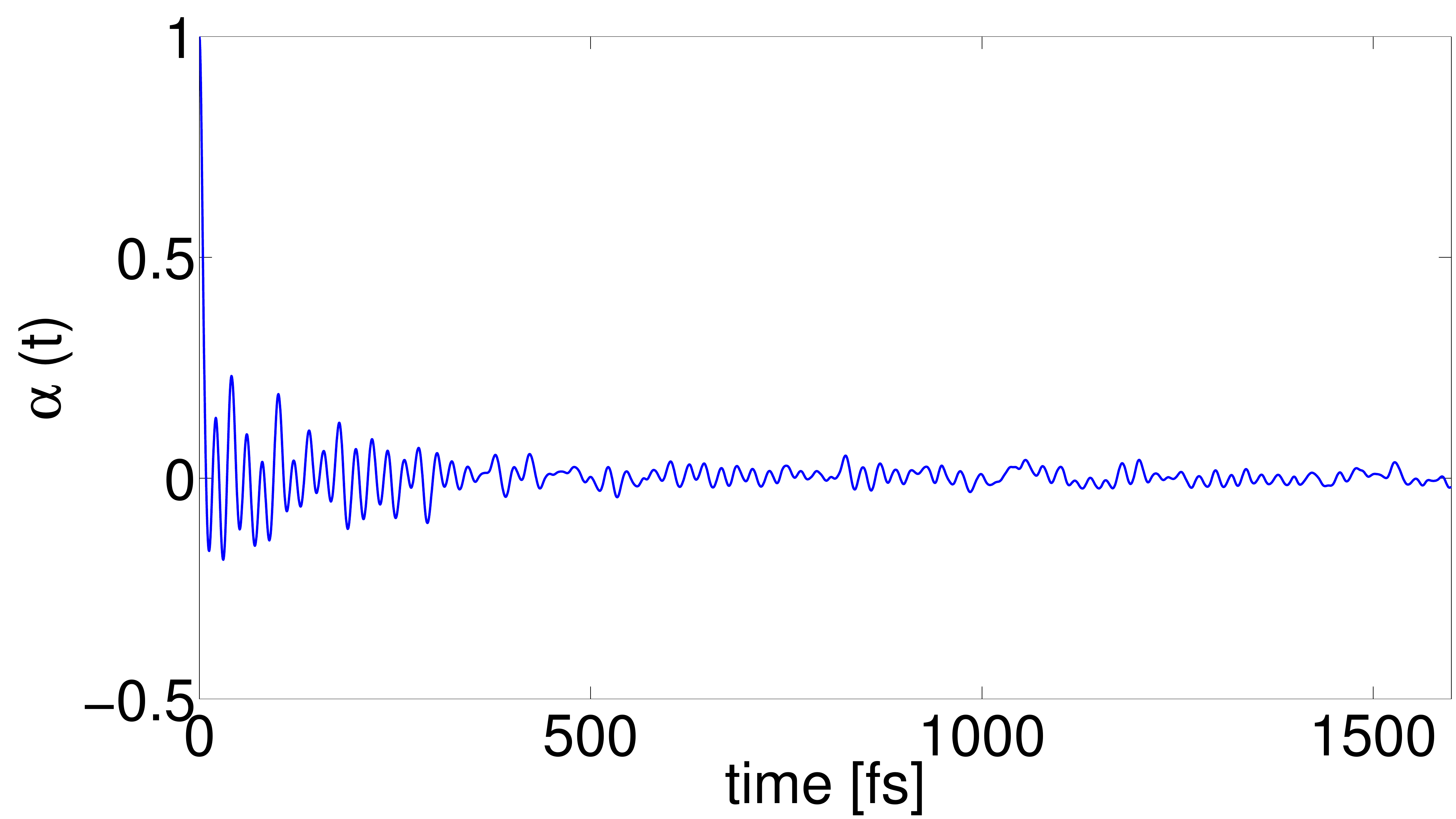}
\par\end{centering}

\begin{centering}
\includegraphics[width=1\columnwidth]{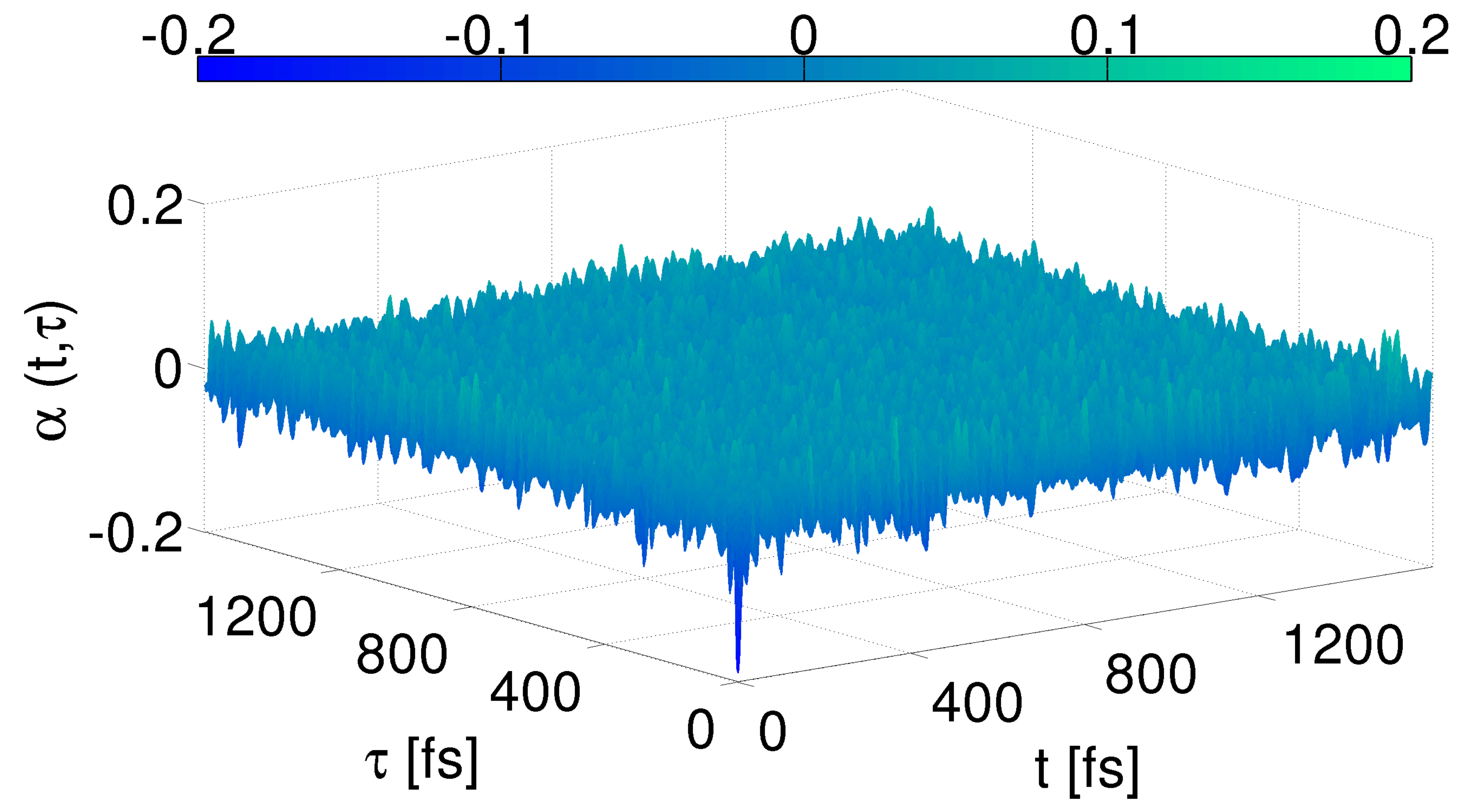}\includegraphics[width=1\columnwidth]{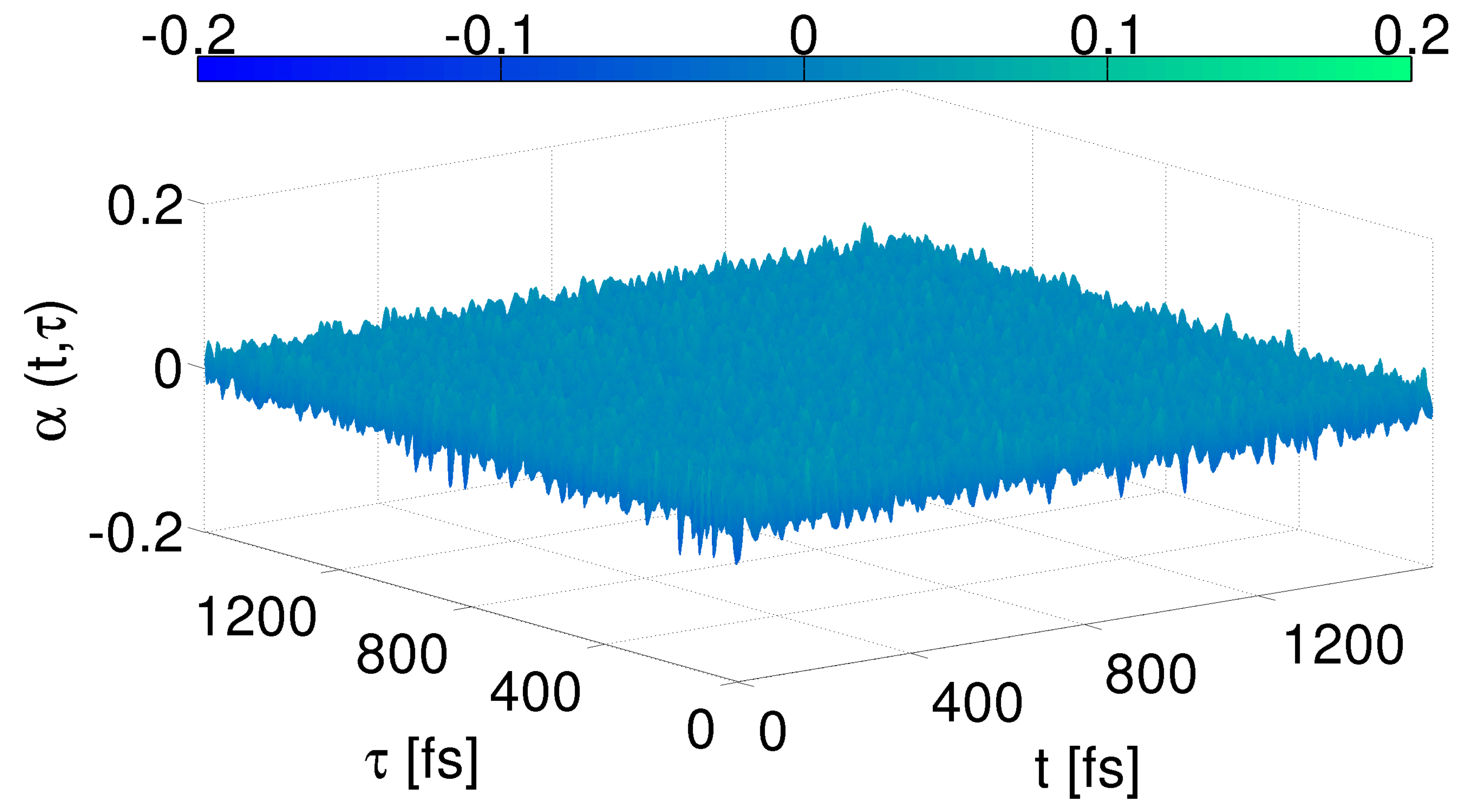}
\par\end{centering}

\caption{Same as for Fig. \ref{fig:corr_s_1} but for site 3 of the FMO complex.
\label{fig: corr_s_3} }
\end{figure*}

\begin{figure*}[H]
\begin{centering}
\includegraphics[width=1\columnwidth]{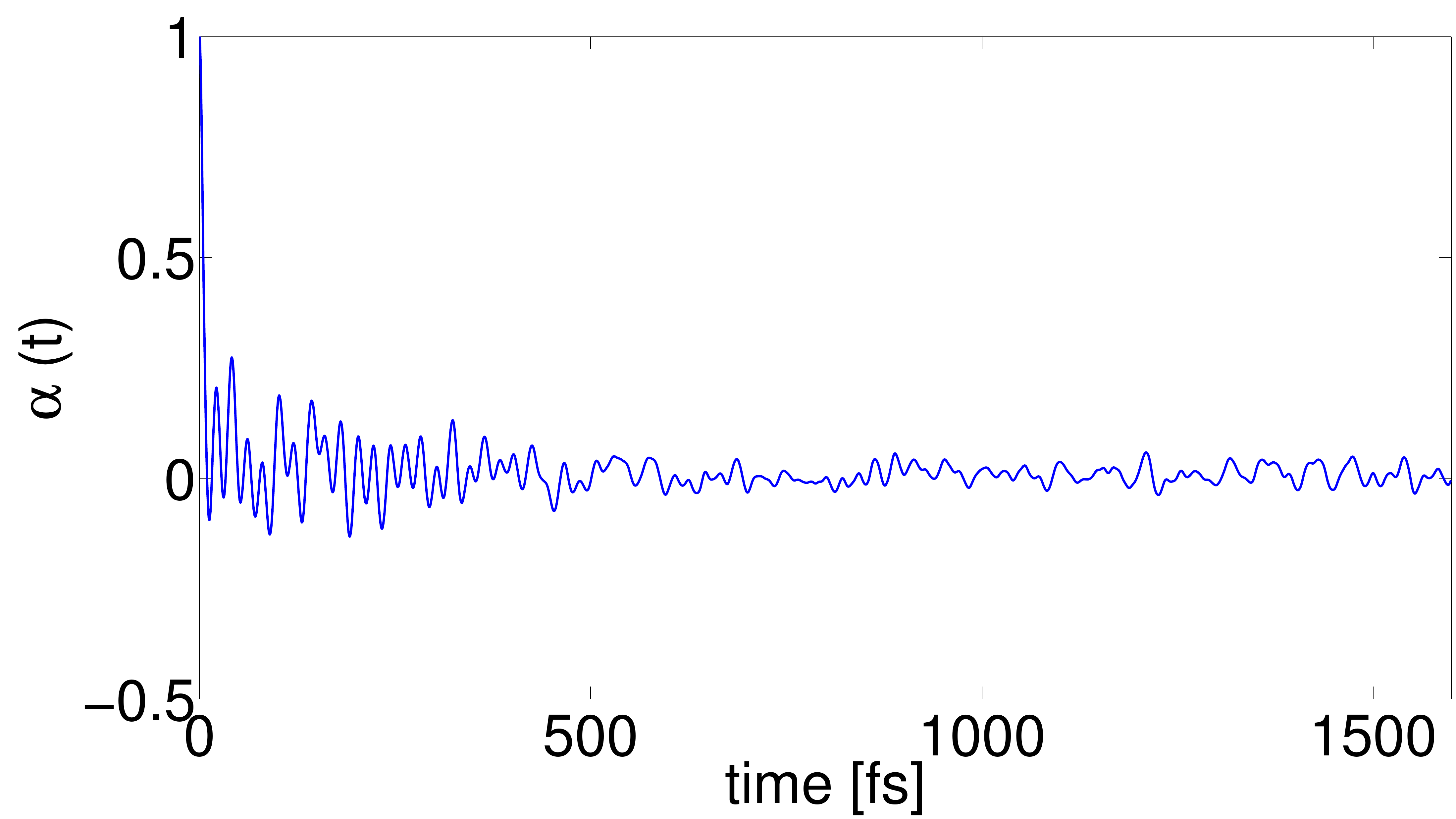}\includegraphics[width=1\columnwidth]{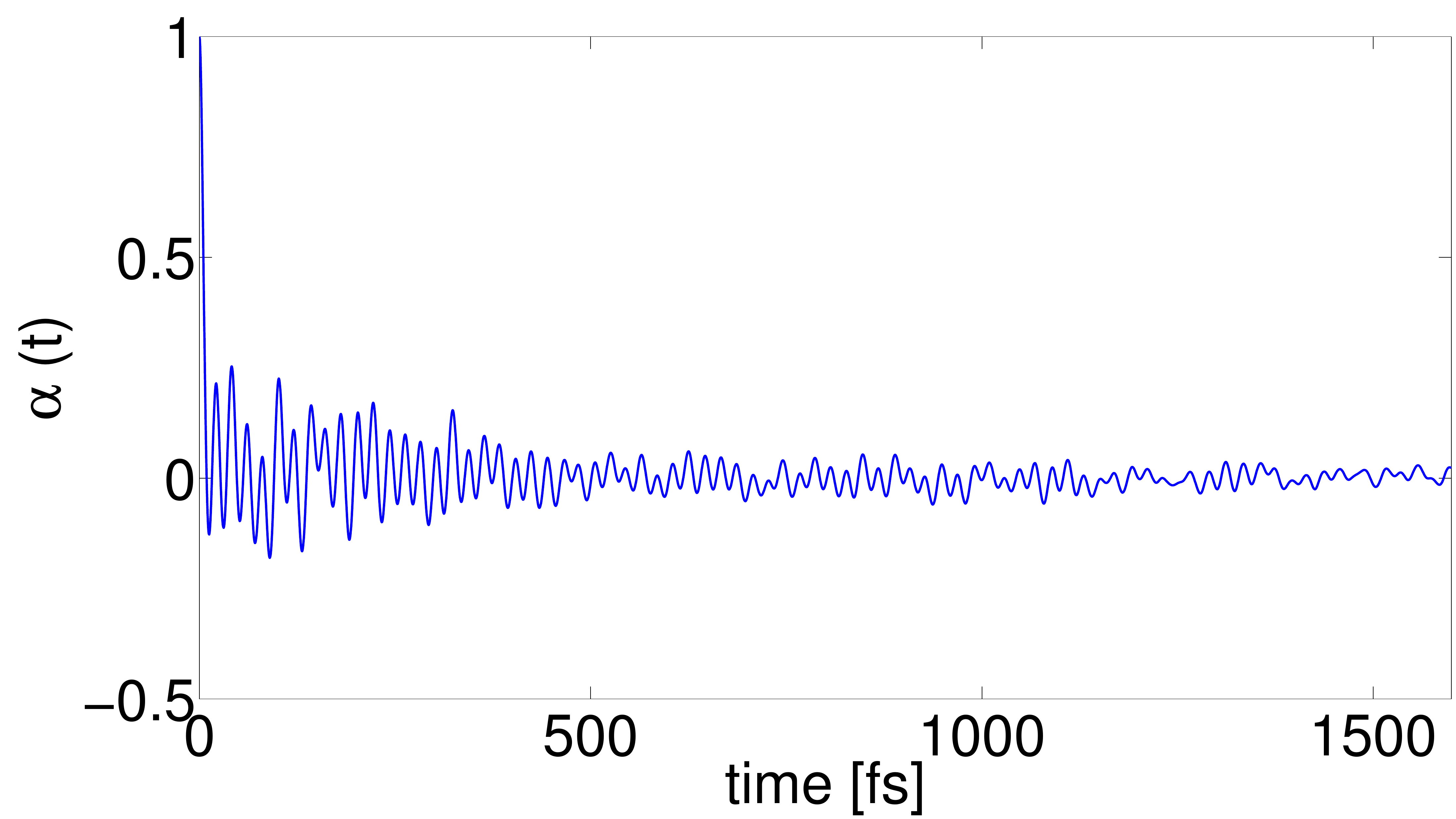}
\par\end{centering}

\begin{centering}
\includegraphics[width=1\columnwidth]{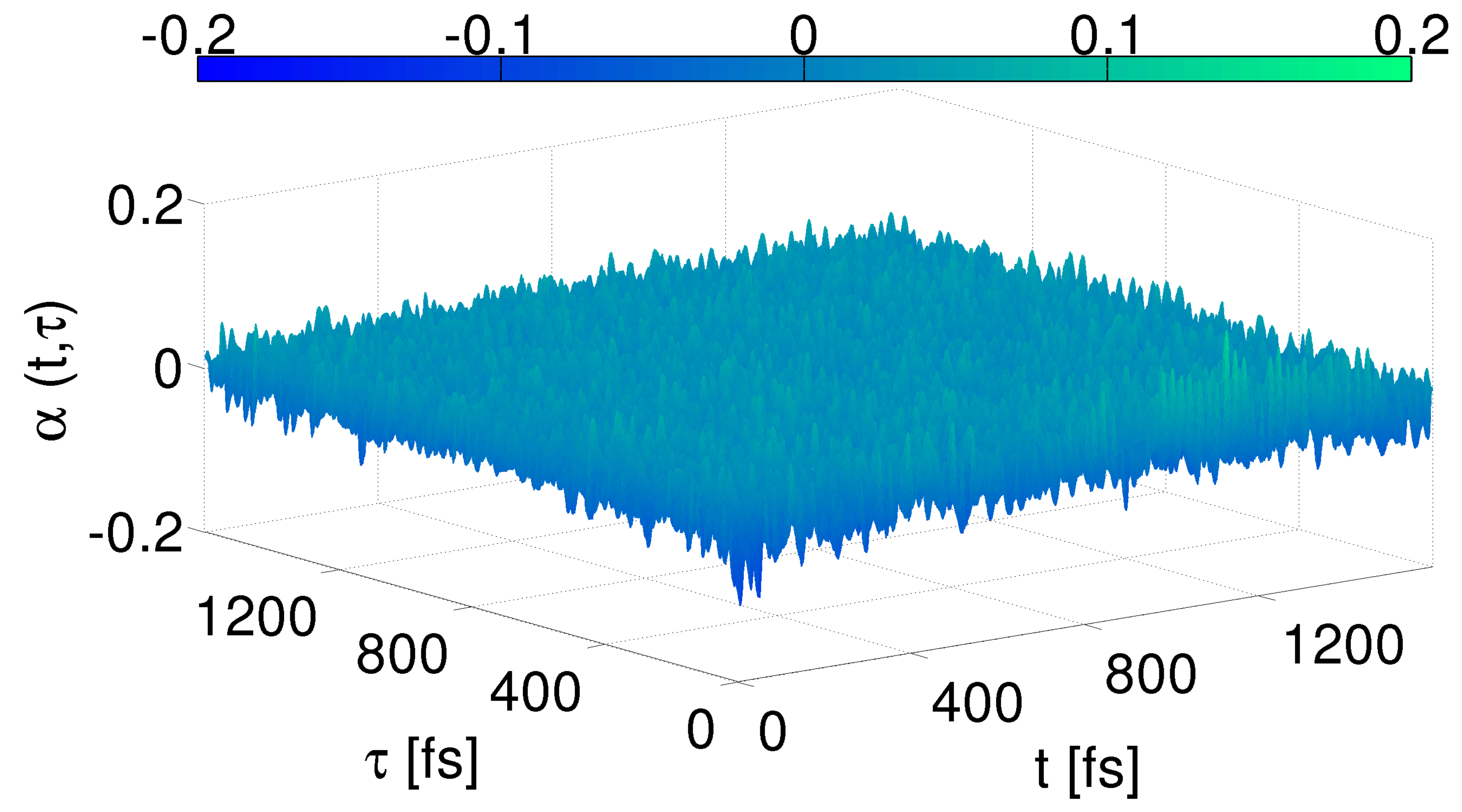}\includegraphics[width=1\columnwidth]{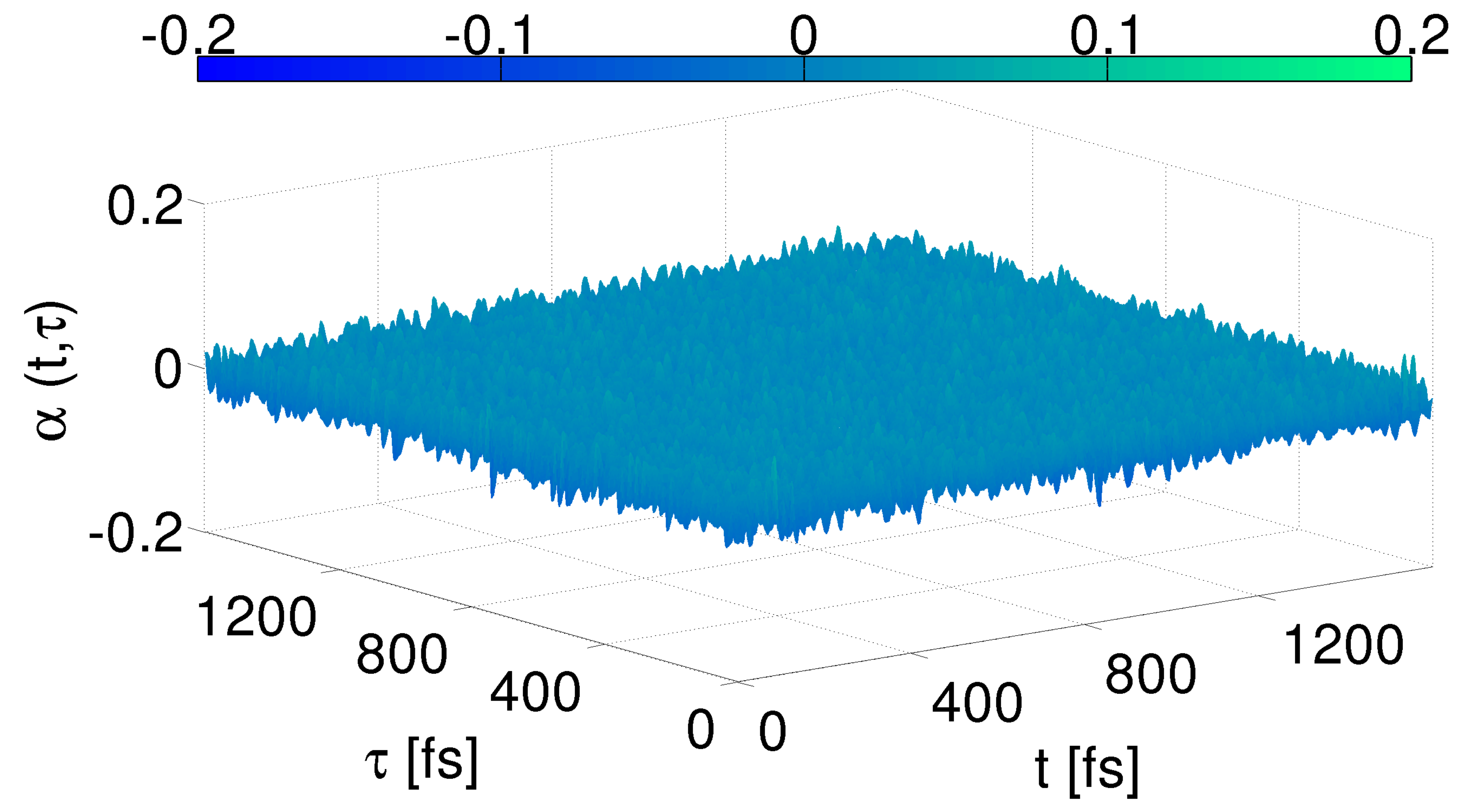}
\par\end{centering}

\caption{Same as for Fig. \ref{fig:corr_s_1} but for site 4 of the FMO complex.
\label{fig: corr_s_4} }
\end{figure*}

\begin{figure*}[H]
\begin{centering}
\includegraphics[width=1\columnwidth]{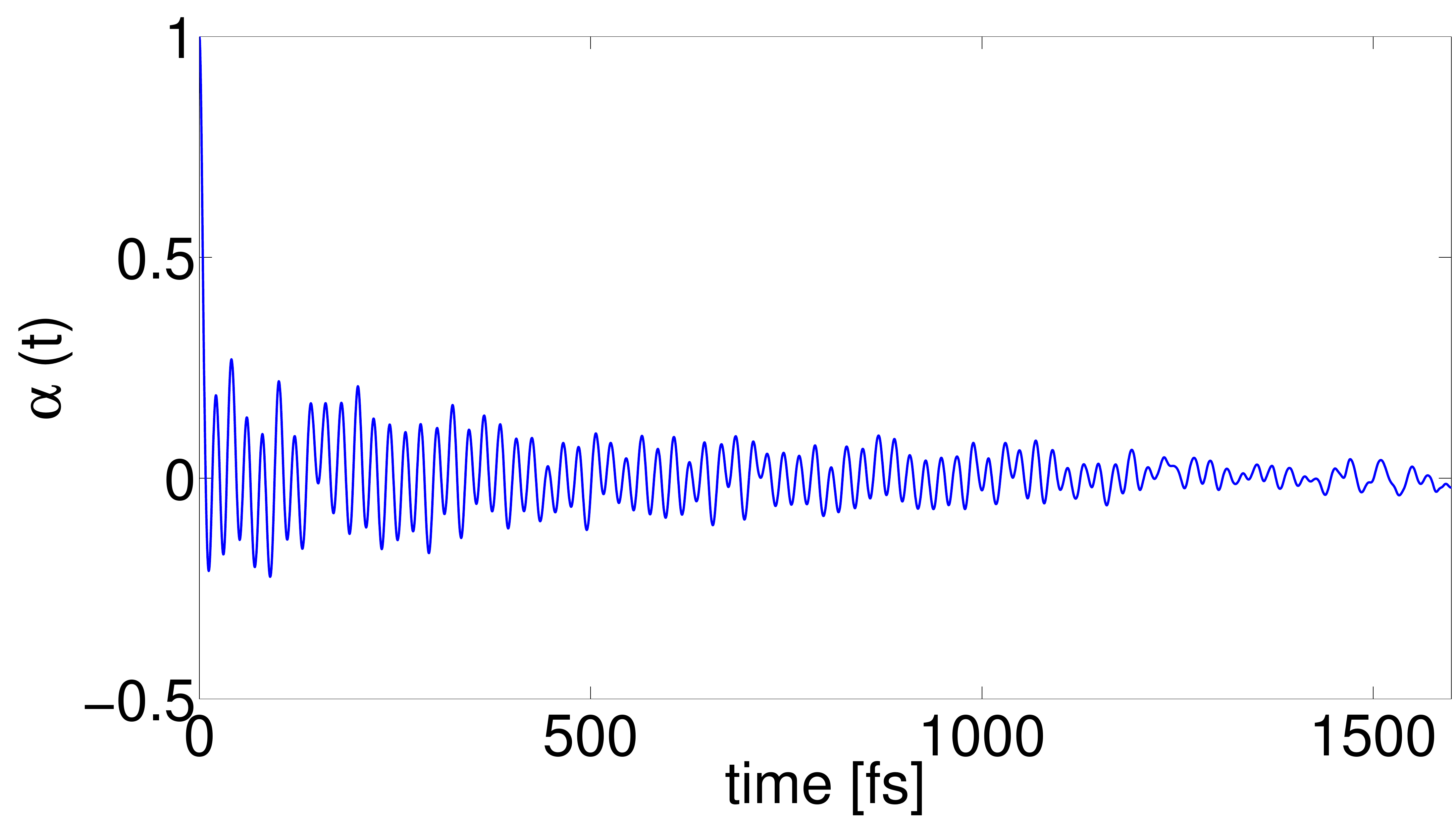}\includegraphics[width=1\columnwidth]{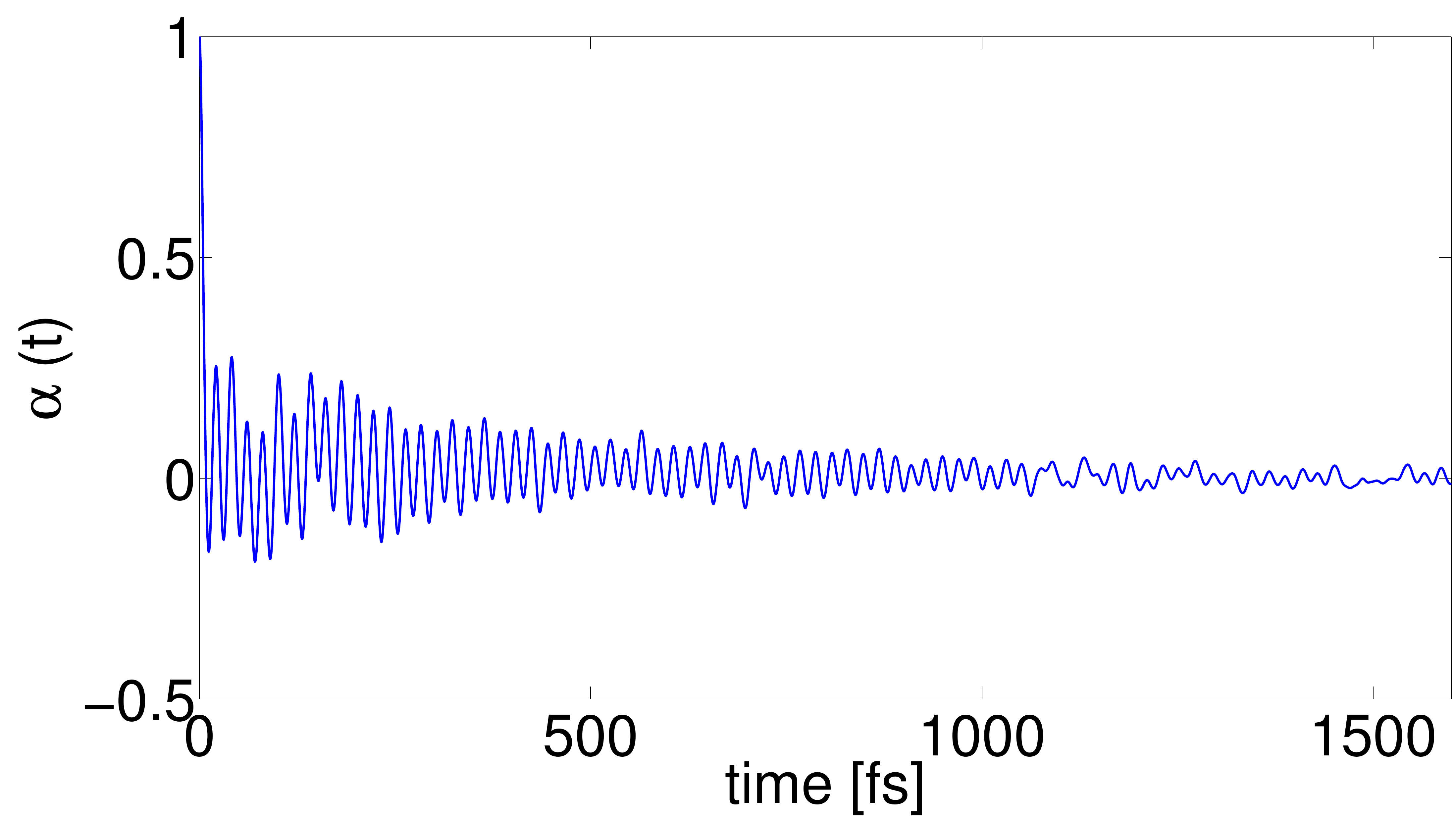}
\par\end{centering}

\begin{centering}
\includegraphics[width=1\columnwidth]{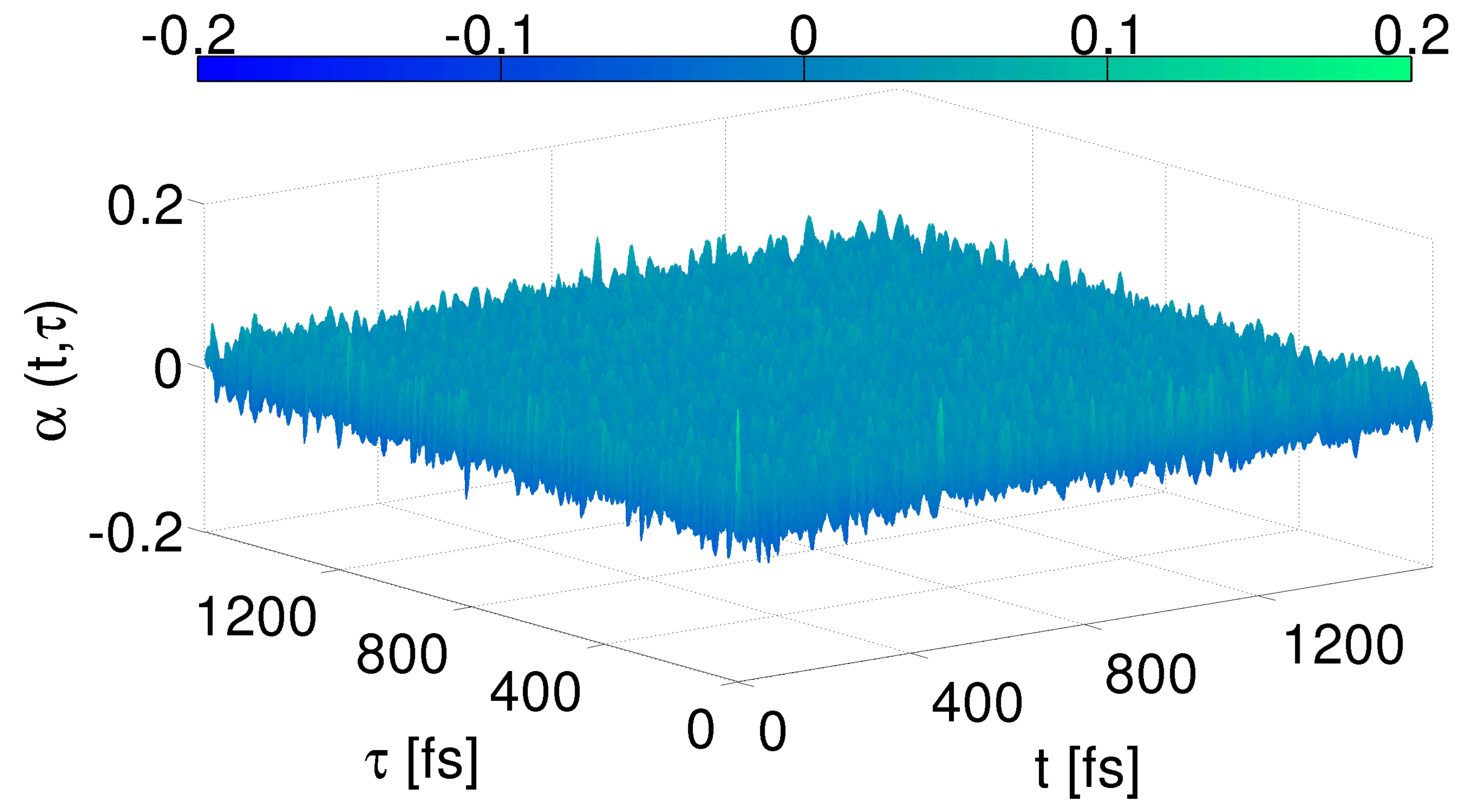}\includegraphics[width=1\columnwidth]{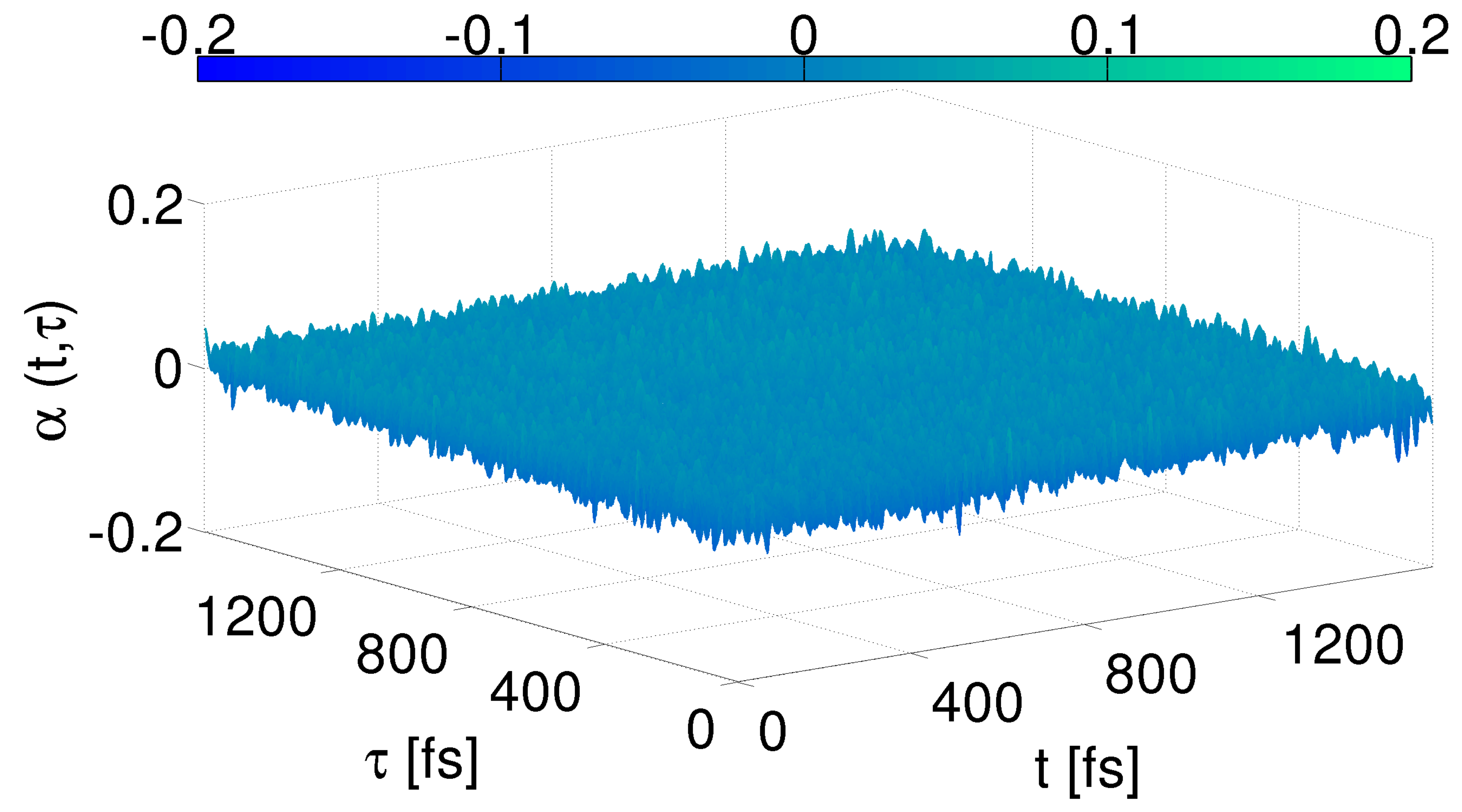}
\par\end{centering}

\caption{Same as for Fig. \ref{fig:corr_s_1} but for site 5 of the FMO complex.
\label{fig:corr_s_5} }
\end{figure*}

\begin{figure*}[H]
\begin{centering}
\includegraphics[width=1\columnwidth]{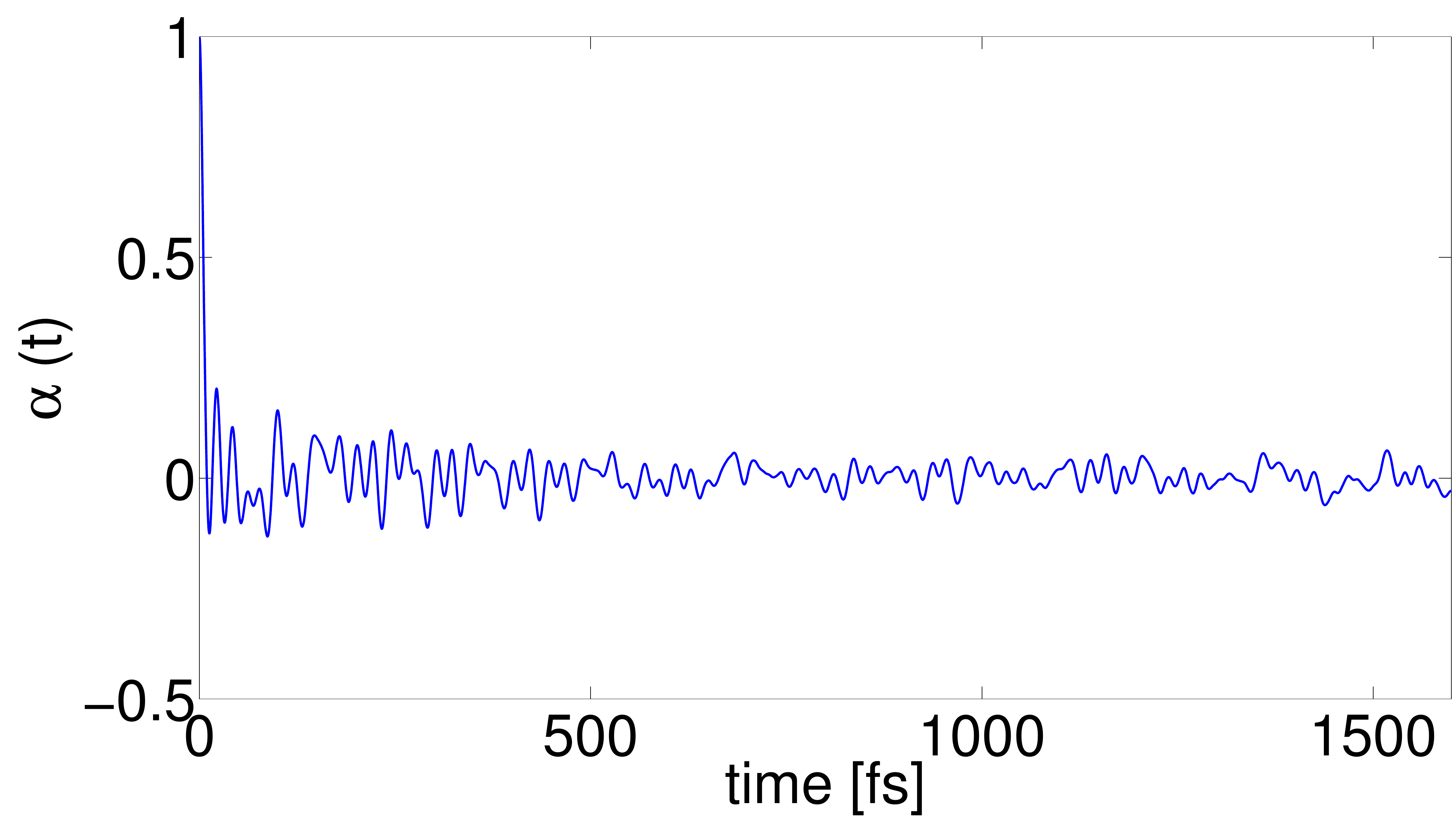}\includegraphics[width=1\columnwidth]{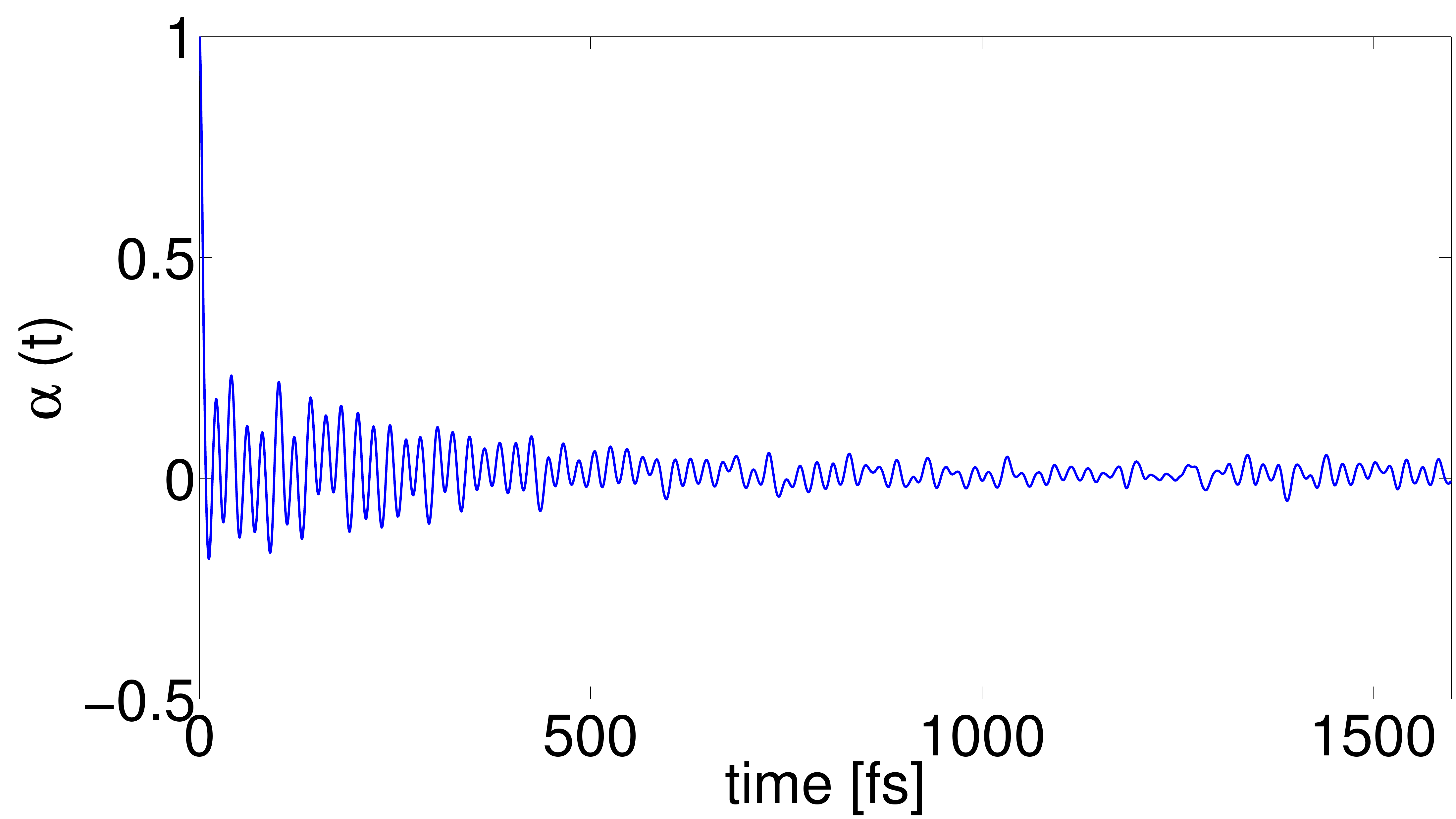}
\par\end{centering}

\begin{centering}
\includegraphics[width=1\columnwidth]{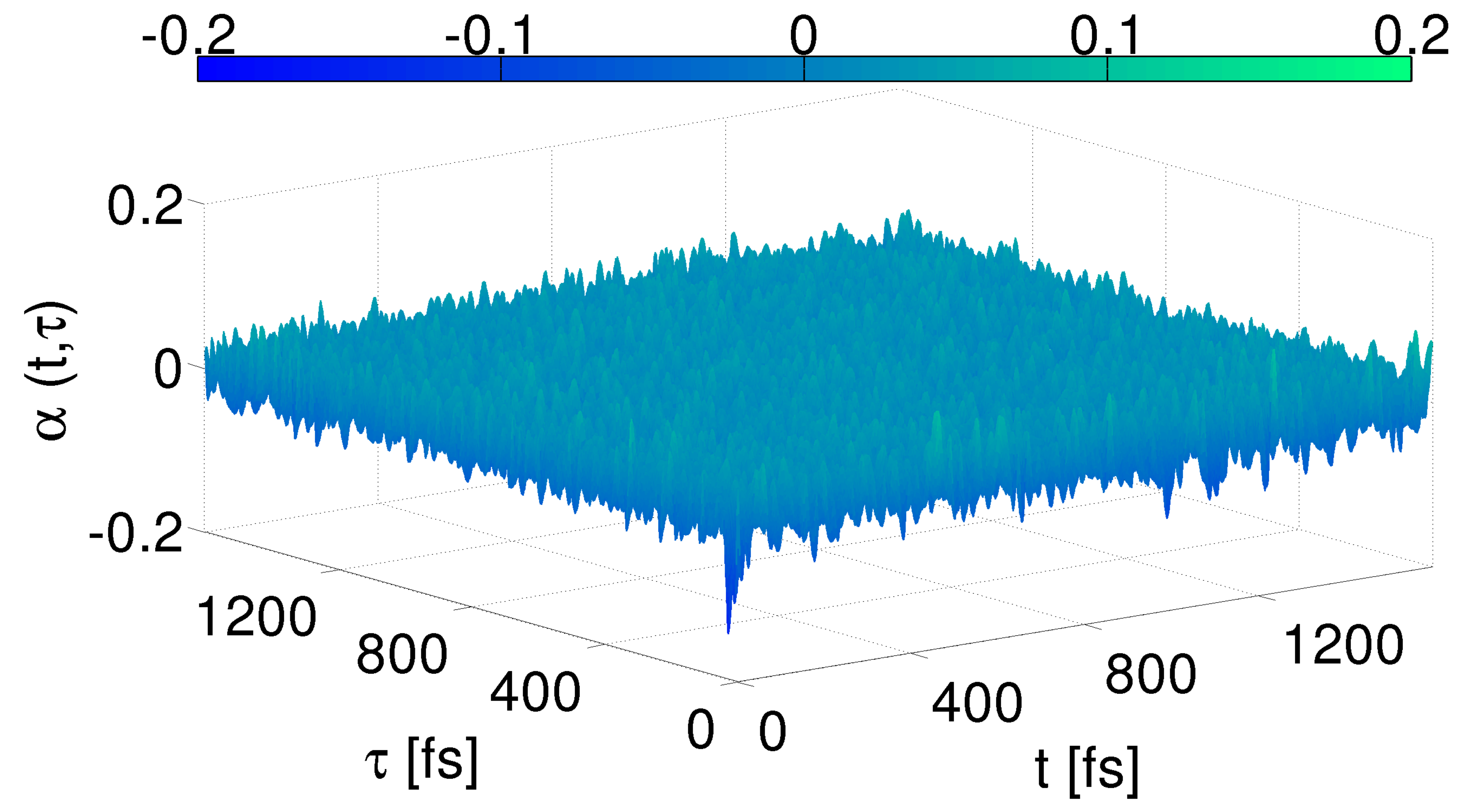}\includegraphics[width=1\columnwidth]{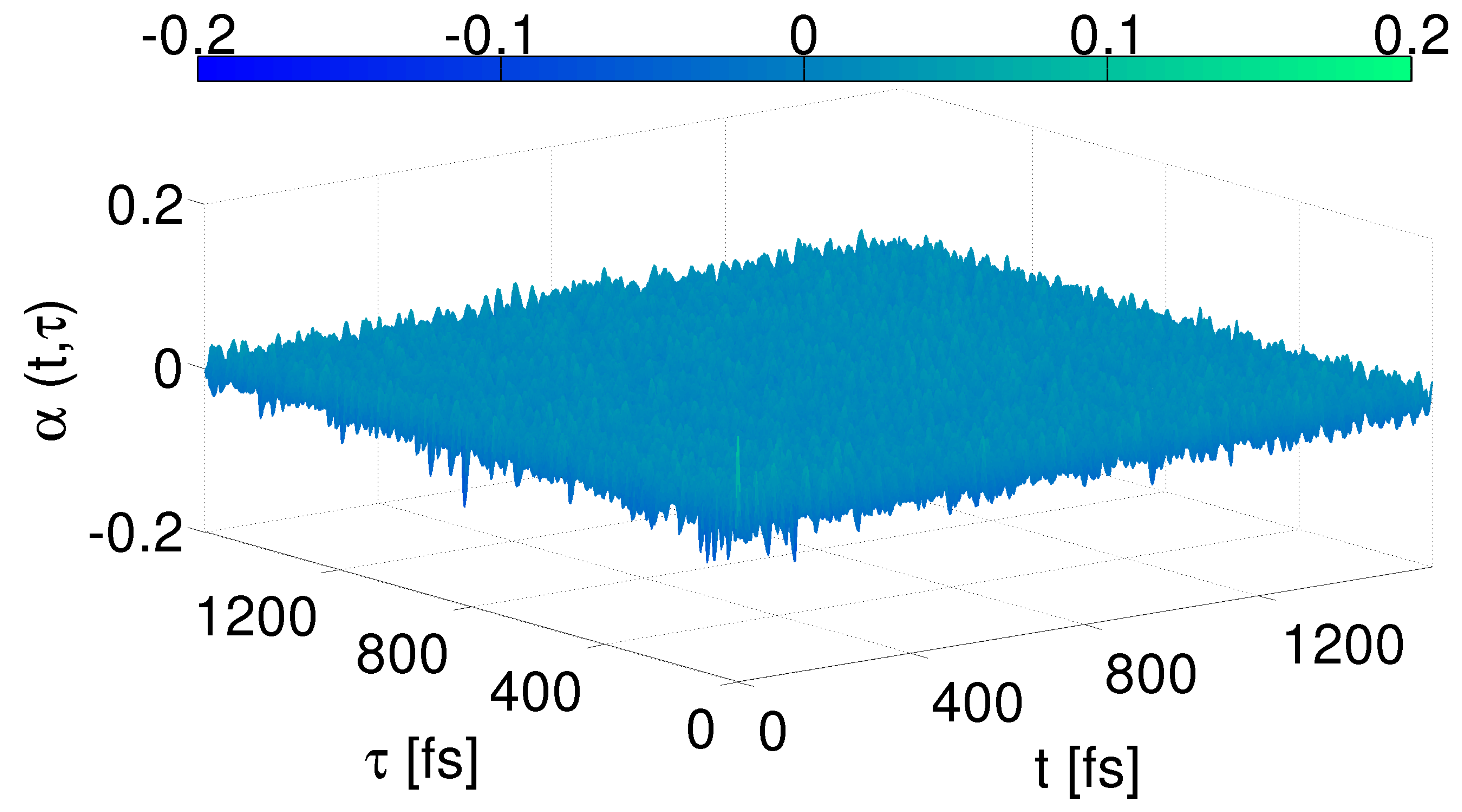}
\par\end{centering}

\caption{Same as for Fig. \ref{fig:corr_s_1} but for site 6 of the FMO complex.
\label{fig:corr_s_6} }
\end{figure*}

\begin{figure*}[H]
\begin{centering}
\includegraphics[width=1\columnwidth]{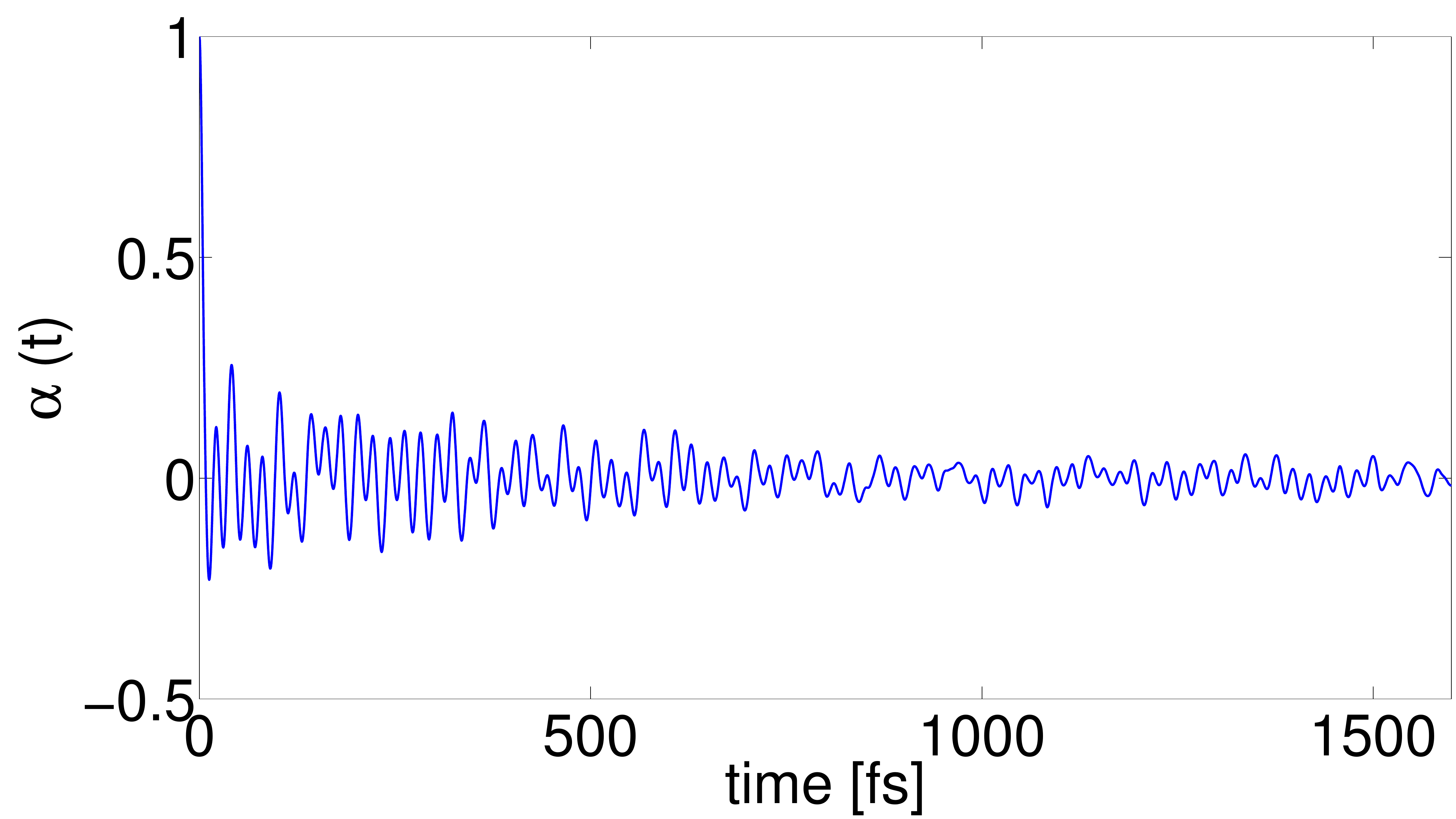}\includegraphics[width=1\columnwidth]{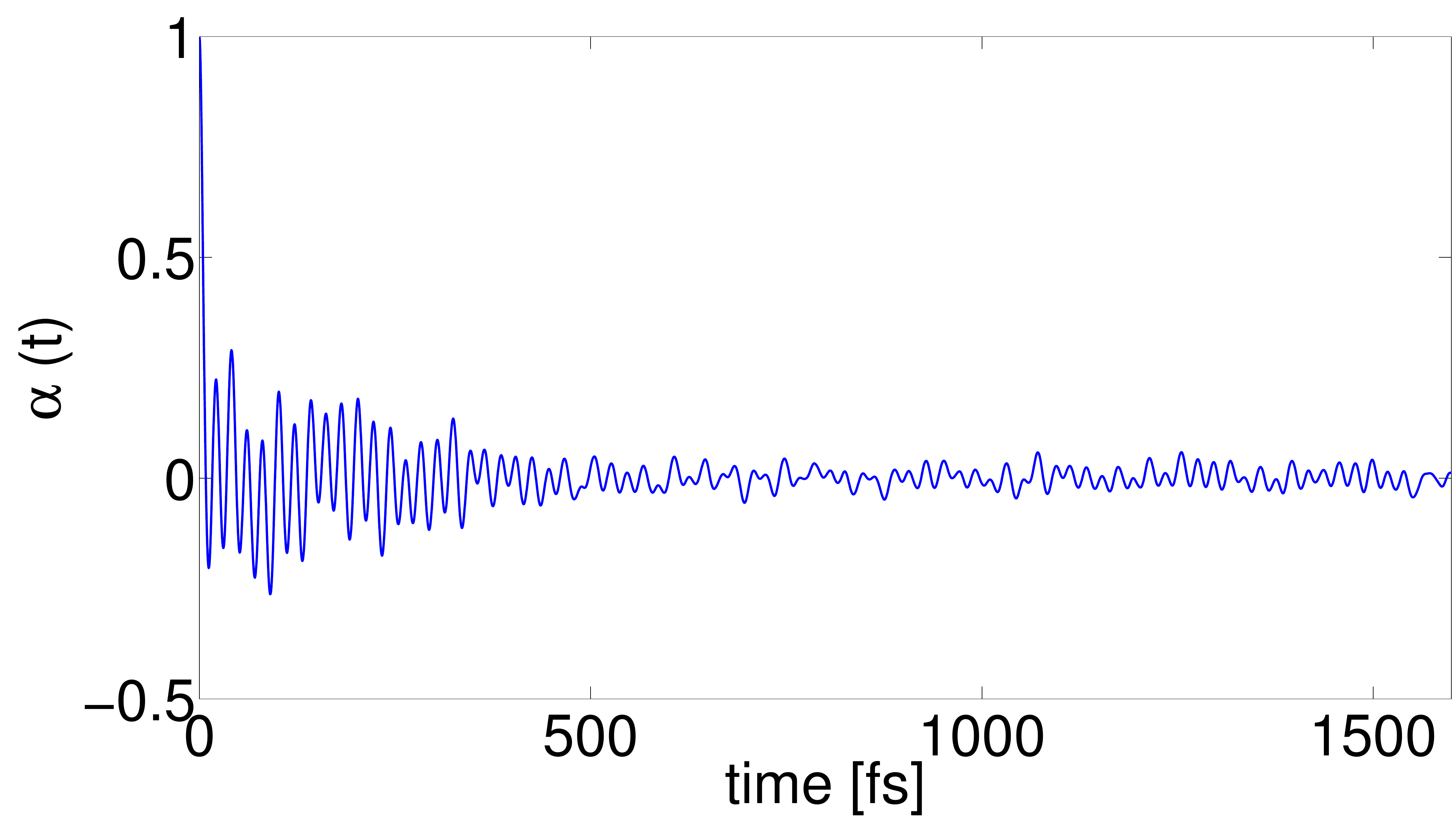}
\par\end{centering}

\begin{centering}
\includegraphics[width=1\columnwidth]{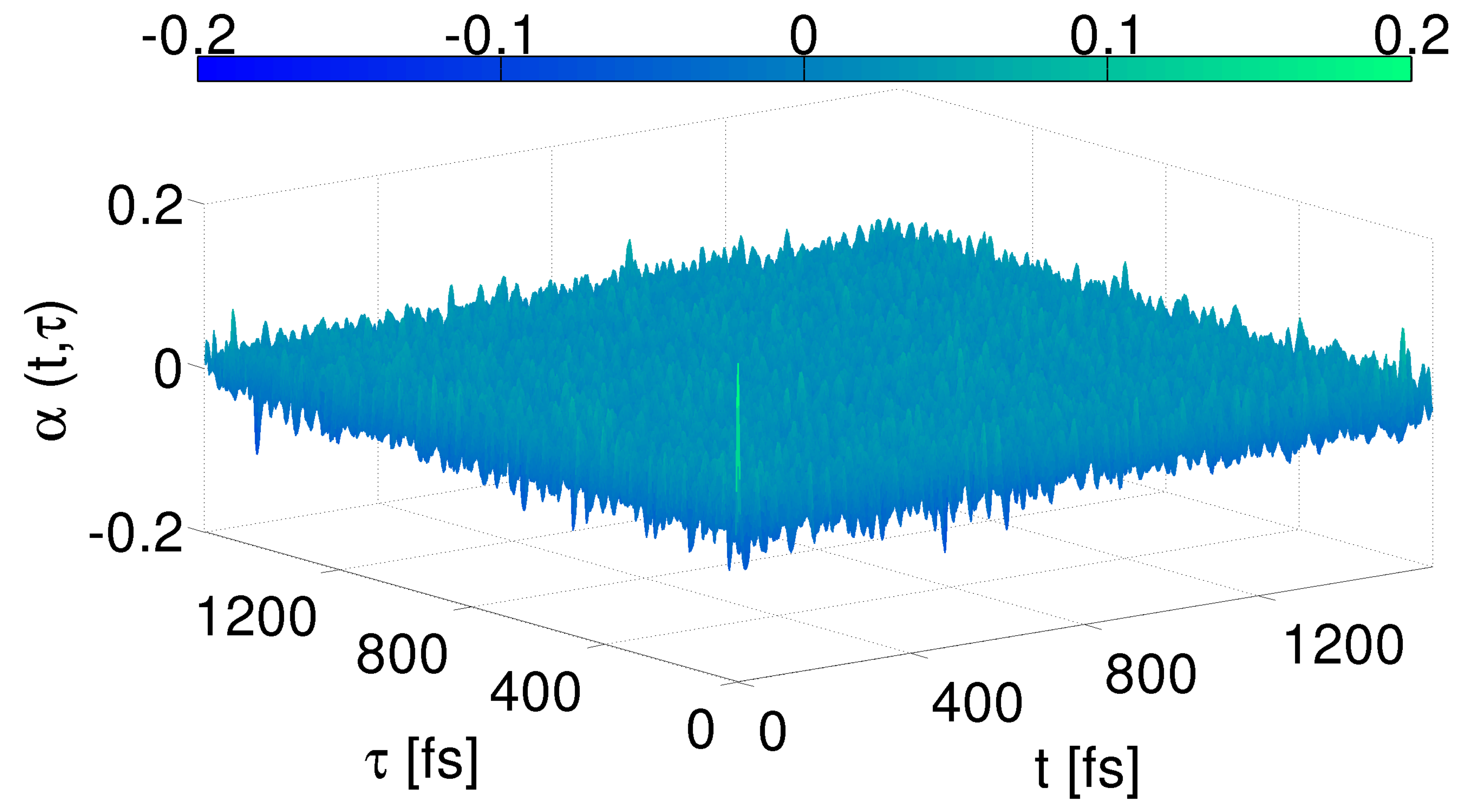}\includegraphics[width=1\columnwidth]{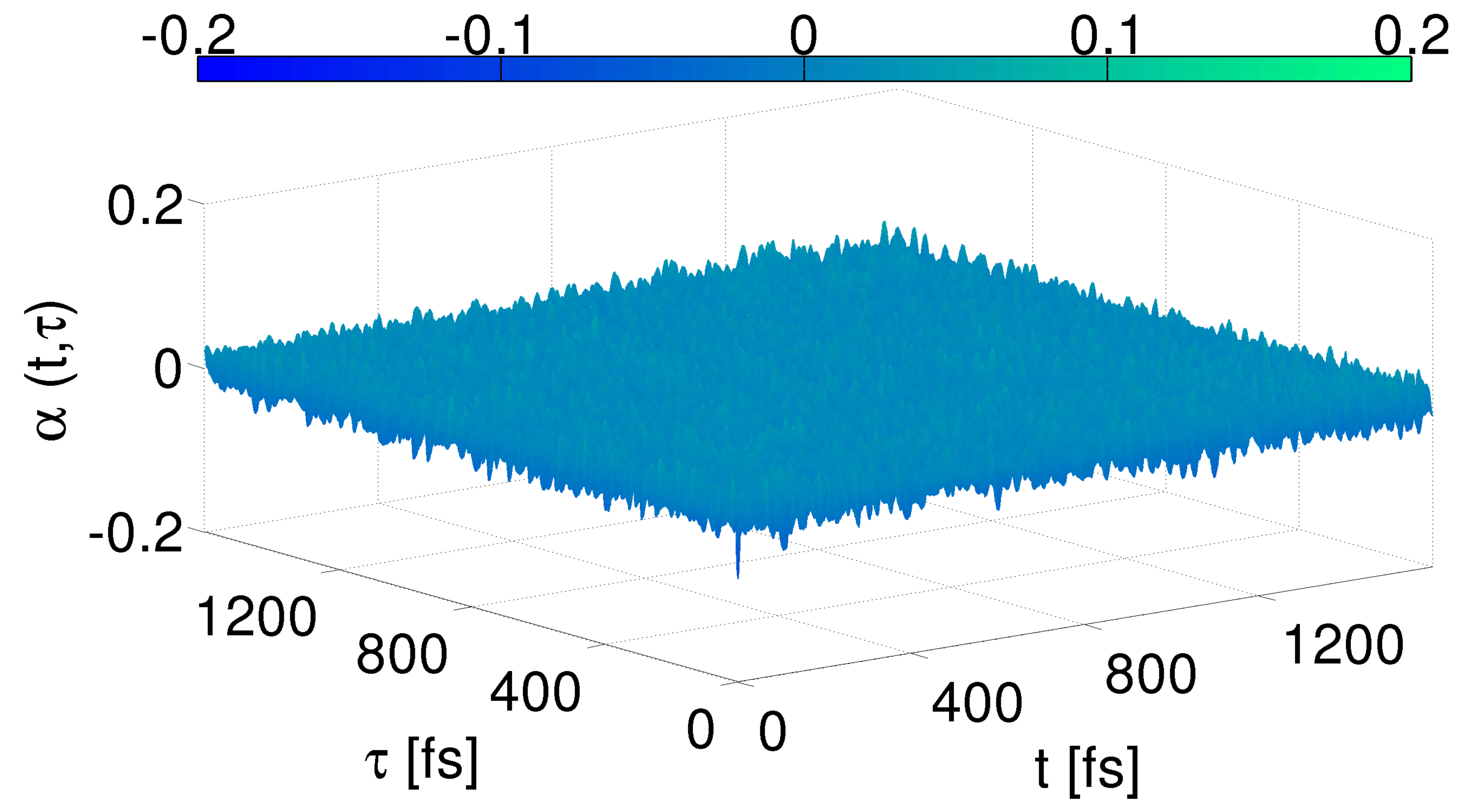}
\par\end{centering}

\caption{Same as for Fig. \ref{fig:corr_s_1} but for site 7 of the FMO complex.
\label{fig: corr_s_7} }
\end{figure*}

In this Section we report the two and three-time correlation functions
for the energy gap fluctuations of the FMO complex as discussed in
the text for all seven sites of the complex and for both temperatures,
77 and 300K. The functions are rescaled by increasing powers of the
standard deviation $s=\sqrt{m^{(2)}}$ as a means of comparison. We
notice, by comparing panels a) and b) to c) and d) in Figures \ref{fig:corr_s_1}-\ref{fig: corr_s_7},
that the three-time correlation function is always much smaller in
amplitude than the corresponding two-time autocorrelation function.
This supports the idea that the harmonic bath and linear coupling
approximations are good for this system. 

\section{Harmonic spectral densities for all site of the fmo complex}

\begin{figure*}[H]
\begin{raggedright}
\includegraphics[clip,angle=-90,width=1\columnwidth]{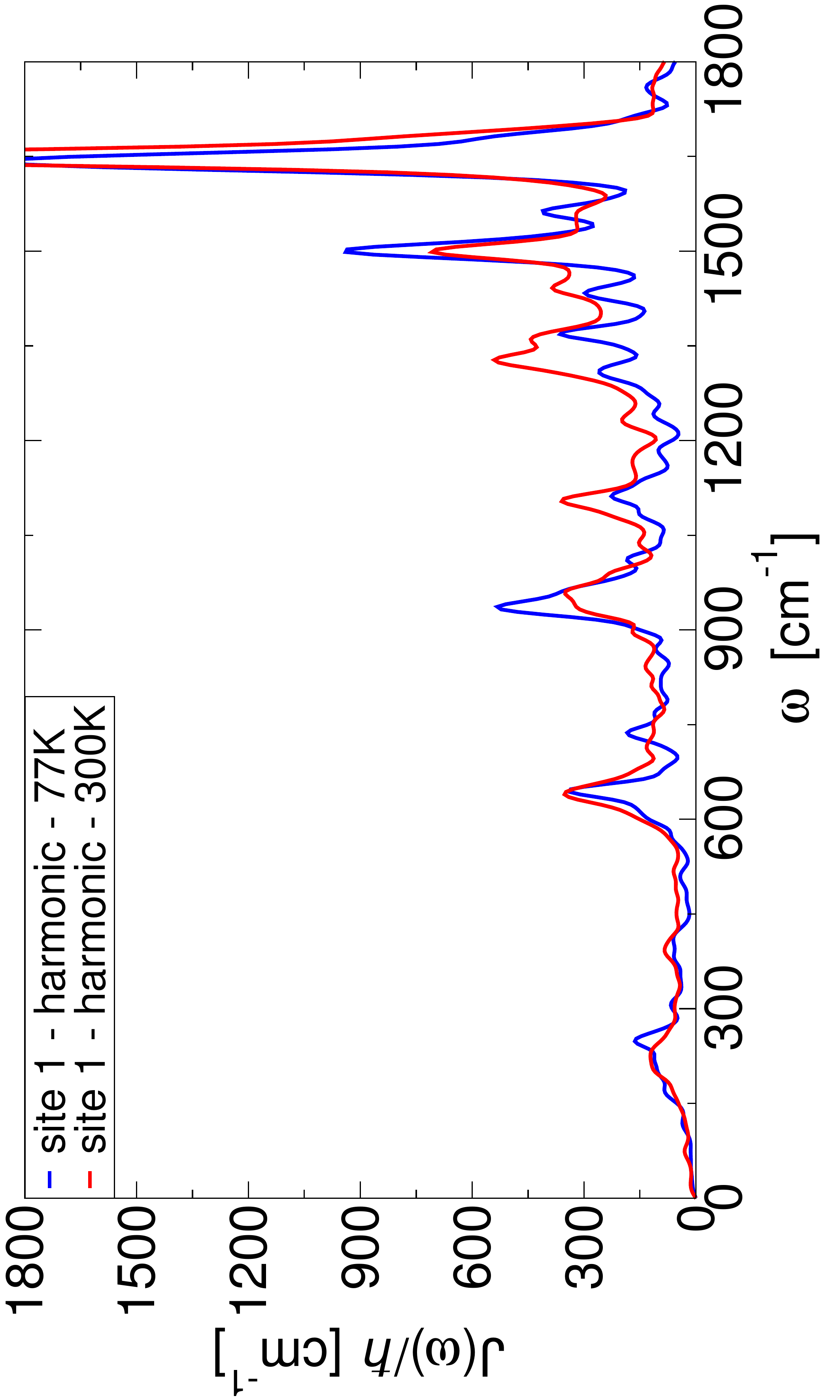}\includegraphics[angle=-90,width=1\columnwidth]{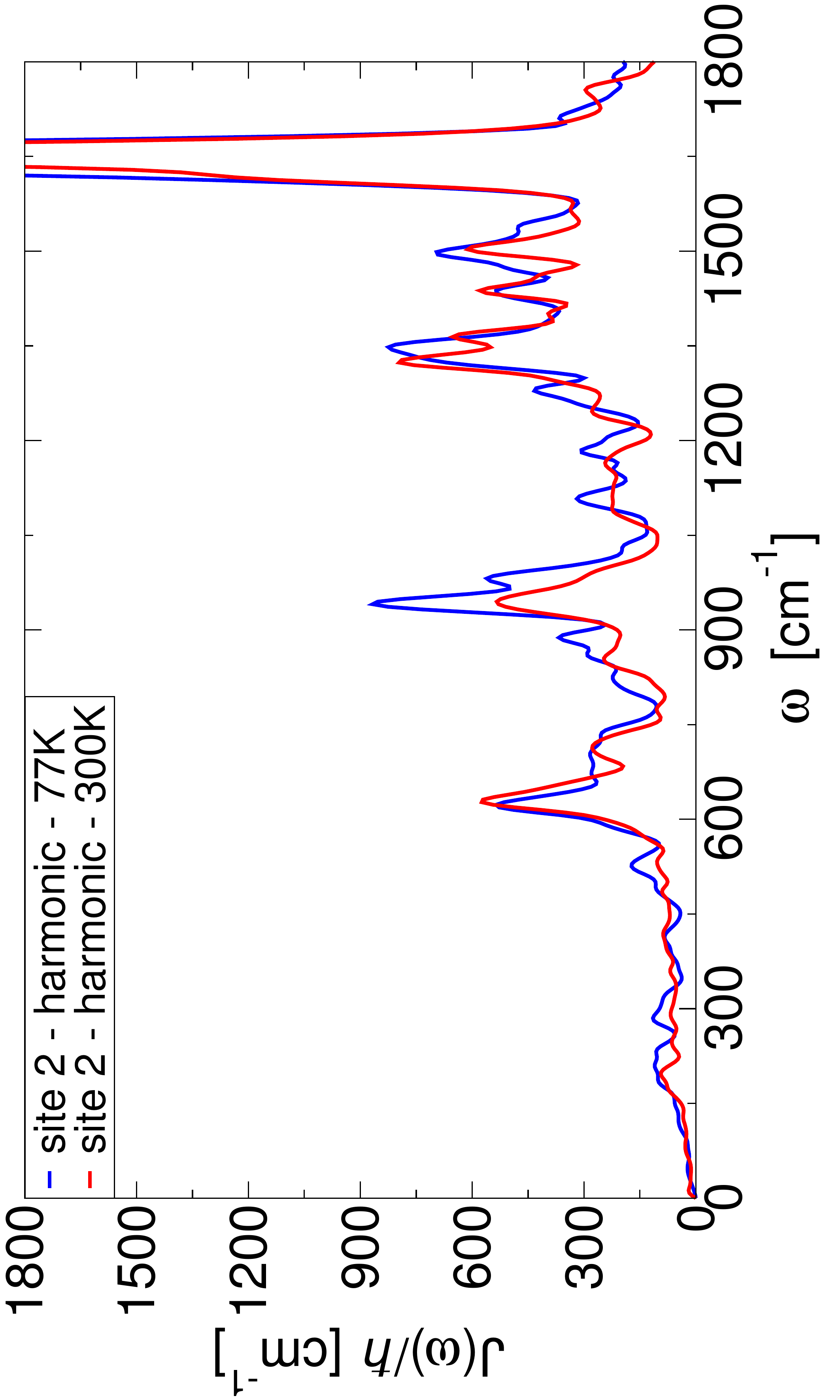}
\par\end{raggedright}

\begin{raggedright}
\includegraphics[angle=-90,width=1\columnwidth]{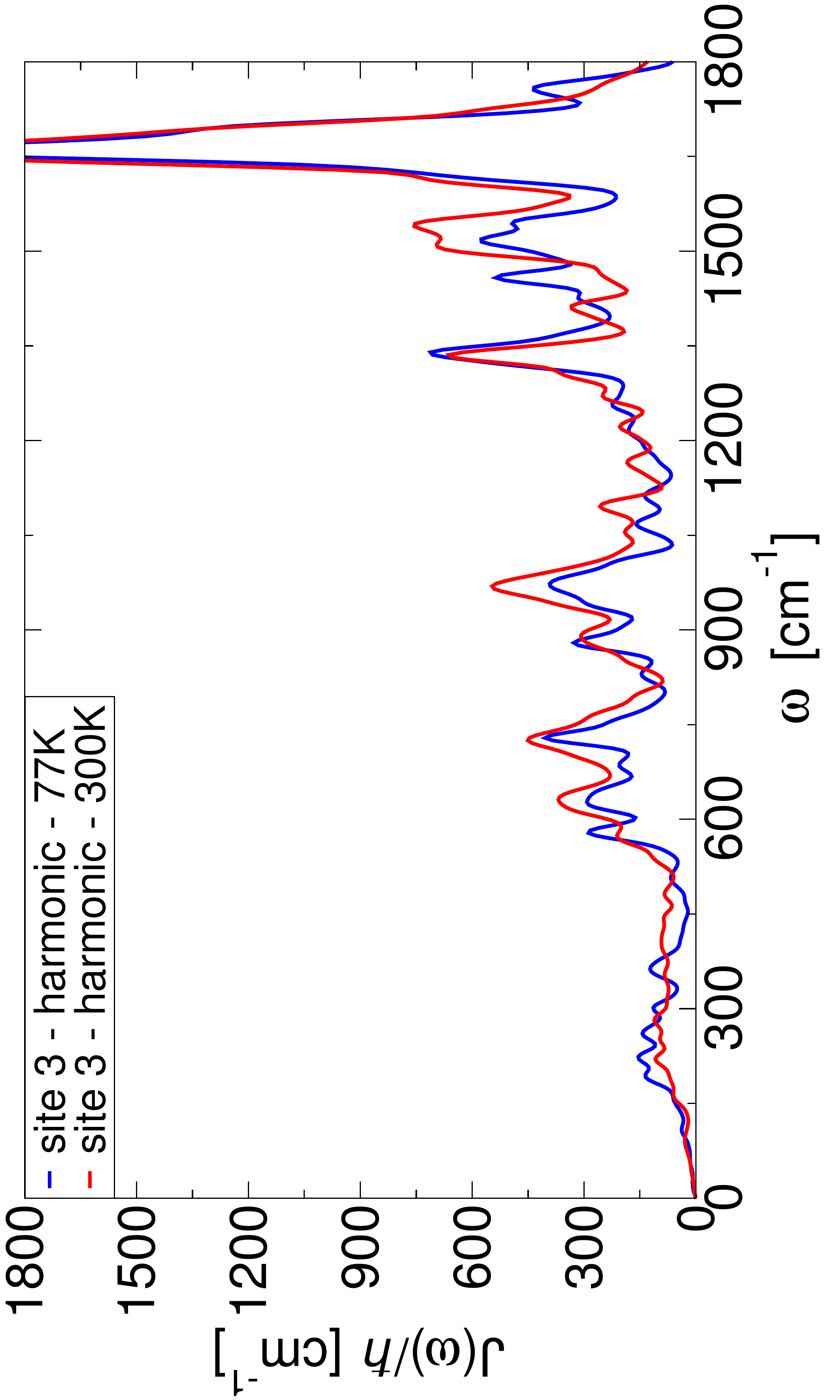}\includegraphics[angle=-90,width=1\columnwidth]{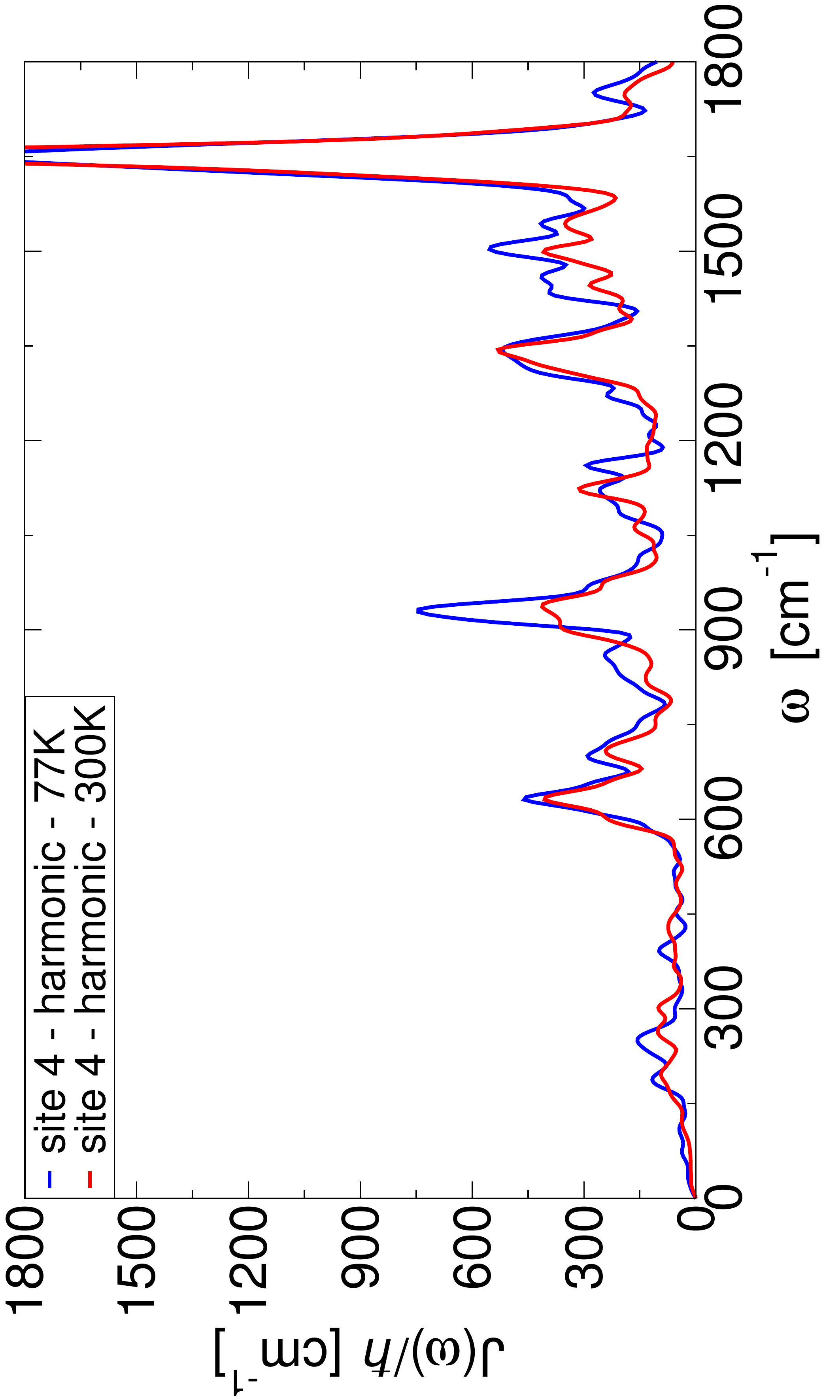}
\par\end{raggedright}

\caption{Comparison of the asymmetric component of temperature-dependent coupling
density $G_{{\rm asym}}(\omega)=J(\omega)\,;$ for sites 1-4 of the
Fenna-Matthews-Olson complex, obtained with the Harmonic approximation
(As described in the text) at 77K and at 300K. Note that, as described
in the text, the spectral density can be obtained by dividing $J(\omega)$
by $\pi$. \label{fig:Harmonic_SD_site17}}
\end{figure*}

\begin{figure*}[H]
\begin{centering}
\includegraphics[angle=-90,width=1\columnwidth]{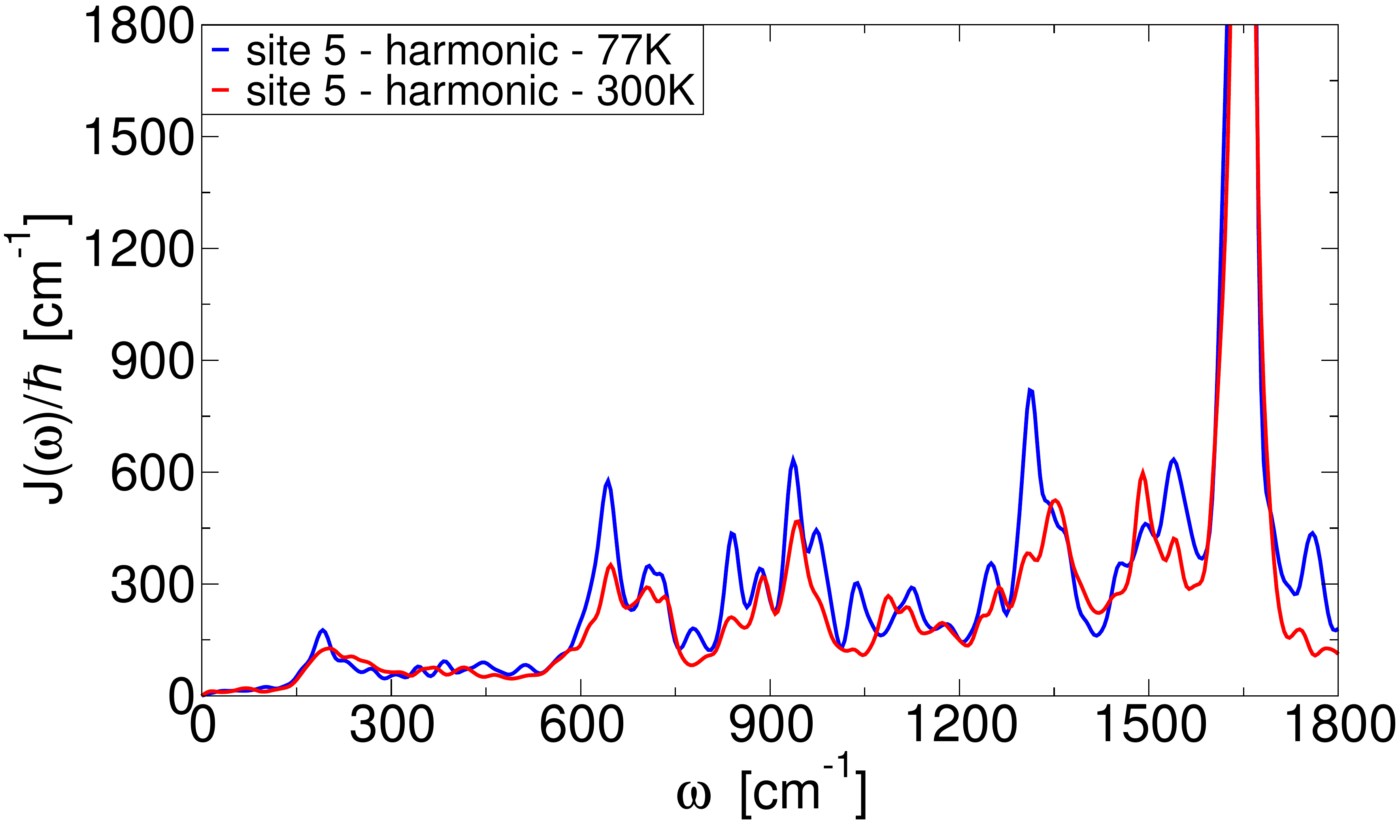}
\par\end{centering}

\begin{centering}
\includegraphics[angle=-90,width=1\columnwidth]{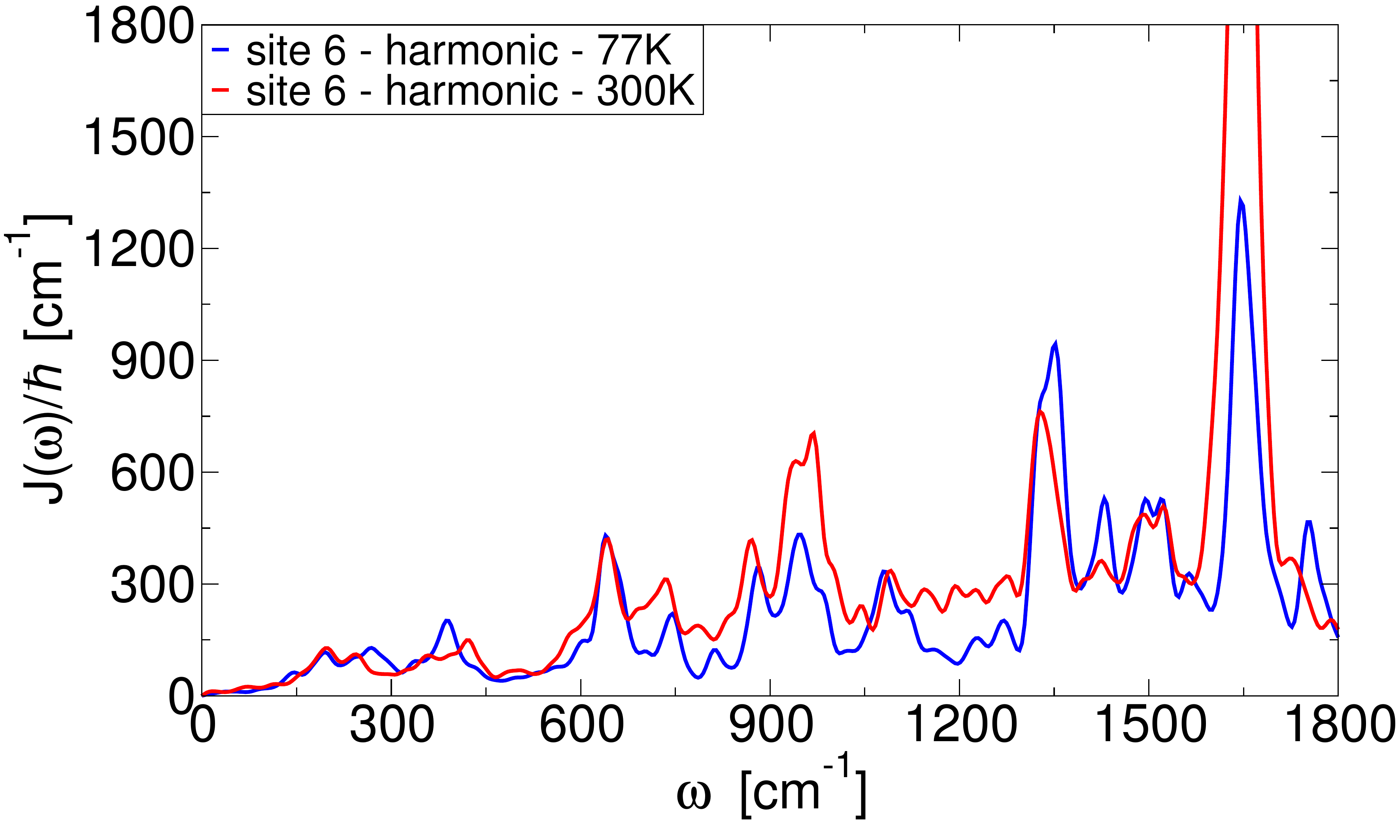}
\par\end{centering}

\begin{centering}
\includegraphics[width=1\columnwidth]{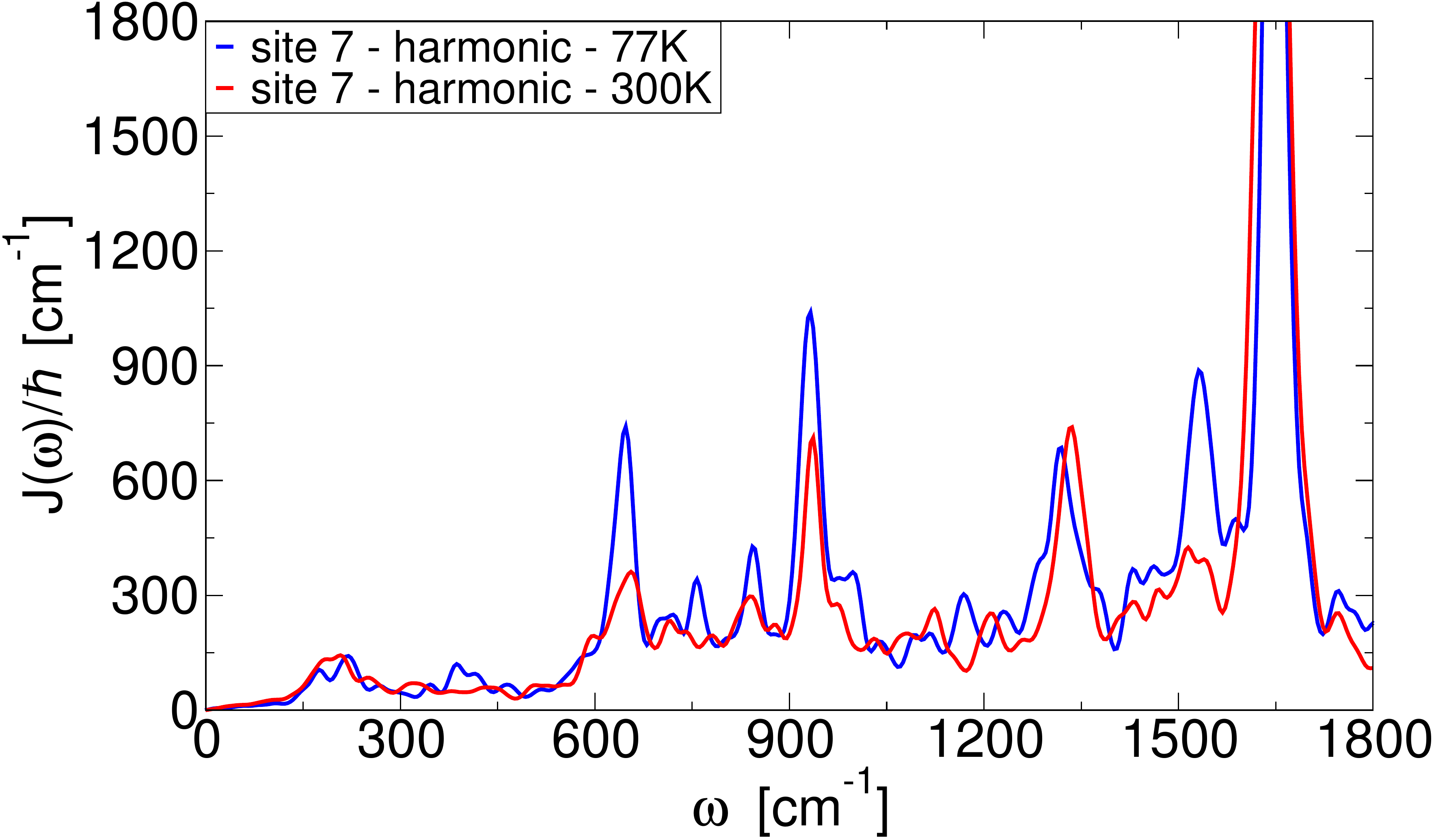}
\par\end{centering}

\caption{Comparison of the asymmetric component of temperature-dependent coupling
density $G_{{\rm asym}}(\omega)=J(\omega)\,;$ for sites 5-7 of the
Fenna-Matthews-Olson complex, obtained with the Harmonic approximation
(As described in the text) at 77K and at 300K. Note that, as described
in the text, the spectral density can be obtained by dividing $J(\omega)$
by $\pi$. \label{fig:Harmonic_SD_site57}}
\end{figure*}

Here, we report $J(\omega)$ the asymmetric component of $G(\omega)$
as described in the text for all temperatures and sites of the FMO
complex. In Figures \ref{fig:Harmonic_SD_site17} and \ref{fig:Harmonic_SD_site57},
we see that the Harmonic approximation gives a roughly temperature
independent quantity for all sites. Further, on this scale the spectral
densities do not differ largely between the different sites. 

\end{appendix}

\bibliographystyle{apsrev4-1}
%
\end{document}